%% file: LP11summary.tex
\documentclass[12pt]{article}
\usepackage{graphicx}

\newcommand\pubnumber{SLAC--PUB--14612}
\newcommand\pubdate{October, 2011}

\def\SLAC{SLAC, Stanford University\\
    2575 Sand Hill Rd.,  Menlo Park, CA 94025  USA}
\def\doeack{\footnote{Work supported by the US Department of Energy,
                     contract DE--AC02--76SF00515.}}

\def\Title#1{\begin{center} {\Large #1 } \end{center}}
\def\Author#1{\begin{center}{ \sc #1} \end{center}}
\def\Address#1{\begin{center}{ \it #1} \end{center}}

\newcommand\pubblock{\rightline{\begin{tabular}{l} \pubnumber\\
         \pubdate \end{tabular}}}
\newenvironment{Abstract}{\begin{quotation} \begin{center}
                       ABSTRACT
     \end{center}\bigskip  }{\end{quotation}}
\newenvironment{Presented}{\begin{quotation} \begin{center} 
             PRESENTED AT\end{center}\bigskip 
      \begin{center}}{\end{center} \end{quotation}}

\def\Acknowledgements{\bigskip  \bigskip \begin{center} \begin{large}
             \bf ACKNOWLEDGEMENTS \end{large}\end{center}}
\input mymacros.tex

\begin{document}
\begin{titlepage}
\pubblock

\vfill
\Title{Summary of Lepton-Photon 2011}
\vfill
\Author{Michael E. Peskin\doeack}
\Address{\SLAC}
\vfill
\begin{Abstract}
In this lecture, I summarize developments presented at the Lepton Photon 2011 
conference and give my perspective on the current situation in high-energy
physics.
\end{Abstract}
\vfill
\begin{Presented}
Lepton Photon 2011\\
Tata Institute for Fundamental Research, Mumbai, India\\
August 21--27, 2011
\end{Presented}
\vfill
\end{titlepage}
\def\thefootnote{\fnsymbol{footnote}}
\setcounter{footnote}{0}
\quad

\tableofcontents

\newpage

\section{Introduction}

I am grateful to the organizers of Lepton Photon 2011 for providing us
a very pleasant and simulating week in Mumbai.  This year's 
 Lepton Photon conference has covered the full range of subjects that fall
within the scope of high-energy physics, including connections to cosmology,
nuclear physics, and atomic physics.  The experiments that were discussed
detect particles ranging in energy from radio frequencies to EeV. Other
speakers thanked the organizers for inviting them to lecture at 
the Lepton Photon conference, but I have to regret that the organizers
of LP11 have given me
an impossible task -- to summarize the results presented and make sense
of them.

This review will then necessarily cover only a subset of what was presented
at the conference.  I will emphasize results in which I have some personal
interest or on which I would like to state a sharp opinion or perspective.   
I apologize to those
speakers whose work is not discussed here.  This review was completed in 
October 2011.   The situation in high-energy physics, especially in 
topics involving TeV energies and LHC, is evolving rapidly.  I am sure that 
some of the opinions expressed here will soon be out date.

This conference is the biennial festival of the enthusiasts of leptons and
photons -- and of quarks and gluons.  This year, we have much to celebrate.
The Large Hadron Collider at CERN is finally launched.  The machine and its
experiments are performing beyond expectations and are carrying us into new 
and unexplored realms of physics at short distances.  At the other extreme,
the links between the two most fundamental sciences, 
particle physics and cosmology, are stronger than ever and are leading us
to new insights at the largest accessible distances, and even beyond.

And yet, we are apprehensive.  Our field seems to be approaching a
definite point of reckoning.  We are headed somewhere at breakneck speed,
but are we racing toward enlightenment, or to disillusionment and chaos?

In reviewing the highlights of LP11, I cannot avoid this question.

First, though, I will discuss some of the wonderful things that we have 
learned in Mumbai.

\section{The Universe}

The cosmological Standard Model -- the flat expanding 
universe containing  5\% baryons, 
20\% dark matter, and 75\% dark energy -- continues to be put on a 
firmer footing.  I refer you to Scott Dodelson's lecture~\cite{Dodelson}
for a review of the current experimental programs aiming to clarify the 
nature of dark energy. For me, dark energy is a total mystery.  It seems to 
be a nonzero cosmological constant.  The mystery is not why this cosmolgical
constant is constant or nonzero, but rather why it is not at least 60 orders
of magnitude larger.  

Another piece of the mystery is the success of inflationary cosmology.
Observations of cosmic structure require that the initial conditions
for our universe contained an almost scale-invariant 
spectrum of adiabatic perturbations, whose parameters were uniform far
outside the horizon.  The theory of inflation predicted such a 
spectrum and continues to
be the only straightforward explanation of it.  Inflation links to the 
dark energy problem in a puzzling way.  Inflation requires that the universe
actually contained a cosmological constant 100 orders of magnitude greater
than the present one early in its history.  The exit from this state of 
exponential expansion and the consequent heating of the universe is what we 
call the Big Bang.  We do not understand how this prior state is related to
the dark energy that we see today.

However mysterious it might be, inflation has predictions that 
agree with observations.  The most striking effects are in the spectrum of 
fluctuations of the cosmic microwave background.  We will soon have much 
more precise observations of the cosmic microwave background from the 
Planck spacecraft.  At LP11, 
Francois Bouchet~\cite{Bouchet} reviewed the design of
Planck and the expectations for its physics program.  However, the main
cosmological results from Planck will not be available until Lepton Photon
2013.

%%%%%%%%%%%%%%%%%%%%%%%%%%%%%%%%%%%%%%%%%%%%%%%%%%%%%%%%%%%%%%%%%%%%%%%%%
\begin{figure}
\begin{center}
\includegraphics[height=2.5in]{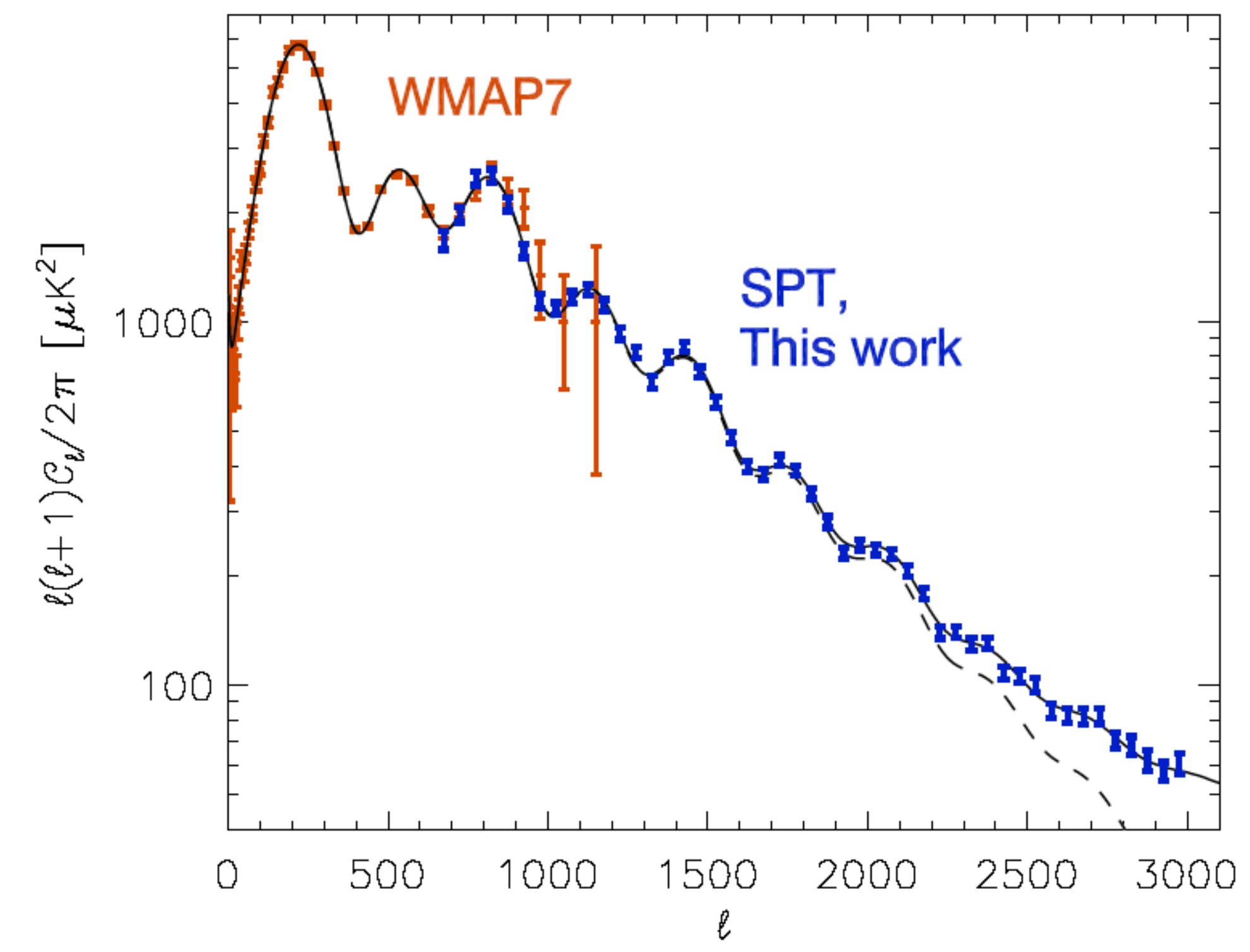}
\caption{The 2011 measurement of the fluctation spectrum of the Cosmic
Microwave Background by the South Pole Telescope~\cite{SPT}.}
\label{fig:SPT}
\end{center}
\end{figure}
%%%%%%%%%%%%%%%%%%%%%%%%%%%%%%%%%%%%%%%%%%%%%%%%%%%%%%%%%%%%%%%%%%%%%%%%%%%

In the meantime, though, we have a new result on the Cosmic Microwave 
Background from the South Pole Telescope~\cite{SPT}, shown in 
Fig.~\ref{fig:SPT}.  The South Pole Telescope has extended the observation
of acoustic peaks in the CMB observed by WMAP and other detectors by another
6 peaks, reinforcing the agreement with astrophysical calculation.  One sees 
by eye that the peaks are not uniform in size.  The first, third, fifth
peaks, and so on, are slightly higher than the general trend; the second and 
fourth peaks are slightly lower.  This observation has a simple and well-known
explanation:  The primordial plasma that radiated the CMB had two components,
a gas of Hydrogen and Helium, and a gas of weakly interacting dark matter.
At the first and the other odd peaks, these components slosh coherently,
at the even peaks, they are sloshing in opposite directions.  This picture
was the basis for the measurement of the dark matter density of the universe
by WMAP~\cite{WMAPthree}, and it remains one of the compelling arguments for 
existence of dark matter. With new results from the Atacama Cosmology
Telescope~\cite{ACT}, Dodelson explained, all aspects of the
current synthesis of cosmology --- the flatness, the baryons, 
the dark matter, the dark 
energy, and the scale-invariant spectrum of perturbations -- are independently
observed and measured in the CMB data~\cite{Dodelson}.

%%%%%%%%%%%%%%%%%%%%%%%%%%%%%%%%%%%%%%%%%%%%%%%%%%%%%%%%%%%%%%%%%%%%%%%%%
\begin{figure}
\begin{center}
\includegraphics[height=1.2in]{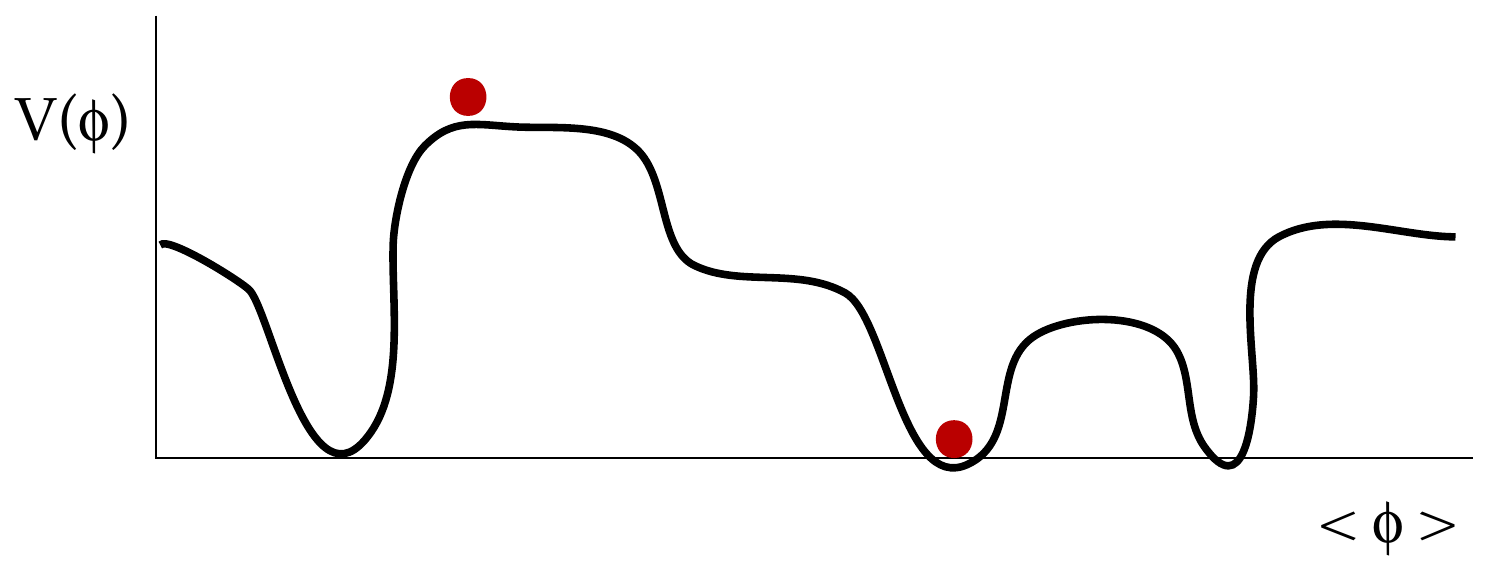}
\caption{Typical scalar field potential of a quantum field theory model of 
inflation.  The energy density at the plateau is $V =  \lambda M^4$,
where $M$ is a very high energy scale, perhaps as large as the Planck mass
$m_\Pl$.  The Hubble constant during inflation is $H_0 = (8\pi \lambda/3)^{1/2}
M^2/m_\Pl$.}
\label{fig:inflate}
\end{center}
\end{figure}
%%%%%%%%%%%%%%%%%%%%%%%%%%%%%%%%%%%%%%%%%%%%%%%%%%%%%%%%%%%%%%%%%%%%%%%%%%%

This is truly excellent, but it comes with a shadow, one emphasized since
the 1980's by Andrei Linde~\cite{Linde}.  To have inflation, there must
be ranges of scalar field values in which the scalar potential is very 
flat, as shown in Fig.~\ref{fig:inflate}.
  Regions of space with such flatness inflate exponentially and
dominate the volume of the universe.  Regions exit inflation by falling
off the plateaus.  If the theory of Nature has multiple vacuum states --
all attractive theories of grand unification have potentials with multiple
minima -- all of these vacua will be populated somewhere in the universe.
It is very important to us which of these minima our local universe sits in.
This choice determines the particles and couplings that we see in particle
physics.  But the perspective of inflation tells us that that choice could be
dictated by {\it history} rather than by {\it necessity}.

The conclusion brings us to the great divide that cut through
nineteenth-century science between physics and chemistry on one side 
and biology and geology on the other.  Physicists, exemplified by Laplace, 
believed that Nature is described by fundamental equations which have a 
unique solution and therefore lead to definite predictions.  Biologists, 
after Darwin, believed that Nature is the logical result of random processes,
with the final solution determined by accidents of history.  Linde argues
that, even for physics, there are basic limits to Laplacian thinking about
the ultimate laws of Nature.  We do not know where these limits will appear,
but, eventually, we must run up against them.

I will return to this point at the end of the lecture.

\section{Cosmic Rays}

The origin of cosmic rays is naturally a subject that should intrigue 
high-energy physicists.  For a long time, however, most research in 
high-energy physics ignored the products of cosmic accelerators, preferring
to concentrate on results from artificial accelerators.  More recently, 
some fascinating problems involving cosmic rays have brought this
subject back into the mainstream.

The first of these problems is the identity of the highest energy 
cosmic rays.  The cosmic ray spectrum above the `knee' at $10^{16}$~eV is 
well known to fall roughly as $dN/dE \sim E^{-3}$. There is no known 
mechanism for producing such high energy cosmic rays within the galaxy, so 
cosmic rays in this range are assumed to be of extragalactic origin.
Ralph Engel~\cite{Engel} explained to us that the story of this region of
the cosmic ray spectrum has changed dramatically over the past 4 years.

\begin{figure}
\begin{center}
\includegraphics[height=2.0in]{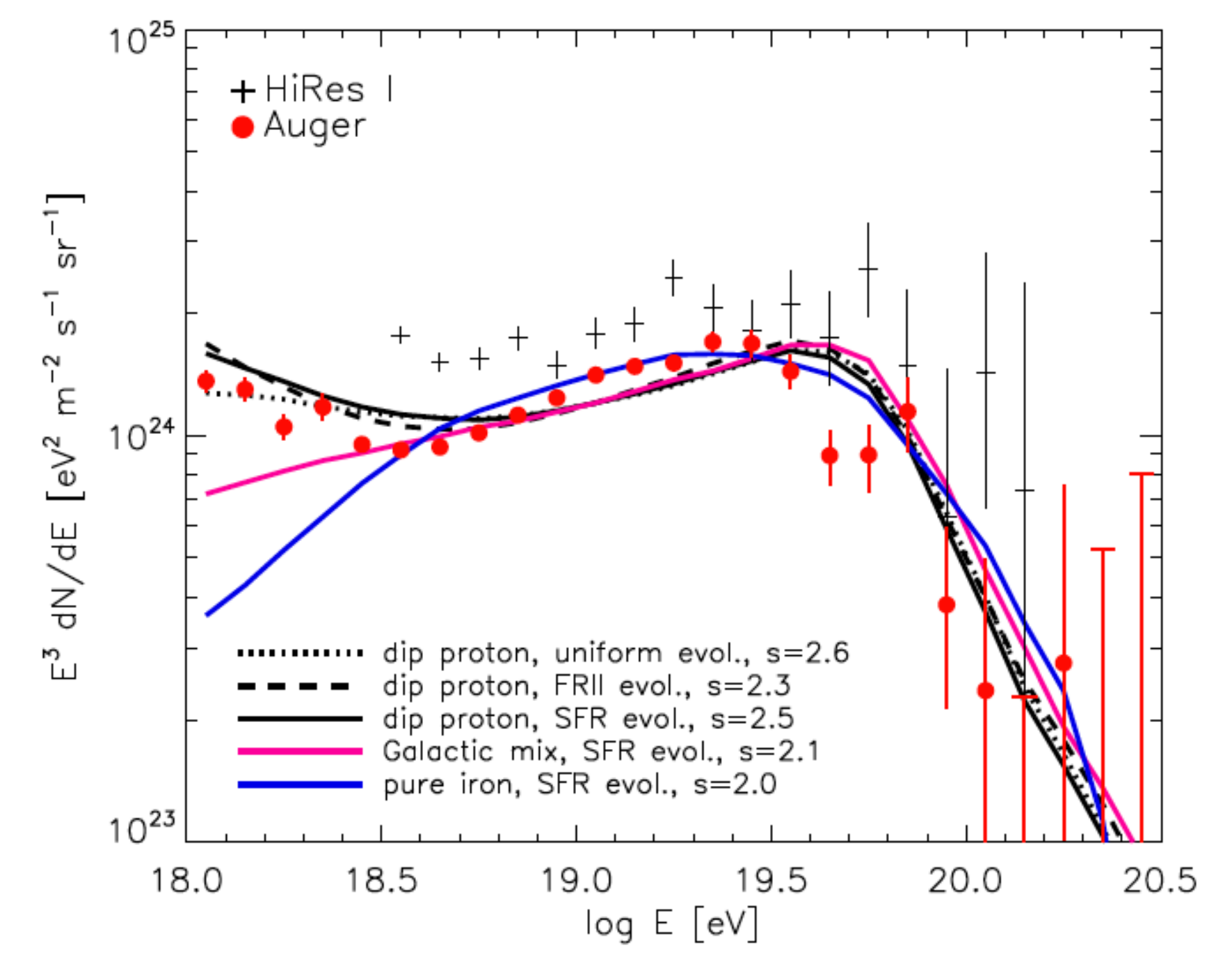}
\caption{The energy spectrum of the highest energy cosmic rays, $E^3 dN/dE$, 
showing the observed cutoff at the highest energies, from the review
\cite{Olinto}.}
\label{fig:GKZ}
\end{center}
\end{figure}

In 2007, the Pierre Auger Collaboration announced evidence for a correlation
between the arrival directions of extremely high energy cosmic rays, above
$10^{18}$~eV, and the directions to nearby Active Galactic
Nuclei~\cite{AugerScience}.  Given the
strength of intergalactic magentic fields, these rays would point back to their
sources only if they were of charge $Z = 1$, that is, protons rather than 
heavy nuclei.  The spectrum of cosmic ray protons is cut off by the 
GKZ effect~\cite{GKZ}:  A sufficiently energetic proton sees 
cosmic microwave background
photons boosted to 300 MeV gamma rays, where they can excite the $\Delta$
resonance and thus cause the proton to lose energy by emission of a pion.
The cutoff in the proton spectrum is predicted to be at $E = 3\times 10^{19}$
GeV.  Indeed, a sharp cutoff in the cosmic ray spectrum is seen at this 
location, as shown in Fig.~\ref{fig:GKZ}.

%%%%%%%%%%%%%%%%%%%%%%%%%%%%%%%%%%%%%%%%%%%%%%%%%%%%%%%%%%%%%%%%%%%%%%%%%
%%%%%%%%%%%%%%%%%%%%%%%%%%%%%%%%%%%%%%%%%%%%%%%%%%%%%%%%%%%%%%%%%%%%%%%%%%%

%%%%%%%%%%%%%%%%%%%%%%%%%%%%%%%%%%%%%%%%%%%%%%%%%%%%%%%%%%%%%%%%%%%%%%%%%
\begin{figure}
\begin{center}
\includegraphics[height=1.6in]{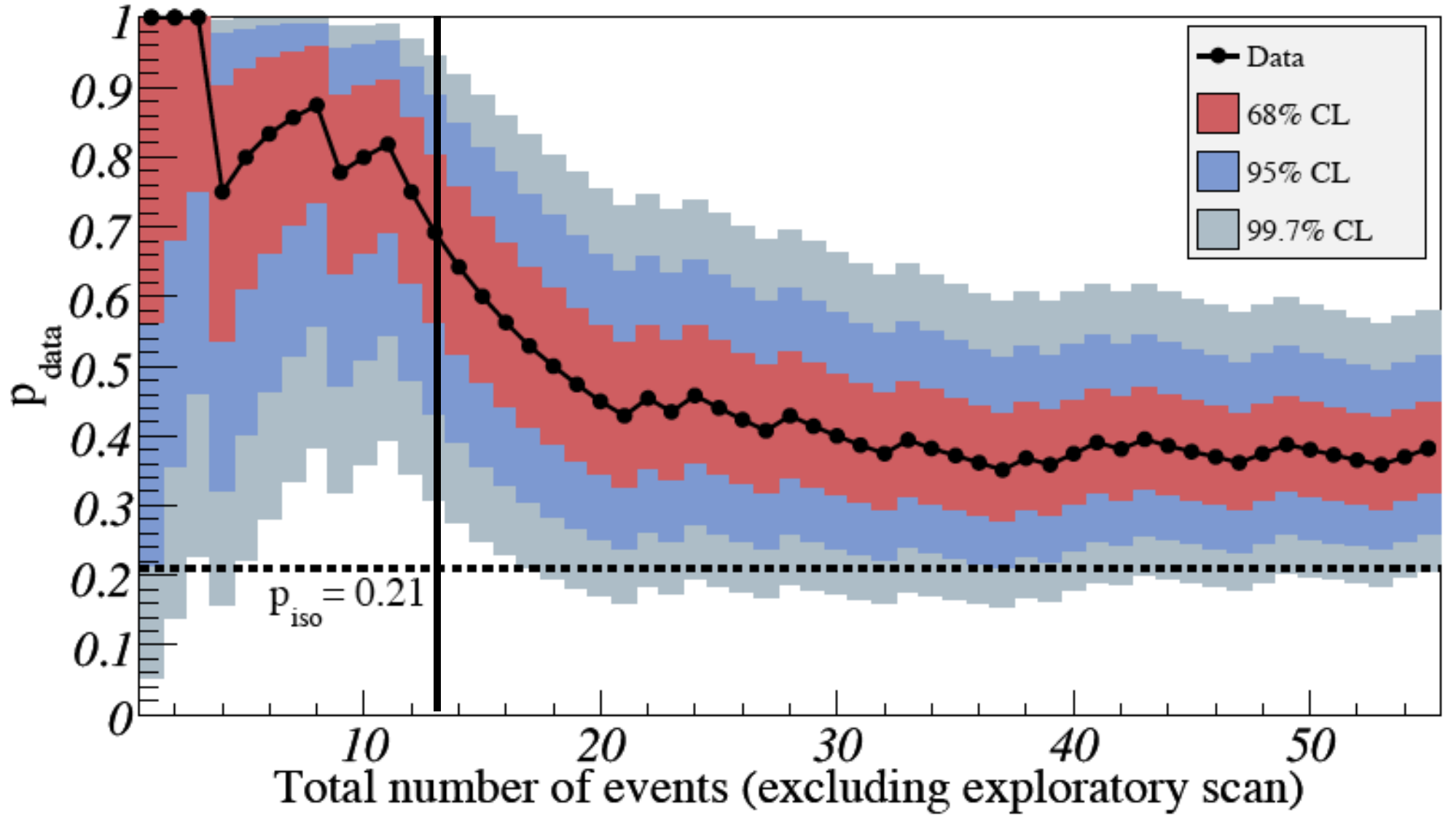}
\caption{The degree of correlation of the arrival directions of 
   cosmic rays above an energy of $10^{18}$~GeV, as observed by the 
 Pierre Auger Collaboration, shown as a function of time~\cite{AugerCorr}.
   The vertical line marks the date of the Auger paper ~\cite{AugerScience}.}
\label{fig:Augerdown}
\end{center}
\end{figure}
%%%%%%%%%%%%%%%%%%%%%%%%%%%%%%%%%%%%%%%%%%%%%%%%%%%%%%%%%%%%%%%%%%%%%%%%%%%
%%%%%%%%%%%%%%%%%%%%%%%%%%%%%%%%%%%%%%%%%%%%%%%%%%%%%%%%%%%%%%%%%%%%%%%%%
\begin{figure}
\begin{center}
\includegraphics[height=1.8in]{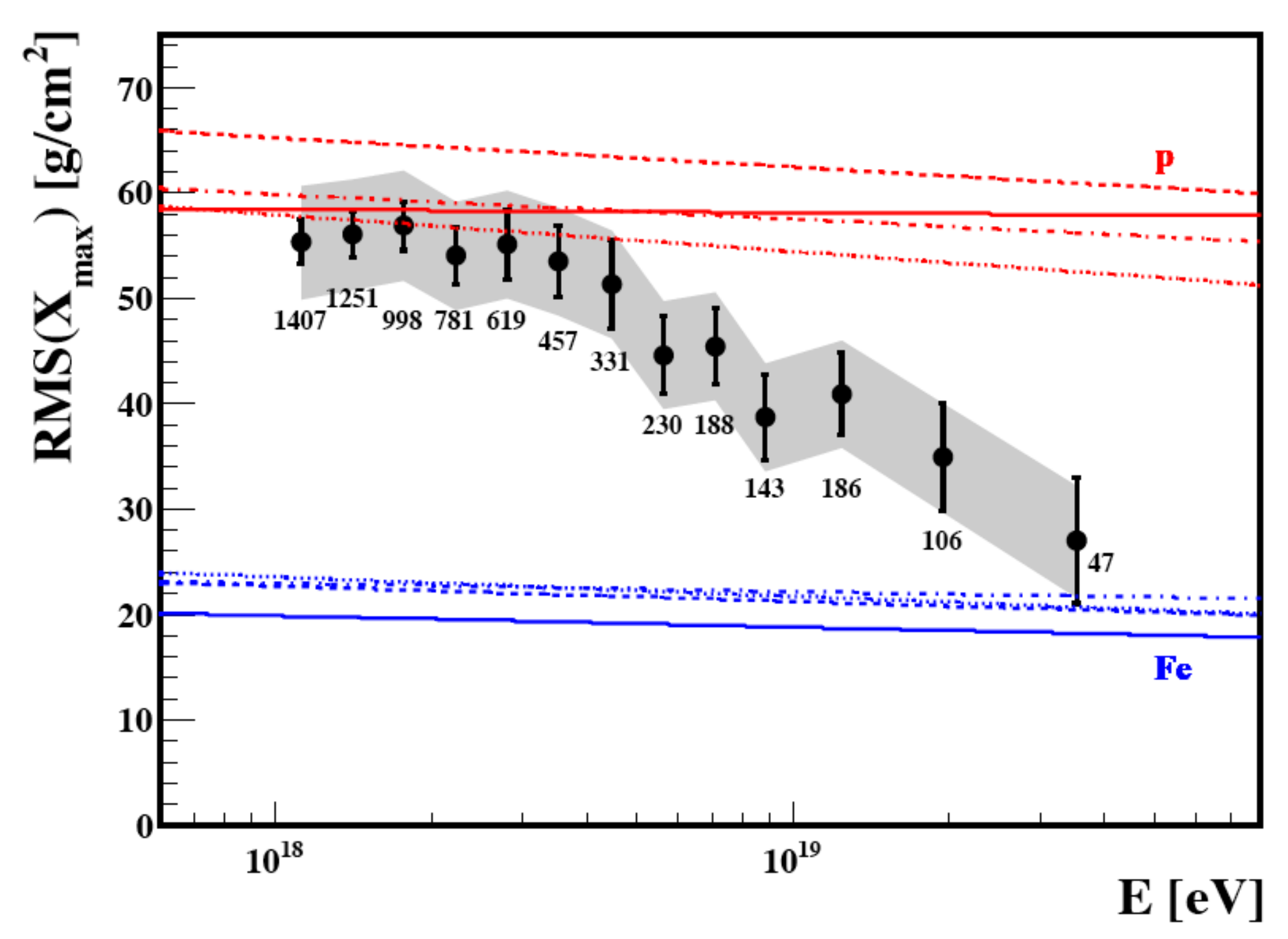}
\caption{Dependence of the RMS fluctuation of the depth of the shower
     maxima of high energy cosmic rays, from~\cite{AugerX}.
  This parameter is predicted to 
     be much smaller for showers initiated by protons than for those
      initiated by heavy nuclei. }
\label{fig:AugerXM}
\end{center}
\end{figure}
%%%%%%%%%%%%%%%%%%%%%%%%%%%%%%%%%%%%%%%%%%%%%%%%%%%%%%%%%%%%%%%%%%%%%%%%%%%

Over the past few years, this beautiful and coherent picture has fallen
apart.  First, the correlation of high-energy cosmic rays with AGNs has 
decreased, as shown in Fig.~\ref{fig:Augerdown}~\cite{AugerCorr}.  
The Telescope Array 
in the Nothern Hemisphere did not support 
the earlier Auger result but rather gave a weak correlation similar to that 
found by Auger now~\cite{Engel}.  The Auger experiment has also tried
to measure the composition of high-energy cosmic rays directly by measuring 
the position in the atmosphere of the shower maximum ($X_{max}$)
and using this to 
estimate the cross section of the species initiating the shower. 
Auger and the HiRes Collaboration disagree on the interpretation of the
results for the shower maximum, even though the actual
measurements are not very different~\cite{AugerX,HiResX}.  However, Auger
also reports a marked decrease in the RMS fluctuation of $X_{max}$ between
$3\times 10^{18}$ and $3\times 10^{19}$ eV, shown in Fig.~\ref{fig:AugerXM}.
Heavy nuclei have a more uniform shower development from event to event,
so this is evidence that the composition of the cosmic rays changes from 
protons to Fe nuclei over this range.

There is no simple theory such as that of  GKZ for the energy cutoff in 
Fe nuclei.  Detailed calculations estimate the cutoff as happening as the
same point as for protons, $3\times 10^{19}$ eV {\it per nucleus}.  The 
energy loss mechanism is the disruption of nuclei  by excitation of the 
giant dipole resonance.  However, because the energies of individual
nucleons  are
lower by a factor 56, the cosmic microwave background is not sufficiently 
energetic to excite this resonance.  The dominant effect is expected to 
come from the Cosmic Infrared Background, which is not of cosmological 
origin.  If the cutoff did result from this mechanism, the highest 
energy cosmic rays reaching earth would be the light nuclei that are the 
results of this disruption process.  A more likely hypothesis is that 
cosmic accelerators, for example, shock fronts in large clusters of 
galaxies, are limited in the energy they can pump into cosmic rays.  If the
energy limit for protons is $10^{18}$ GeV, the energy limit for Fe will be
26 times higher, and Fe nuclei will dominate at the highest 
energies~\cite{Engel}.

To resolve this problem, we not only need higher-statistics measurements of 
cosmic rays air showers but also a more accurate theory of the  proton-nucleus
and nucleus-nucleus cross section at very high energy.  LHC data on the
proton-proton cross sections at 7 and 14 TeV and on the multiplicity 
distribution in minimum-bias events will provide important information for
modelling these effects.

\begin{figure}
\begin{center}
\includegraphics[height=2.1in]{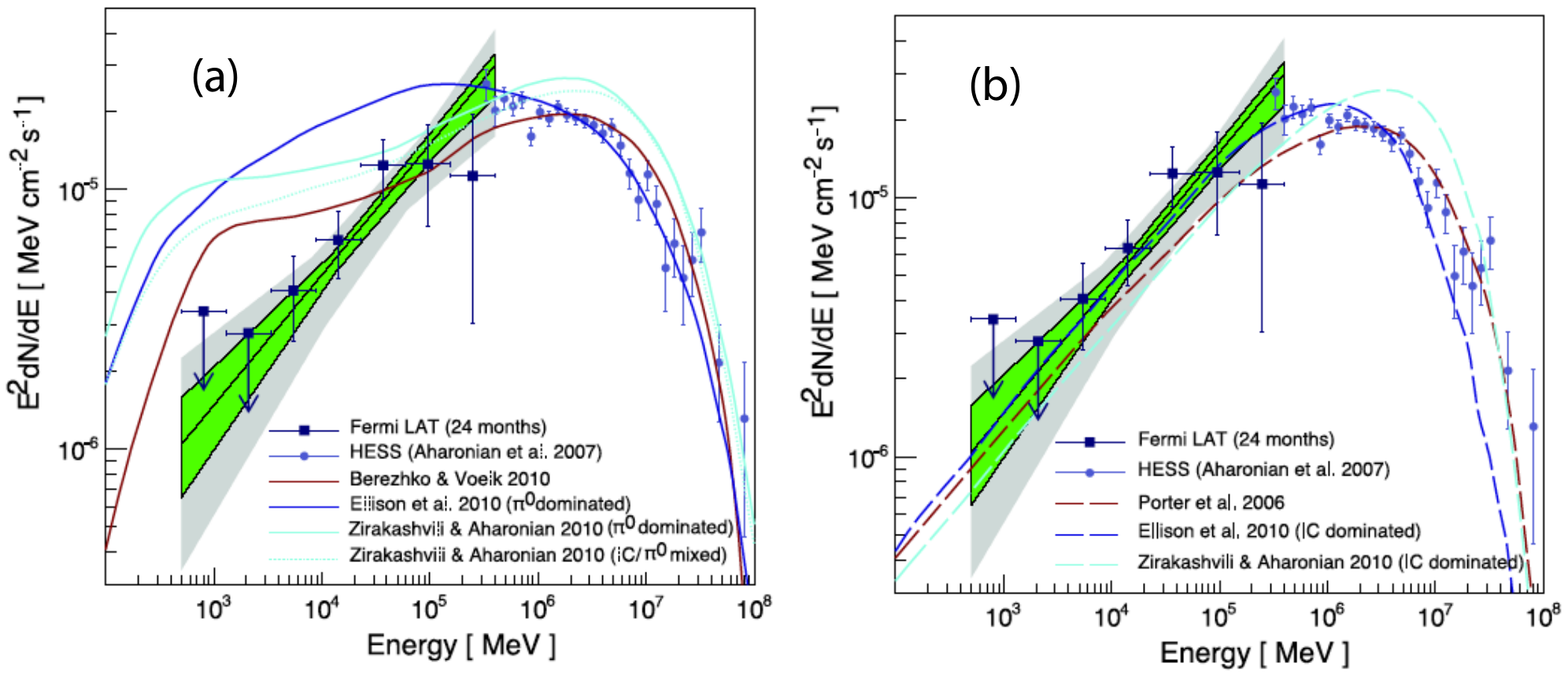}
\caption{Test of the spectrum of gamma rays from the supernova
  remnant RX~J1713.--3946 measured by the Fermi GST and the Hess Telescope
against models in which (a) protons  and (b) electrons are the primary
species
   accelerated by this object, 
    from \cite{FermiSNR}.}
\label{fig:Fermi}
\end{center}
\end{figure}

As this situation has become murkier, our understanding of lower-energy 
cosmic rays created by galactic sources has become clearer.  An interesting
type of source to study is a supernova remnant lying close to a 
molecular cloud.  Cosmic rays from the source irradiate the cloud, producing
gamma rays that point back to the cloud
 and identify the system. As Simona Murgia
explained to us~\cite{Murgia},  the spectrum of
these gamma rays is characteristically different for
sources in which the primary radiation is protons, which produce gammas by 
$\pi^0$ decay, or electrons, which 
produce gammas by inverse Compton scattering.
The difference is especially clear in the GeV energy range that can be 
studied
by the Fermi Gamma-ray Space Telescope.  For example, Fig.~\ref{fig:Fermi}
shows the Fermi gamma ray spectrum of the supernova remnant RX J1713.--3946,
together with HESS data on the same object~\cite{FermiSNR,HessSNR}.  
The observations clearly 
favor a model in which the primary species accelerated is electrons.   Fermi
has also identified sources for which the primary species accelerated is 
protons.

%%%%%%%%%%%%%%%%%%%%%%%%%%%%%%%%%%%%%%%%%%%%%%%%%%%%%%%%%%%%%%%%%%%%%%%%%
%%%%%%%%%%%%%%%%%%%%%%%%%%%%%%%%%%%%%%%%%%%%%%%%%%%%%%%%%%%%%%%%%%%%%%%%%%%

\begin{figure}
\begin{center}
\includegraphics[height=1.8in]{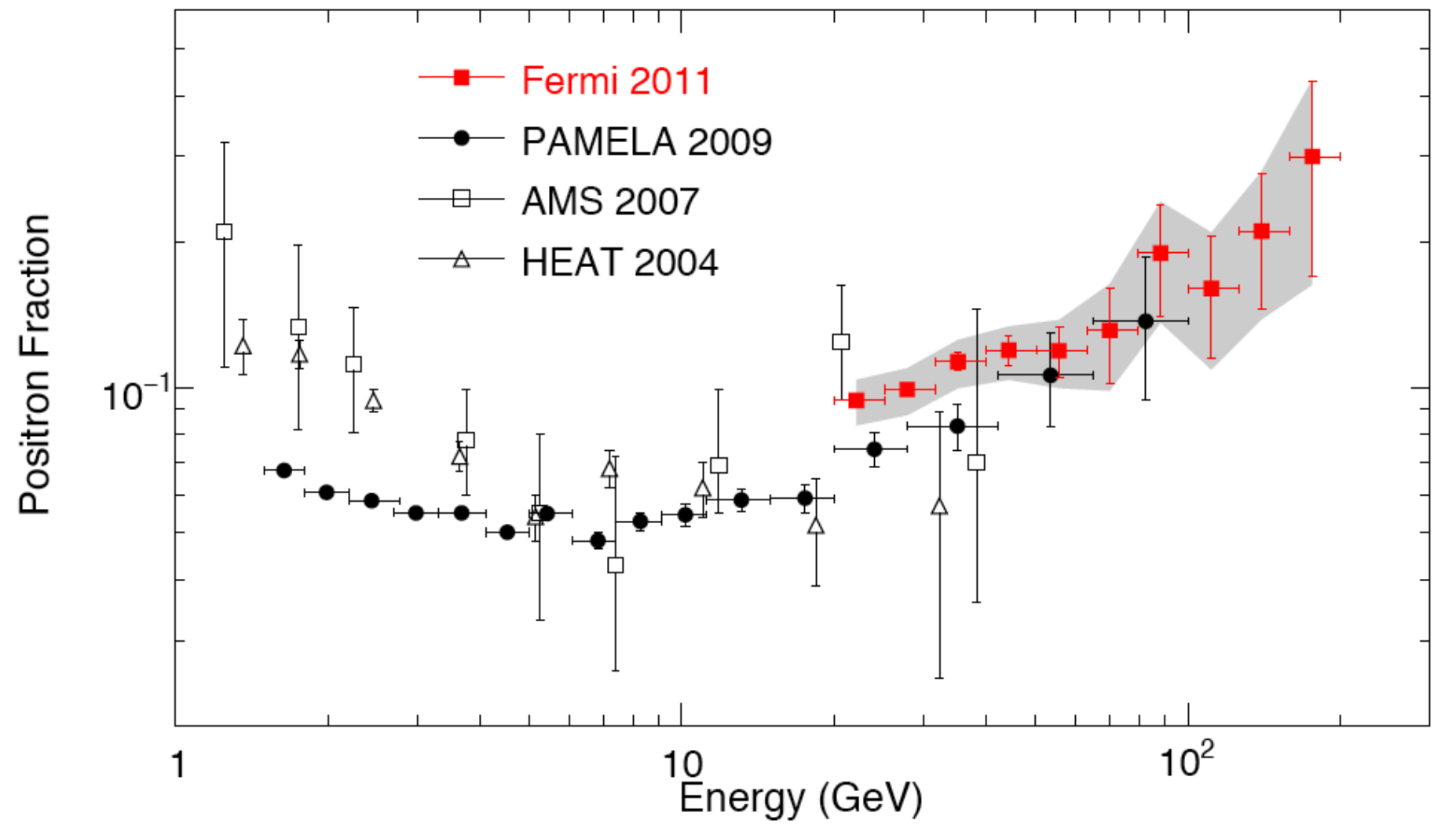}
\caption{Comparision of the 2011 measurement of the $e^+/e^-$ ratio in 
     cosmic rays by the Fermi GST to the earlier measurement by the 
     PAMELA satellite~\cite{Pamela}, 
    from \cite{PamelaF}.}
\label{fig:Pamela}
\end{center}
\end{figure}

There is one more very interesting result this year from Fermi.  Despite the 
fact that Fermi is a non-magnetic detector, the Fermi LAT Collaboration was
able to measure the position/electron ratio in cosmic rays by comparing 
time intervals of the orbit in which the Earth's magnetic field dominantly 
throws positively charged particles into the satellite to time intervals in 
which the Earth's magnetic field dominantly selects negatively charged
particles~\cite{PamelaF}.  The results confirm the celebrated results from 
PAMELA~\cite{Pamela} showing
a rise in the $e^+/e^-$ ratio  in the 100 GeV region;
 see Fig.~\ref{fig:Pamela}.

%%%%%%%%%%%%%%%%%%%%%%%%%%%%%%%%%%%%%%%%%%%%%%%%%%%%%%%%%%%%%%%%%%%%%%%%%
%%%%%%%%%%%%%%%%%%%%%%%%%%%%%%%%%%%%%%%%%%%%%%%%%%%%%%%%%%%%%%%%%%%%%%%%%%%

\section{Heavy Quark Physics}

In the study of CP violation, we are coming to the end of an era, with 
the BaBar and Belle experiments preparing their final results.  It is 
hard to remember that, as recently as 1999, we had hardly any evidence on 
the mechanism of CP violation in hadronic weak interactions.  In that
year, we only knew that the value of the $K^0$ decay parameter $\epsilon'$
was nonzero, and that there was a first indication of CP violation in 
$B\to J/\psi K^0_S$ from CDF.  

\begin{figure}
\begin{center}
\includegraphics[height=3.3in]{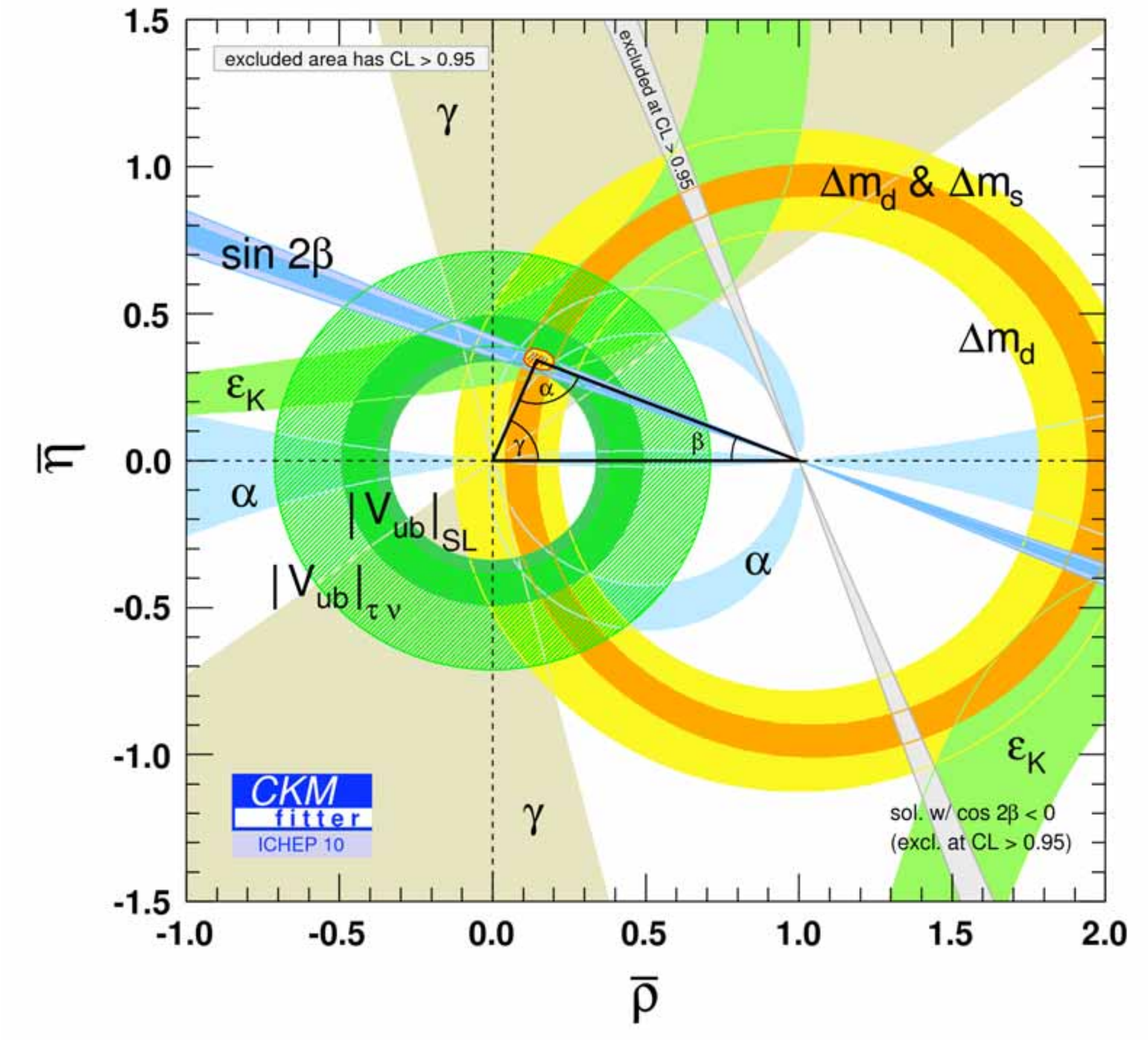}
\caption{Current status of the global fit to the parameters $(\rho,\eta)$ of
the CKM modelCKM, from \cite{CKMfitter}.  The bands show the primary 
contributing measurements.}
\label{fig:CKMglobal}
\end{center}
\end{figure}

The KEK and SLAC B-factories have totally changed
this situation.  We now have a long list of CP-violating effects measured
in $B$ meson decays.  Up to a few discrepancies and tensions, all of this 
data is explained by a single CP-violating parameter, 
the Kobayashi-Maskawa phase in the weak interaction mixing matrix. The 
current status of determinations of the Cabibbo-Kobayashi-Maskawa matrix
was beautifully reviewed at this conference by Tim Gershon~\cite{Gershon}.
 The overall situation is summarized by the global fit to the parameters
$(\rho,\eta)$ of this mixing matrix by the CKMfitter group shown in 
Fig.~\ref{fig:CKMglobal}~\cite{CKMfitter}.  In this figure, we see that 
six different high-precision data sets from $B$ decay and the $\epsilon$
parameter from the $K^0$ system are mutually consistent with a specific
parameter choice $\rho \approx 0.2$, $\eta \approx 0.3$.  
   
%%%%%%%%%%%%%%%%%%%%%%%%%%%%%%%%%%%%%%%%%%%%%%%%%%%%%%%%%%%%%%%%%%%%%%%%%
%%%%%%%%%%%%%%%%%%%%%%%%%%%%%%%%%%%%%%%%%%%%%%%%%%%%%%%%%%%%%%%%%%%%%%%%%%%

\begin{figure}
\begin{center}
\includegraphics[height=2.3in]{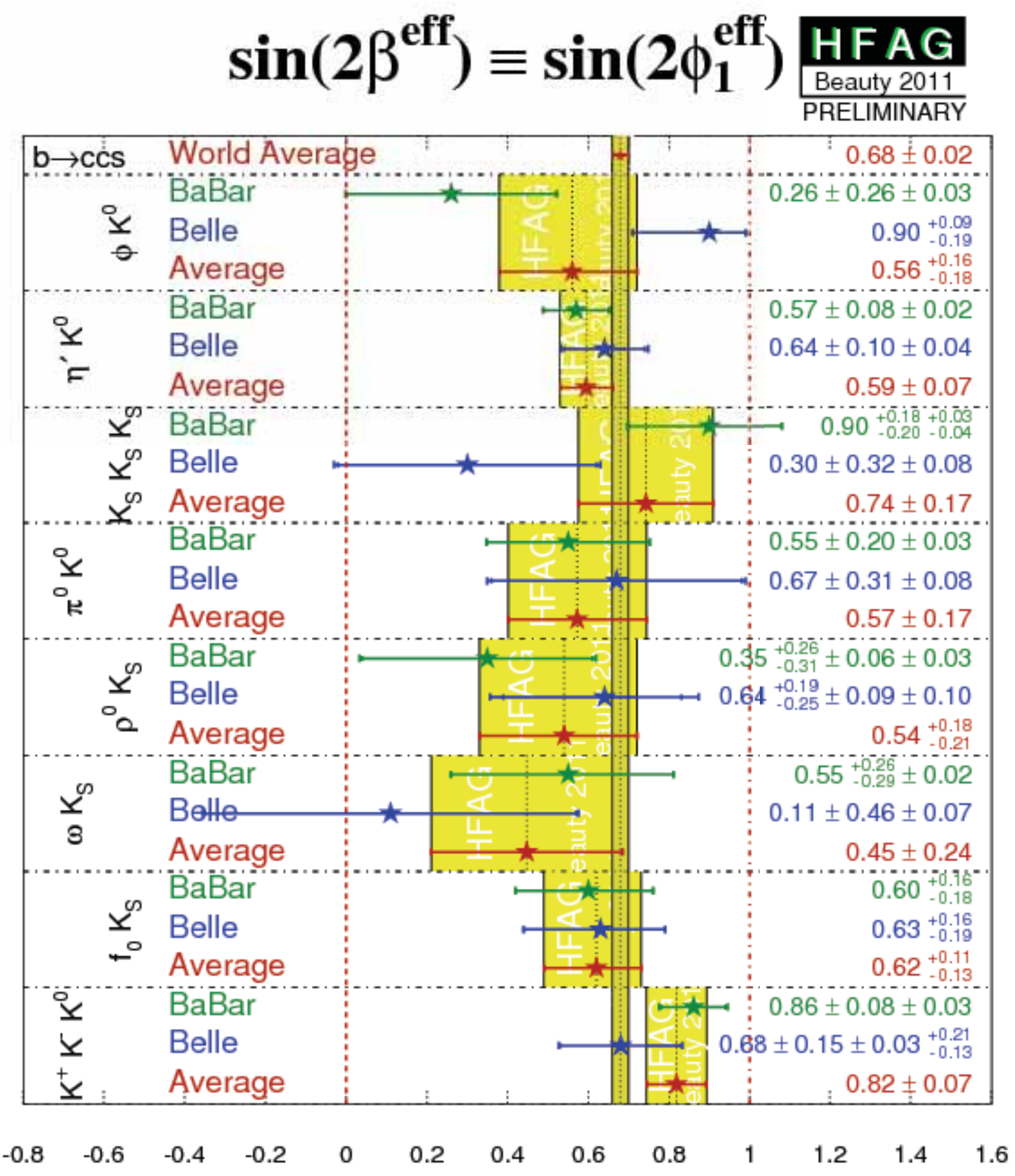}
\caption{Current status of determinations of $\sin 2\beta_{eff}$ 
governing the time-dependent CP asymmetry in $b\to s\bar s s$ decays, 
fom~\cite{HFAGpenguin}.  The results should be compared to the prediction
that this parameter should agree closely with the parameter $\sin2\beta$
measured in $b\to c\bar c s$ decays.}
\label{fig:Bpenguin}
\end{center}
\end{figure}

Over the course of the B factory program, a number of measurements that
seemed to indicate anomalies have come into line as statistics have 
accumulated.  An example is shown in Fig.~\ref{fig:Bpenguin}.  The 
amplitude of time-dependent CP violation in $b\to s\bar s s$ penguin decays
is expected in the CKM to agree with the amplitude $\sin 2\beta$ observed
in $B\to J/\psi K^0_S$.  At the 2003 Lepton Photon conference, there 
seemed to be a large discrepancy between these values~\cite{prevsss}, 
but now the B factory
measurements are in good agreement with the well-determined value 
$\sin 2 \beta = 0.68$~\cite{HFAGpenguin}.

%%%%%%%%%%%%%%%%%%%%%%%%%%%%%%%%%%%%%%%%%%%%%%%%%%%%%%%%%%%%%%%%%%%%%%%%%
%%%%%%%%%%%%%%%%%%%%%%%%%%%%%%%%%%%%%%%%%%%%%%%%%%%%%%%%%%%%%%%%%%%%%%%%%%%

\begin{figure}
\begin{center}
\includegraphics[height=1.6in]{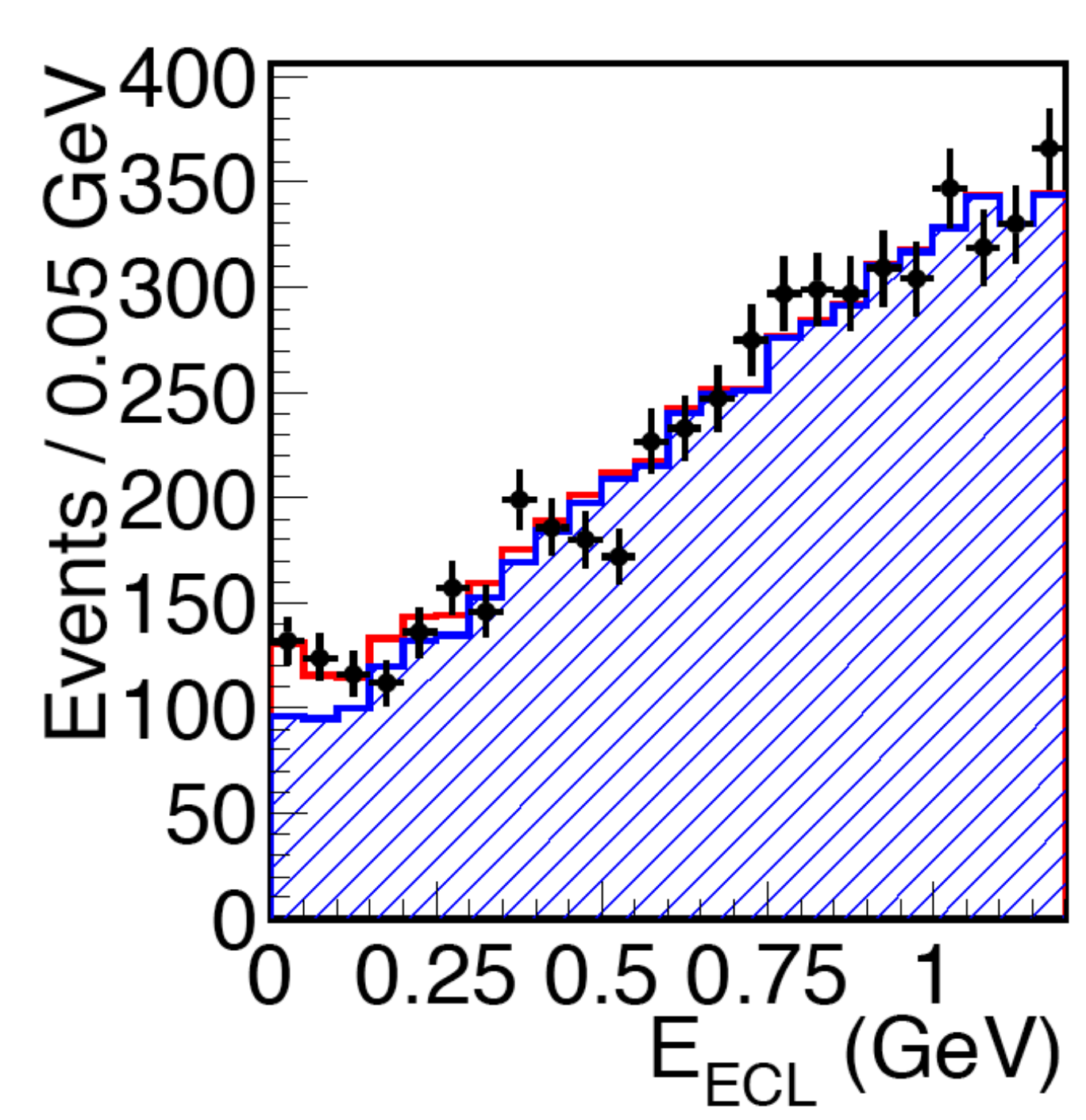}
\caption{Signal (red) and estimated background (blue) in one of the 
Belle measurements of the rate of $B\to \tau \nu$, from 
 \cite{BelleBtaunu}.}
\label{fig:Btaunu}
\end{center}
\end{figure}

One discrepancy that still remains is in the rate of the decay $B\to \tau \nu$.
This is an interesting channel, since it could receive a contribution from 
the tree-level decay of a $b$ quark through charged Higgs boson exchange.
The measurement of this decay was reviewed at this conference by 
Younghoon Kwon~\cite{Kwon}.
The current rate is  2.6~$\sigma$ larger than the prediction of the global
CKM fit.   A similiar, though still less signficant, excess is 
seen in $B\to D\tau\nu$.  These are  very difficult processes to detect.
One of the final fits from the Belle $B\to \tau \nu$ analysis is shown
in Fig.~\ref{fig:Btaunu}~\cite{BelleBtaunu}.  However, there are independent
and compatible results from Belle and 
BaBar~\cite{BelleDtaunu,BaBartaunu,BaBarDtaunu}.  The measurement of 
$B\to \tau\nu$ is extraordinarily difficult for the LHCb experiment, but
it might be possible for that experiment to study $B\to D\tau\nu$ with
higher statistics.

%%%%%%%%%%%%%%%%%%%%%%%%%%%%%%%%%%%%%%%%%%%%%%%%%%%%%%%%%%%%%%%%%%%%%%%%%
%%%%%%%%%%%%%%%%%%%%%%%%%%%%%%%%%%%%%%%%%%%%%%%%%%%%%%%%%%%%%%%%%%%%%%%%%%%

\begin{figure}
\begin{center}
\includegraphics[height=3.6in]{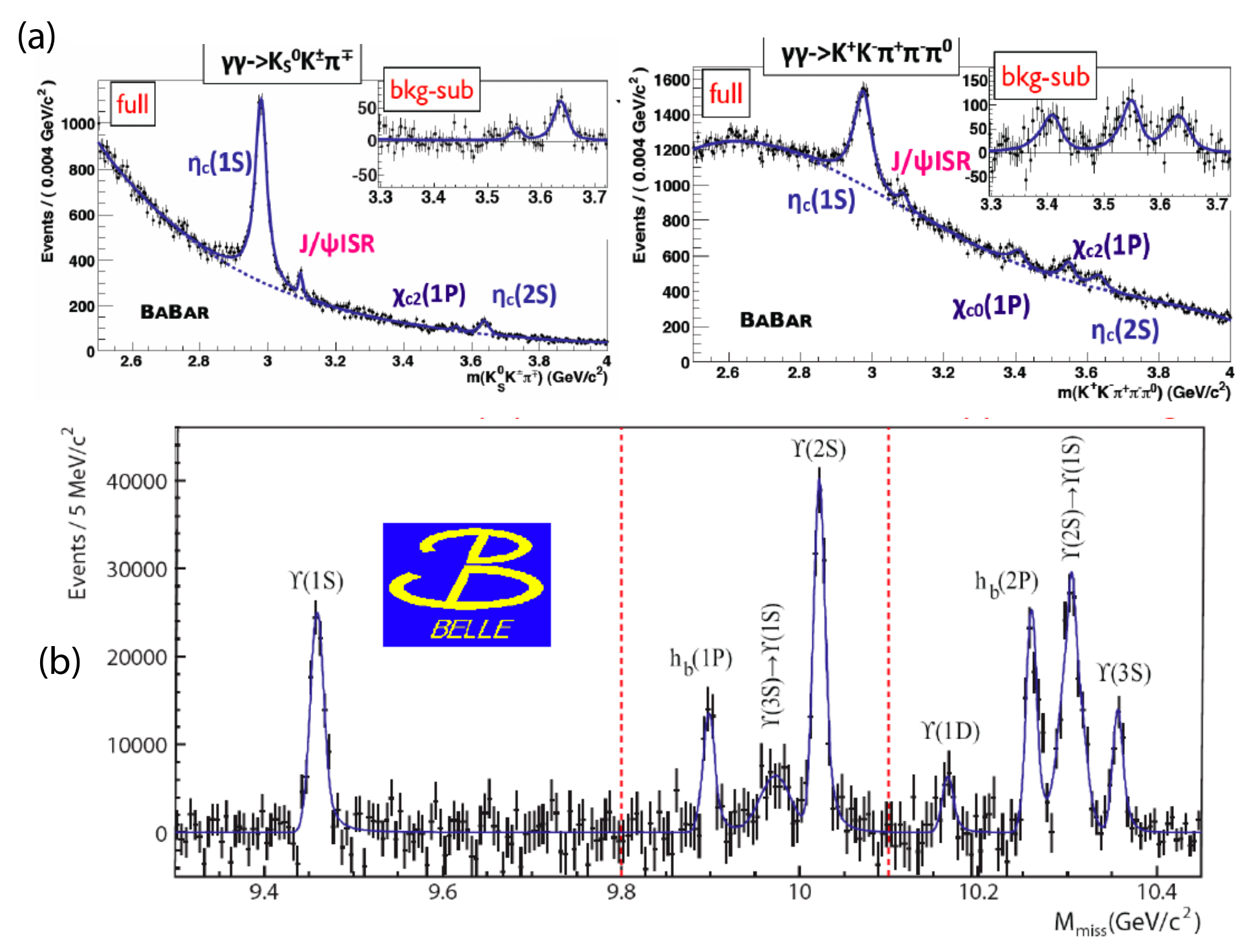}
\caption{(a) The BaBar measurement of the mass distribution in two 
$K\bar K$ final states in $\gamma\gamma$ collisions, showing the 
$C= +1$ states of the charmonium system, from \cite{BaBarspec,Li}.  (b)
The Belle measurement of the missing mass distribution in 
$\Upsilon($5S)
$\to \pi^+\pi^-+X$, showing the $C = -1$ states of the $\Upsilon$ 
spectrum, including the P-wave state $h_b($1P), $h_b($2P), 
from \cite{Bellespec,Li}.}
\label{fig:spectroscopy}
\end{center}
\end{figure}

The B factories have also contributed important results in other domains,
including the discovery of $D^0$--$\bar D^0$ mixing and the exotic states
$X$, $Y$, $Z$ of the charmonium system.  Hai-Bo Li spoke at LP11 on the
current status of hadron spectroscopy, presenting many new results from the
B factories, KLOE, and BES~\cite{Li}.  I do not have space to do justice 
to the details, but I would like to present two particularly beautiful 
plots in Fig.~\ref{fig:spectroscopy}.  The top plot show the reconstruction
by BaBar of two final states of $\eta_C$ decay, observed in $\gamma\gamma$
collisions~\cite{BaBarspec}.
 Both $\eta_c($1S) and $\eta_c($2S) are clearly visible.  The 
bottom plot shows the missing mass spectrum in $\Upsilon($5S)
$\to \pi^+\pi^-+X$,
as observed by Belle~\cite{Bellespec}.  
The entire spectrum of $C= -1$ $b\bar b$ states
appears, including the 1D state and the first two $C$ odd P states.
We will miss such striking and illustrative plots as the major 
results in heavy quark physics come increasingly from hadron colliders.

%%%%%%%%%%%%%%%%%%%%%%%%%%%%%%%%%%%%%%%%%%%%%%%%%%%%%%%%%%%%%%%%%%%%%%%%%
%%%%%%%%%%%%%%%%%%%%%%%%%%%%%%%%%%%%%%%%%%%%%%%%%%%%%%%%%%%%%%%%%%%%%%%%%%%

I now turn to the one truly anomalous result in $B$ physics, the observation
by the D\O\ experiment of a nontrivial asymmetry between $\mu^+\mu^+$ and 
$\mu^-\mu^-$ production in proton-antiproton collisions.  The kinematics of
the muons suggests that the anomaly is dominated by $B$ and $B_s$ meson 
decays. In the CKM model, such inclusive, time-integrated asymmetries
are predicted to have values very close to zero.  However, at LP11, 
Rick van Kooten~\cite{vanKooten} reported an asymmetry
\beq
      A^b_{s\ell} = (- 0.787 \pm 0.172 \pm 0.093)\ \% \ , 
\eeq{Absl}
a value 3.9$\sigma$ discrepant from the small CKM 
expectation~\cite{DzeroAbsl}.

 The measurement by D\O\ is elegant.  Among other features, it uses the
fact that D\O\ frequently reverses the polarity of its magnet to collect
independent data samples that cancel some systematic errors.  Unfortunately, 
there is a systematic asymmetry, also at the 1\%\  level, due to
the fact that the D\O\ detector 
is made of matter rather than antimatter. This is more
difficult to reverse and cancel.  By separating the data into two samples
by impact parameter, D\O\ shows that the bulk of the effect is associated
with small impact parameter.  This means that the anomaly is mainly due to
$B_s$ decays---or to uncancelled prompt backgrounds.

Is it reasonable that there are no anomalies in the $B_d$ system while
order-1 anomalies appear in the $B_s$ system?  Theory favors the idea that
the $B_d$ system should be more sensitive to new physics.  But this is 
only a generic statement.  It is possible that the BaBar and 
Belle experiments were unlucky, while the LHCb experiment will be 
extremely lucky.

%%%%%%%%%%%%%%%%%%%%%%%%%%%%%%%%%%%%%%%%%%%%%%%%%%%%%%%%%%%%%%%%%%%%%%%%%
\begin{figure}
\begin{center}
\includegraphics[height=1.4in]{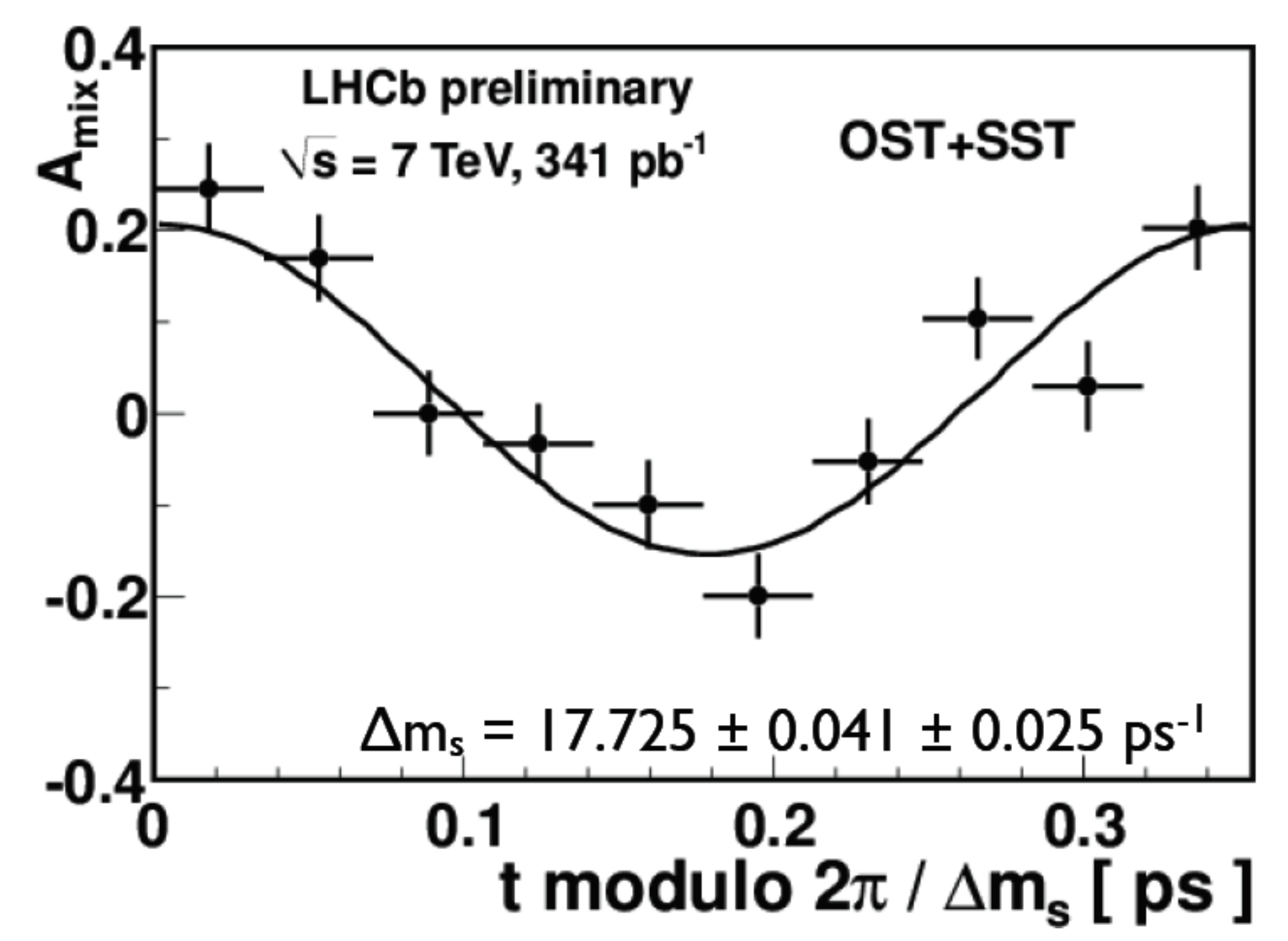}
\caption{The particle-antiparticle oscillation in $B_s$--$\bar B_s$ mixing,
as observed by the LHCb experiment~\cite{Raven}.}
\label{fig:LHCbmix}
\end{center}
\end{figure}
%%%%%%%%%%%%%%%%%%%%%%%%%%%%%%%%%%%%%%%%%%%%%%%%%%%%%%%%%%%%%%%%%%%%%%%%%%%
%%%%%%%%%%%%%%%%%%%%%%%%%%%%%%%%%%%%%%%%%%%%%%%%%%%%%%%%%%%%%%%%%%%%%%%%%
\begin{figure}
\begin{center}
\includegraphics[height=2.1in]{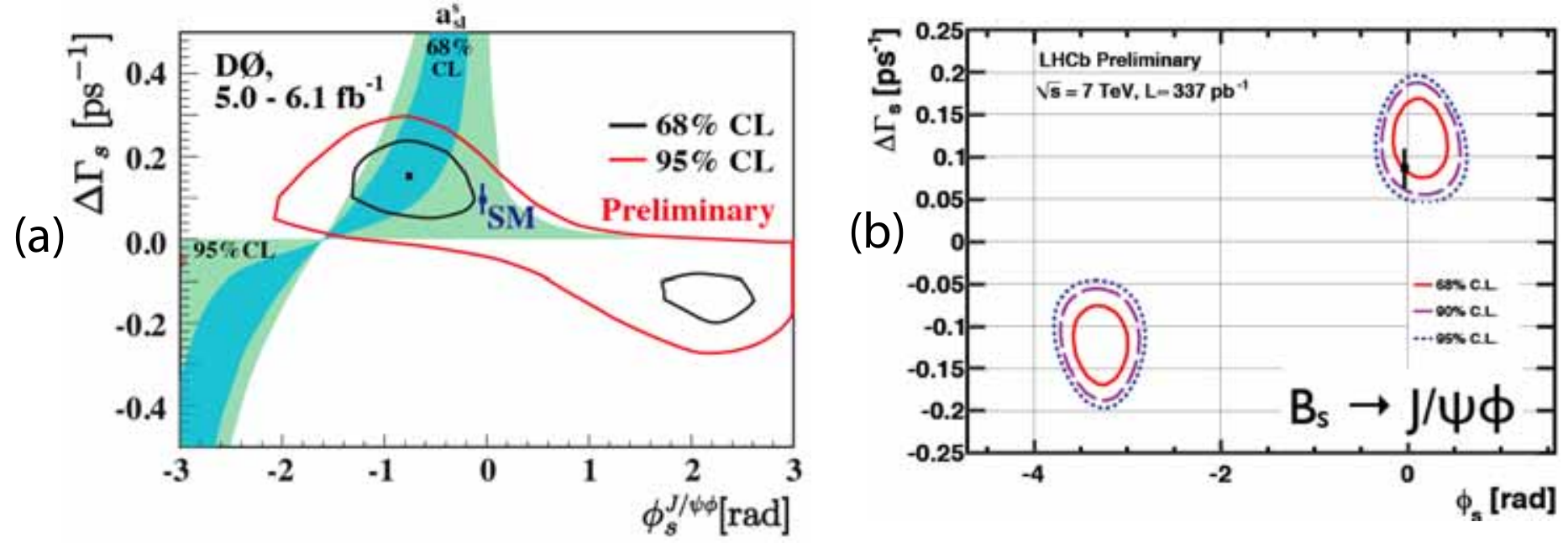}
\caption{(a) Regions of the ($\phi_s,\Delta \Gamma_s$) parameter space
for $B_s$--$\bar B_s$ mixing consistent with measurements of the mixing
by the D\O\ experiment (contours), and consistent with the D\O\ value of 
$A^b_{sl}$, from \cite{DzeroBs}.  (b) Regions of the  
($\phi_s,\Delta \Gamma_s$) parameter space
for $B_s$--$\bar B_s$ mixing consistent with the new LHCb measurement,
from ~\cite{LHCbmix}.} 
\label{fig:testBs}
\end{center}
\end{figure}
%%%%%%%%%%%%%%%%%%%%%%%%%%%%%%%%%%%%%%%%%%%%%%%%%%%%%%%%%%%%%%%%%%%%%%%%%%%
At LP11, Gerhard Raven presented the first results from LHCb~\cite{Raven}, 
including a
remarkable new measurement of the $B_s$--$\bar B_s$ mixing parameters.  The
time-dependent mixing amplitude is shown in Fig.~\ref{fig:LHCbmix}.  For 
those of us who are used to seeing $B_s$--$\bar B_s$ mixing as a rapid
 oscillation that must be Fourier-analyzed to be observed, this is a 
very impressive improvement of the state of the art~\cite{LHCbmix}.
  From these results,
LHCb has extracted values of the mixing phase $\phi_s$ and the off-diagonal
width $\Delta\Gamma$.  If the origin of the D\O\ anomaly were to come 
from $B_s$--$\bar B_s$ mixing---though it could also arise from 
a non-standard direct decay amplitude---the values of these parameters 
should be
in the region shaded blue in Fig.~\ref{fig:testBs}(a)~\cite{DzeroBs}. 
  Instead, LHCb 
data favors the parameter region shown in Fig.~\ref{fig:testBs}(b). The
result is consistent with the CKM model expectation.  Please note also that
the scale in the LHCb plot is magnified by about a factor 2 from that in 
the D\O\ plot.  Over the next year, we will see whether other studies of 
the $B_s$ system will reveal another source of non-standard behavior or 
will refute D\O's claim of an anomaly.

\section{Neutrinos}

There are  many open problems in the area of neutrino masses and mixings.
I will concentrate on only one issue, but one that is especially relevant
to the new developments presented at LP11: {\it  Has $\theta_{13}$ been 
observed?}
This is a crucial question for the future of neutrino physics.  All 
CP-violating effects in neutrino oscillations are multiplied by 
$\sin\theta_{13}$, so all prospects for measuring CP violation in neutrino
oscillations depend on this parameter being large enough.

Up to now, we have only had limits on $\theta_{13}$, 
\beq
     \sin^2\theta_{13} < 0.12 \qquad \mbox{(95\% \ conf.)}
\eeq{stot}
from the Chooz~\cite{Chooz} and MINOS~\cite{MINOSlim} experiments.  At LP11,
however, we saw tantalyzing evidence for $\theta_{13}\neq 0$ from MINOS
and T2K.

The T2K result, described at LP11 by Hiro Tanaka~\cite{Tanaka},
 is quite elegant and persuasive.  The experiment runs a muon neutrino 
beam from the JPARC laboratory in Tokai to Kamiokande, where events are 
examined for electron appearance in charged-current reactions.
The experiment uses an off-axis beam
that provides a neutrino spectrum of low energy with a relatively narrow
energy spread.  The angle and energy are tuned to put the first oscillation
maximum at the location of the Super-Kamiokande detector.  The beam has
measured small $\nu_e$ contamination and the solar neutrino oscillation
mixing $\theta_{12}$ makes only a 
minor contribution, so it is only necessary to demonstrate
that electrons are actually being produced above background.  

%%%%%%%%%%%%%%%%%%%%%%%%%%%%%%%%%%%%%%%%%%%%%%%%%%%%%%%%%%%%%%%%%%%%%%%%%
\begin{figure}
\begin{center}
\includegraphics[height=1.8in]{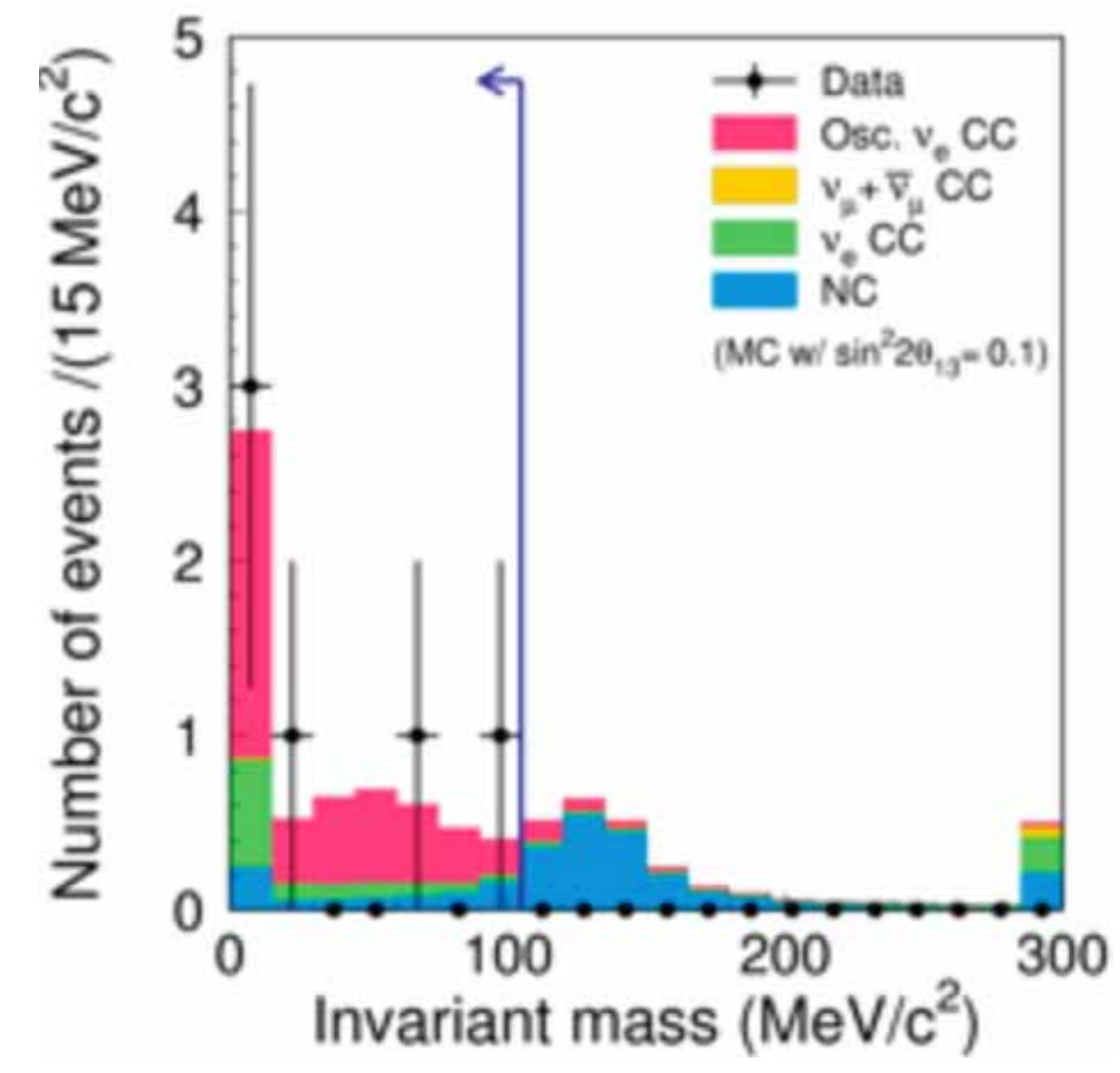}
\caption{Invariant masses obtained for events in the T2K sample classified
as single-ring (candidate $e^-$)
 events, then reanalyzed as two-ring (two-photon) events,
from \cite{Tanaka}.}
\label{fig:TtwoK}
\end{center}
\end{figure}
%%%%%%%%%%%%%%%%%%%%%%%%%%%%%%%%%%%%%%%%%%%%%%%%%%%%%%%%%%%%%%%%%%%%%%%%%%%

The most 
dangerous background is the neutral-current process $\nu N \to \nu \pi^0 N$,
with the $\pi^0$ decay giving an electromagnetic shower similar to that
of an electron.  Fortunately, the $\pi^0$s are produced at relatively
low momentum ($p_{\pi^0} < 200$ MeV), so most $\pi^0$ decays produce two 
distinct Cherenkov rings.  This signature is cleanly separated from the
single-ring electron events in Super-K.  The background rejection claimed
is
\beq
      18 \ \mbox{events} \to 6\ \mbox{single-ring} 
                              \to 0.6\ \mbox{final\ background}
\eeq{TtwoKbkgd}
The total background expected, including also events from $\nu_e$ contamination
in the beam, is $1.5\pm 0.3$ events, and 6 events
are observed~\cite{TtwoK}.  
The invariant mass distribution of the electromagetic 
component observed for the 6 events is shown in Fig.~\ref{fig:TtwoK}; 
the pink region shows the prediction for the best-fit oscillation model.

The total background is nontrivial; the 6 events could be the result of
a fluctuation.  The p-value for $\theta_{13} = 0$ is 0.7\%, corresponding to 
exclusion at 2.5 $\sigma$.

The MINOS experiment has reported a similar result~\cite{MINOS}, 
described at LP11 in the 
lecture by Jenny Thomas~\cite{Thomas}.  MINOS is not at the oscillation 
maximum and has no clean technique for $e/\pi^0$  separation but does 
benefit from much higher statistics.  Using a multivariate discriminant to 
select a region with highest probability for electron appearance, MINOS
observes 62 events with $49.5\pm 7.0 \pm 2.8$ expected background.  There is
a small excess, giving a p-value for $\theta_{13} = 0$ of 11\%, 
corresponding to 
exclusion at 1.2 $\sigma$.

In his summary of the complete world situation for neutrino mixing, 
Thomas Schwetz-Mangold described a global fit to the data, adding all 
sources of evidence, that excludes
$\theta_{13} = 0$ at 3.2 $\sigma$~\cite{Schwetz,Schwetzfit}.  
The lecture was followed by an animated
discussion of what the criteria should be for discovery in neutrino physics.
The chair, Young-Kee Kim, asked me to include my opinion on this issue
in the summary talk.

My position is quite conservative.  I take seriously the good advice 
given by Jenny Thomas in her lecture that 2\%, even 1\%, is a substantial
probability for a fluctuation that should not be ignored.  My advice is:
\begin{enumerate}
\item Even to announce ``evidence for ...'', we must demand $3\sigma$ 
 observation in a single analysis with systematics under the control of 
 one experimental group.  Only when this criterion is reached should 
  corroborating evidence be piled on.
\item A $3 \sigma$ observation still needs corroboration.  The criterion
 for ``discovery'' should be $5 \sigma$, as it typically is in collider
   experiments.
\end{enumerate}

I was eager to report the observation of $\theta_{13}\neq 0$ in my 
summary talk, but
we are not there yet.  In the next two years, we are expecting additional
data from MINOS, a new large data set from T2K, and the first results from 
three reactor experiments---Double Chooz, RENO, and Daya Bay~\cite{Luk}.
If the current hints are correct, we will have discovery of $\theta_{13}$
by the time of Lepton-Photon 2013.  At that meeting, I hope, a luckier 
community will celebrate this achievement.

I was grateful for another piece of news in Tanaka's talk.  JPARC, KEK, and
Kamiokande have survived the Tohoku earthquake in good order and are 
preparing for new experiments in 2012.  Our whole community sends these
labs our best hopes for the future.

\section{Muons}

Andreas Hoecker reviewed a variety of experiments on the charged 
leptons~\cite{Hoecker}.
I will briefly discuss two of these.

\begin{figure}
\begin{center}
\includegraphics[height=2.2in]{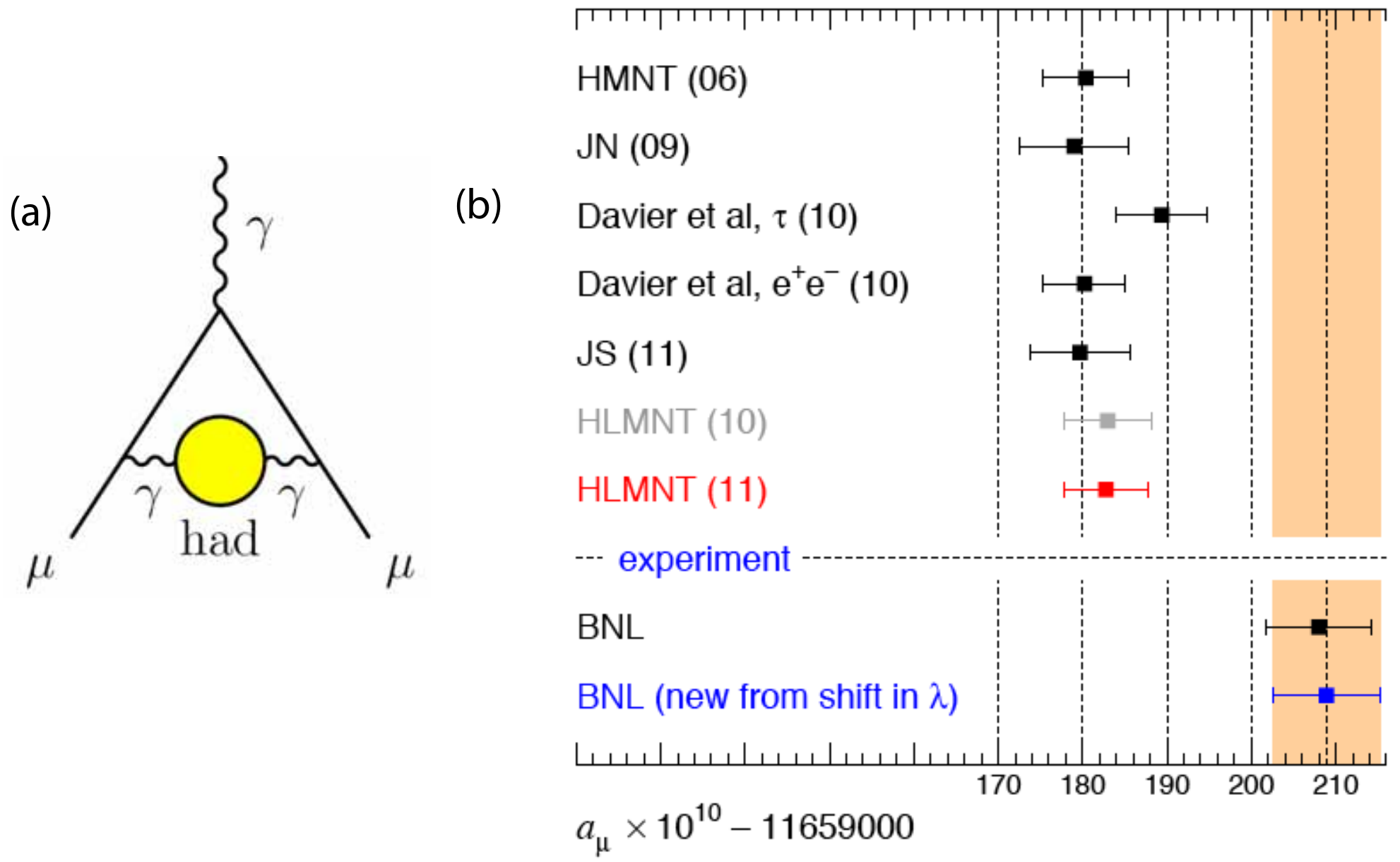}
\caption{(a) The lowest order hadronic vacuum polarization contribution
to the muon $(g-2)$. (b) 
The last few significant figures of the predictions of the 
muon $(g-2)$ by the groups \cite{DHMZ,JS,Hagiwara}, and comparision
to the experimental measurement~\cite{BNLgmt}, from \cite{Hagiwara}.}
\label{fig:LOH}
\end{center}
\end{figure}

The first is the measurement of the muon $(g-2)$.  The actual value of 
the muon $(g-2)$ was measured with better than a part per billion precision
by an elegant BNL experiment that presented its final results in 
2006~\cite{BNLgmt}.  The experimental result deviates from the Standard
Model expectation by more than 3 parts per billion, well outside the 
experimental error.
 However, the interpretation of that experiment has 
been clouded by the fact that the Standard Model prediction for the
muon $(g-2)$ depends on low-energy hadronic amplitudes that are difficult
to evaluate accurately.  The most troublesome of these is the simple
lowest-order hadronic vacuum polarization diagram shown in 
Fig.~\ref{fig:LOH}(a).  It is not possible to compute the required hadronic
amplitude from QCD, so this amplitude must be obtained from a dispersion
integral over  the measured cross section for the process $\ee \to$ hadrons.
To match the accuracy of the BNL experiment, the
 $\ee$ cross section must be given to an accuracy of parts per mil.
Unfortunately, most of the world's data comes from experiments that were not
designed for this level of accuracy.

A new development in the past two years is the addition of two high-precision
data sets from the radiative return process $\ee\to \gamma + \pi^+\pi^-$, one
from BaBar~\cite{BaBarreturn}, one from KLOE~\cite{KLOEreturn}.
Three teams of experts have tried to turn the corpus of $\ee$ data into a 
precision determination of the hadronic vacuum polarization 
diagram~\cite{DHMZ,JS,Hagiwara}.  The final results of the analyses are 
shown in Fig.~\ref{fig:LOH}(b).  The final value of the hadronic
vacuum polarization contributions is about $(685\pm 4)\times 10^{-10}$, with
some variation in the last significant figure among the three groups.
About half of the error comes from the uncertainly in QED radiative
corrections that must be applied to the measurements.  An additional error
of $2\times 10^{-10}$ is estimated for the hadronic light-by-light-scattering
diagram.

%%%%%%%%%%%%%%%%%%%%%%%%%%%%%%%%%%%%%%%%%%%%%%%%%%%%%%%%%%%%%%%%%%%%%%%%%
%%%%%%%%%%%%%%%%%%%%%%%%%%%%%%%%%%%%%%%%%%%%%%%%%%%%%%%%%%%%%%%%%%%%%%%%%%%

The results of these analyses show a persistent discrepancy between the 
BNL result and Standard Model theory, the level of 3.3--3.6 $\sigma$.
If this effect is real, it could indicate interesting effects from 
beyond the Standard Model, due to light sleptons or other exotic lepton
partners.  However, the threshold for claiming this as a 
discovery is high.  There is
now a proposal to repeat the muon $(g-2)$ experiment at Fermilab~\cite{Fgmt}.
However, without a method to increase the precision on the hadronic 
matrix elements, it is not clear to me that that a higher statistics 
measurement of $(g-2)$ will improve our level of understanding.

In the area of lepton flavor violation, the MEG experiment on $\mu\to e \gamma$
has announced a 
new result, adding a new data set from 2010 with double the statistics of 
their previous 2009 result~\cite{MEG}.  
For the 2010 run, the apparatus was improved
in several ways, most notably, with sharper time resolution.  A cluster of 
candidate events found in 2009 was not confirmed in the 2010 data.  My 
interpretation is that this is actually good news, giving a real prospect 
that the MEG experiment can explore for this lepton flavor violating process
down to a branching ratio of $10^{-13}$.

\section{The Large Hadron Collider}

I echo the sentiments of many speakers at LP11 by congratulating Steve
Myers and his team for a very successful year of operation of the 
CERN Large Hadron Collider at 7 TeV.  The LHC experiments have now
accumulated large and interesting data sets.  I will describe results
from that data in the remainder of this lecture.  First, though, I would
like to make two more general observations about the LHC.

The first of these follows Thomas Friedman's phrase, 
``the world is flat''~\cite{Friedman}.  The construction and 
commissioning of the LHC experiments has been a global effort.  Now we
are seeing the fruits of that process.  The experitise on how the 
experiments work is distributed worldwide.  The experimental data is 
available to all members of each detector
collaboration through a global computer
network.  And, we are all connected by 21st-century communication systems.
The distribution of data and expertise implies a new era for high-energy
physics.  Experimenters no longer need to travel to the accelerator.  Working
in their home labs, they {\it are already} at the 
accelerator.  Any group, anywhere,
could make the crucial breakthrough that  leads to a discovery.  All that is
needed is eager students and patient faculty.

\begin{figure}
\begin{center}
\includegraphics[height=4.0in]{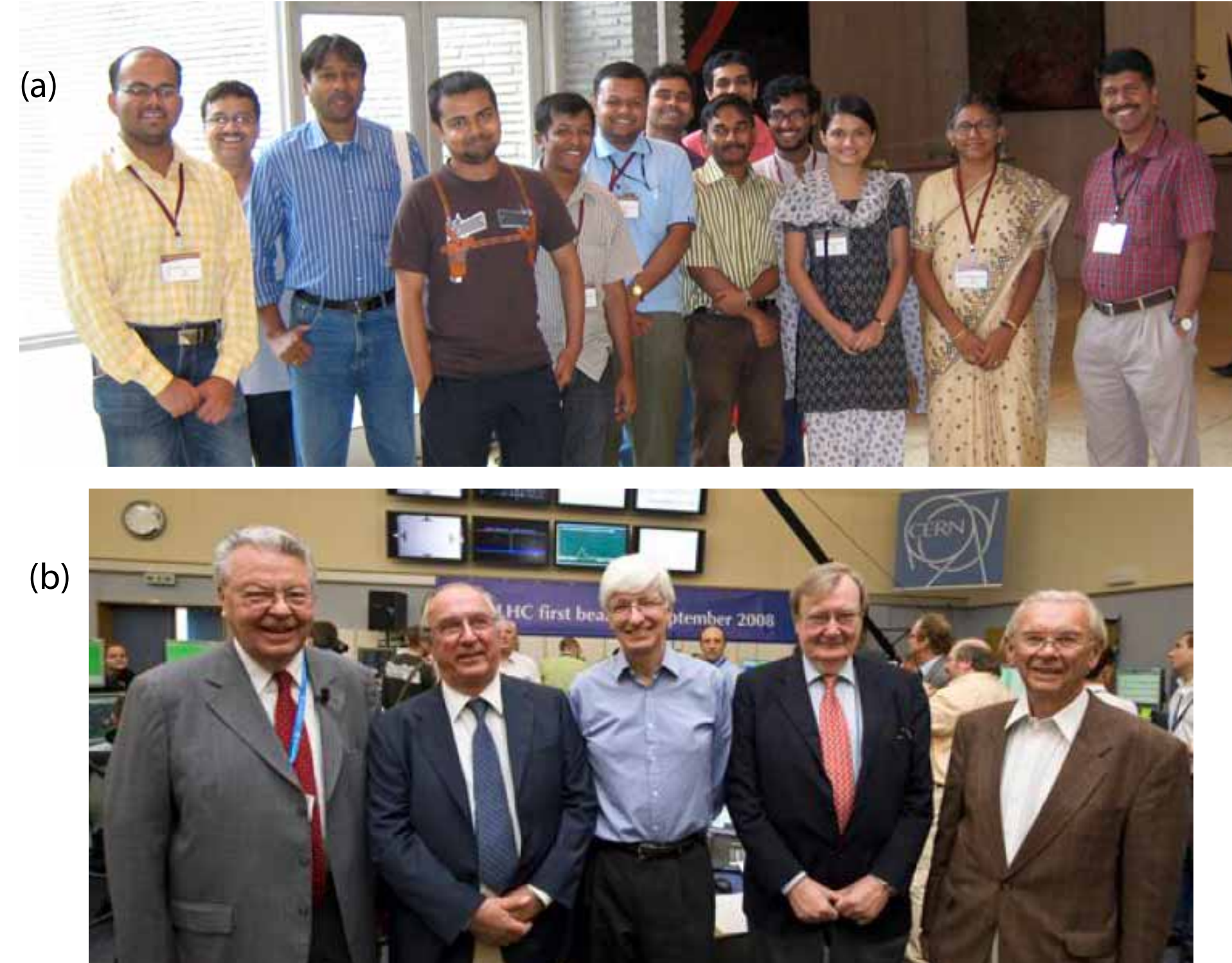}
\caption{(a) The CMS group of the Tata Institute for Fundamental Research,
key participants, in the flat world, in the CMS supersymmetry searches.
(b) The five Directors General of CERN who presided over the design and 
construction of the LHC, (from the left) Robert Aymar, Luciano Maiani, Chris
Llewellyn-Smith, Carlo Rubbia, and Herwig Schopper, the creators of our 
flat world, from~\cite{LHCpix}.}
\label{fig:LHCgroup}
\end{center}
\end{figure}

As an illustration, I show in Fig.~\ref{fig:LHCgroup}(a) the CMS 
group of the Tata Institute for Fundamental Research.  
This is one of many groups around the world 
working aggressively to mine the data from the LHC.  The Lepton Photon 
meetings have traditionally been located at the electron accelerator
laboratories.  The hosting of LP11 in Mumbai is a new direction that 
celebrates this state of affairs.

However, there is another side to the story.  A flat world revolves around
a center.  For the LHC, the center is CERN.  It is of course true that a 
major accelerator project requires billions of dollars, forefront technological
expertise, and a huge construction project.  But an enterprise the size of 
the LHC requires even more.

%%%%%%%%%%%%%%%%%%%%%%%%%%%%%%%%%%%%%%%%%%%%%%%%%%%%%%%%%%%%%%%%%%%%%%%%%
%%%%%%%%%%%%%%%%%%%%%%%%%%%%%%%%%%%%%%%%%%%%%%%%%%%%%%%%%%%%%%%%%%%%%%%%%%%

The LHC was imagined in the early 1980's. The first data is arriving only 
now.  The project has then required an institution whose goals could 
be coherent over all of that time---a period of more than a generation.
Throughout this period, CERN has worked constantly, on a global scale,
 with governments, with the scientific community, and with the public to 
move the LHC project forward to its realization.
   It is a unique  achievement.  Our whole 
community must be grateful to CERN as an institution for making it possible.

As an illustration, I show in Fig.~\ref{fig:LHCgroup}(b) the picture 
from the LHC First Beam Day in 2008~\cite{LHCpix}
 of the five DGs of CERN who presided
over the design and construction of the LHC.

%%%%%%%%%%%%%%%%%%%%%%%%%%%%%%%%%%%%%%%%%%%%%%%%%%%%%%%%%%%%%%%%%%%%%%%%%
\begin{figure}
\begin{center}
\includegraphics[height=4.4in]{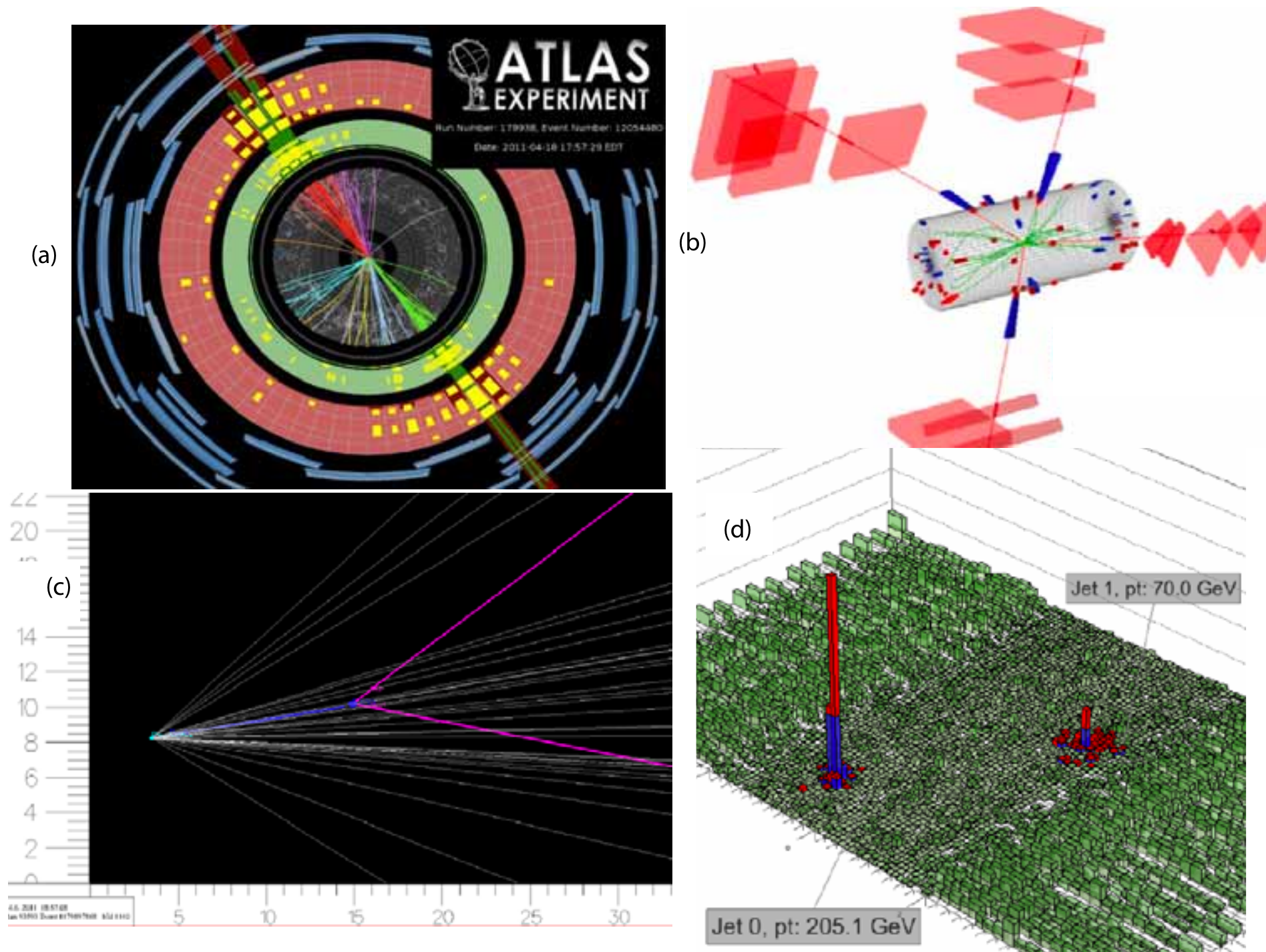}
\caption{A montage of LHC events shown at LP11: 
(a) the highest-mass jet pair observed
so far ($m_{jj} = 4$~TeV),  (b) a candidate $h^0 \to Z^0Z^0\to 4\mu$ event 
from CMS, (c) a candidate $B_s \to \mu^+\mu^- $ event from LHCb, 
(d) an event display from the heavy ion program showing quenching of a 
jet of transverse momentum 200~GeV.}
\label{fig:LHCmontage}
\end{center}
\end{figure}
%%%%%%%%%%%%%%%%%%%%%%%%%%%%%%%%%%%%%%%%%%%%%%%%%%%%%%%%%%%%%%%%%%%%%%%%%%%

Frederick Bordry described to us the commissioning and current status of the
LHC, which is now running routinely at luminosities above 
$2\times 10^{33}$/cm$^2$/sec~\cite{Bordry}.  As of LP11, the 
ATLAS and CMS experiments
had collected over 2.4 fb$^{-1}$ of data.  Many of the analyses shown at
LP11 were based on more than 1.0 fb$^{-1}$ of data.
In Fig.~\ref{fig:LHCmontage}, I show a collection of images from the first
year of the LHC.  These include, from ATLAS, the highest-mass jet pair
recorded to date, with $m(jj) = 4.0$ TeV,  from CMS, the first $Z^0Z^0$ to 
four muon event, from LHCb, a candidate event for $B_s \to \mu^+\mu^-$, and, 
from the heavy ion program, an event display from CMS showing quenching of
a jet of transverse momentum 200 GeV.

%%%%%%%%%%%%%%%%%%%%%%%%%%%%%%%%%%%%%%%%%%%%%%%%%%%%%%%%%%%%%%%%%%%%%%%%%
\begin{figure}
\begin{center}
\includegraphics[height=2.2in]{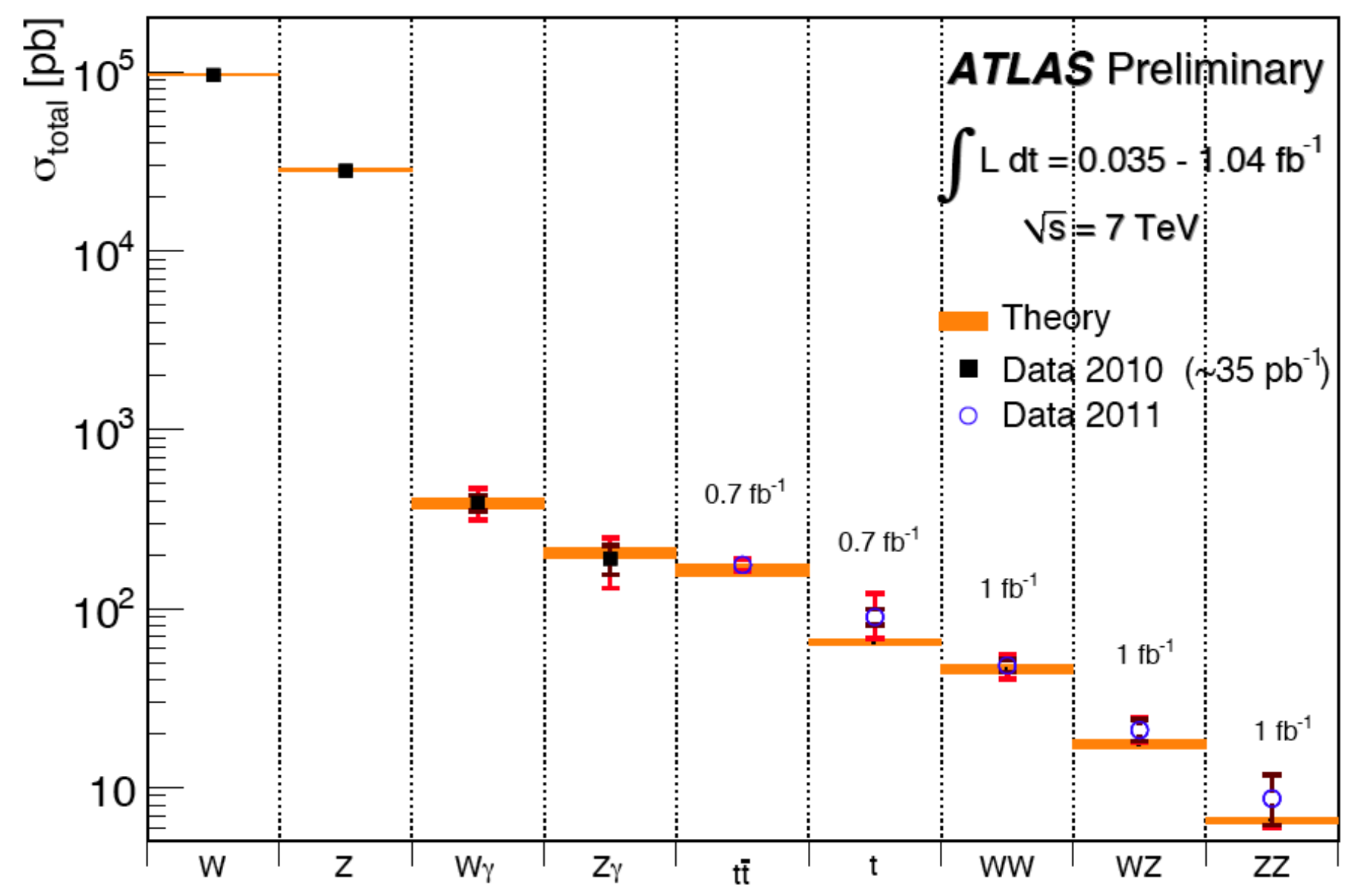}
\caption{Rediscovery of the Standard Model: ATLAS measurements of 
   cross sections for production of 2 massive Standard Model 
  particles~\cite{ATLASredis}.}
\label{fig:rediscover}
\end{center}
\end{figure}
%%%%%%%%%%%%%%%%%%%%%%%%%%%%%%%%%%%%%%%%%%%%%%%%%%%%%%%%%%%%%%%%%%%%%%%%%%%

Fig.~\ref{fig:rediscover}, from ATLAS~\cite{ATLASredis}, shows the current 
status of the rediscovery of the Standard Model.  Cross sections for 
all of the key two-body
processes have now been measured, all the way down to $Z^0Z^0$ production.
The stage is set for the discovery of new processes not included in the 
Standard Model.

\begin{figure}
\begin{center}
\includegraphics[height=2.2in]{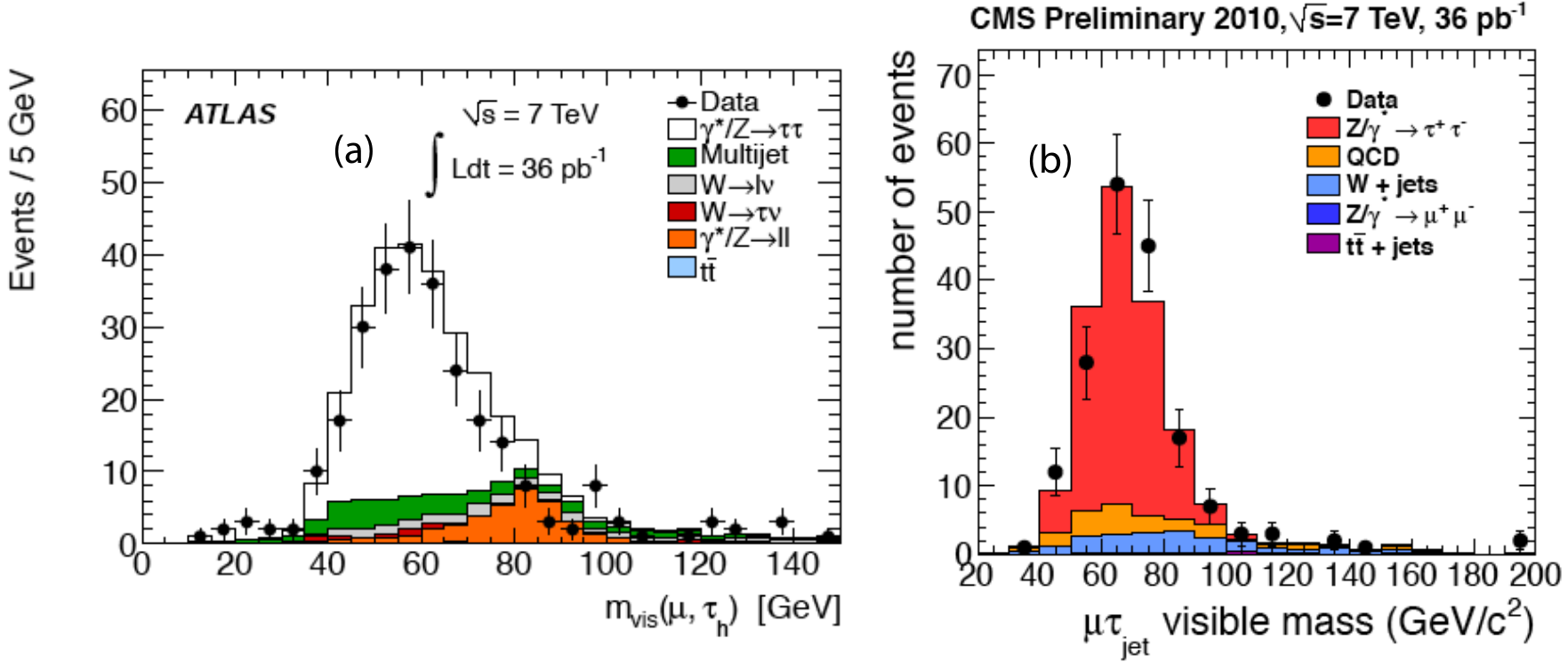}
\caption{From the poster session at LP11: Reconstruction of the $Z^0$ 
resonance in the decay $Z^0\to \tau^+\tau^-$, with one $\tau$ decaying 
to a muon and 
the other decaying hadronically~\cite{ATLAStauposter,CMStauposter}.}
\label{fig:tau}
\end{center}
\end{figure}

I found a pleasant surprise in the poster session of LP11.  A majority of
posters from ATLAS and CMS showed analyses involving the $\tau$ lepton.
Indeed, it is finally true that the $\tau$, the last major physics object
to be developed at the LHC, has come of age.  Fig.~\ref{fig:tau} shows
an important calibration plot for the $\tau$, the reconstruction of the 
$Z^0$ in $Z^0\to \tau^+\tau^-$, with one muonic and one hadronic $\tau$,
from the posters presented by the graduate students Michael Trottier-McDonald
and Raman Khurana~\cite{ATLAStauposter,CMStauposter}.

%%%%%%%%%%%%%%%%%%%%%%%%%%%%%%%%%%%%%%%%%%%%%%%%%%%%%%%%%%%%%%%%%%%%%%%%%
%%%%%%%%%%%%%%%%%%%%%%%%%%%%%%%%%%%%%%%%%%%%%%%%%%%%%%%%%%%%%%%%%%%%%%%%%%%

\section{Quantum Chromodynamics}

The startup of the LHC has triggered tremendous progress in many aspects of
QCD.  Although, already for many years, the data from the Fermilab Tevatron
has called for very sophisiticated calculations in perturbative QCD, the 
higher energy of the LHC and the increased precision of its detectors has
raised the standard for the required level of QCD predictions.  Theorists 
have answered with a stream of new methods and results.  At LP11, Jim Pilcher
reviewed the remarkable progress in 
theory and experiment~\cite{Pilcher}. 
  I will now highlight a few topics from
that discussion.

%%%%%%%%%%%%%%%%%%%%%%%%%%%%%%%%%%%%%%%%%%%%%%%%%%%%%%%%%%%%%%%%%%%%%%%%%
\begin{figure}
\begin{center}
\includegraphics[height=3.0in]{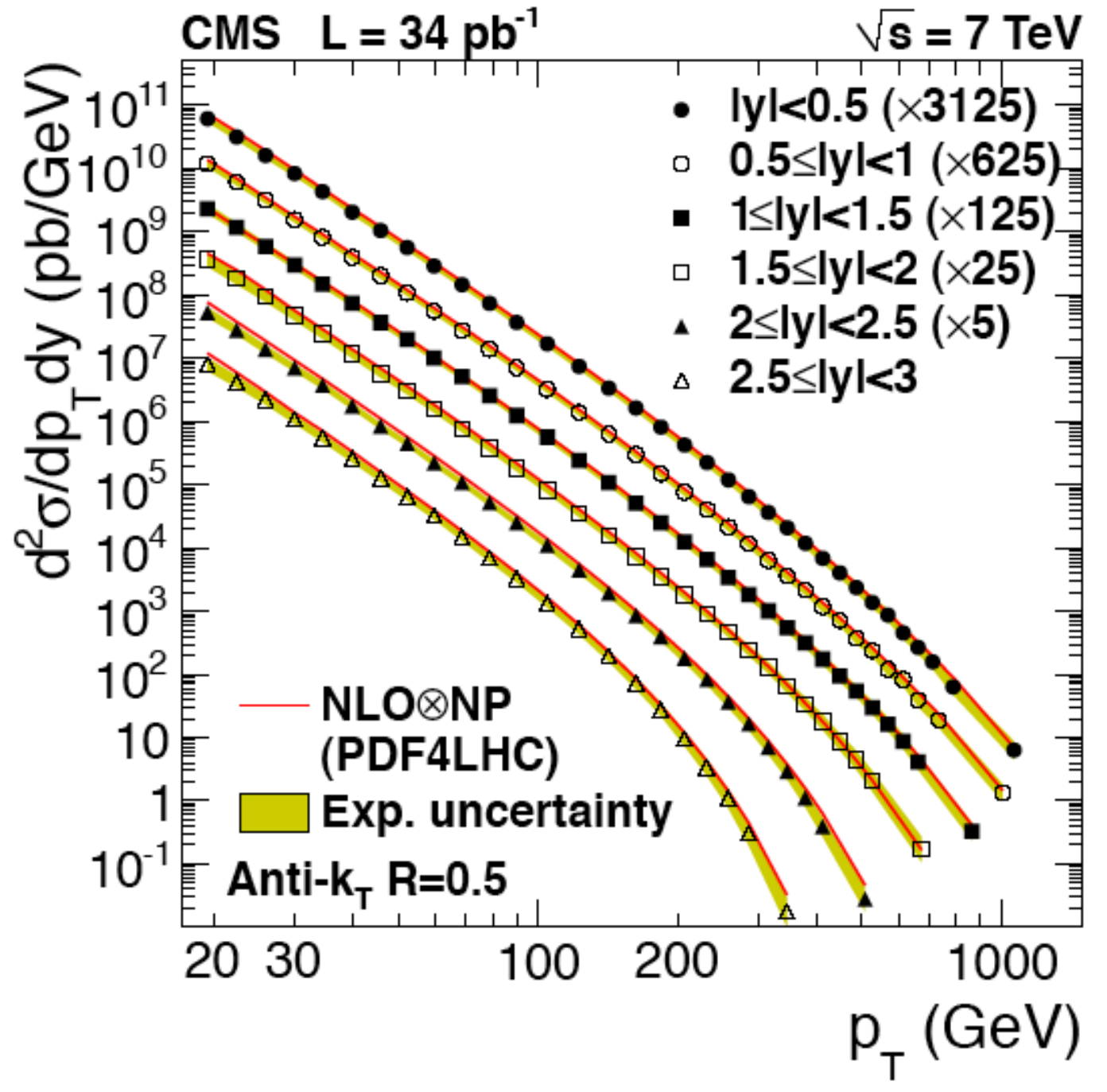}
\caption{Comparison of NLO QCD theory and experiment for the inclusive
jet cross section at 7 TeV, from \cite{CMSjets}.}
\label{fig:CMSjets}
\end{center}
\end{figure}
%%%%%%%%%%%%%%%%%%%%%%%%%%%%%%%%%%%%%%%%%%%%%%%%%%%%%%%%%%%%%%%%%%%%%%%%%%%
As an introduction, I show in Fig.~\ref{fig:CMSjets} the measurement by CMS
of the inclusive jet cross section at 7 TeV in bands of rapidity up to $|y| = 
3$~\cite{CMSjets}.  The detailed agreement of theory and experiment is 
obscured somewhat by the very compressed log scale on the vertical axis.
Nevertheless, the level of understanding achieved already at this early
stage of the LHC program is very impressive.

%%%%%%%%%%%%%%%%%%%%%%%%%%%%%%%%%%%%%%%%%%%%%%%%%%%%%%%%%%%%%%%%%%%%%%%%%
\begin{figure}
\begin{center}
\includegraphics[height=2.0in]{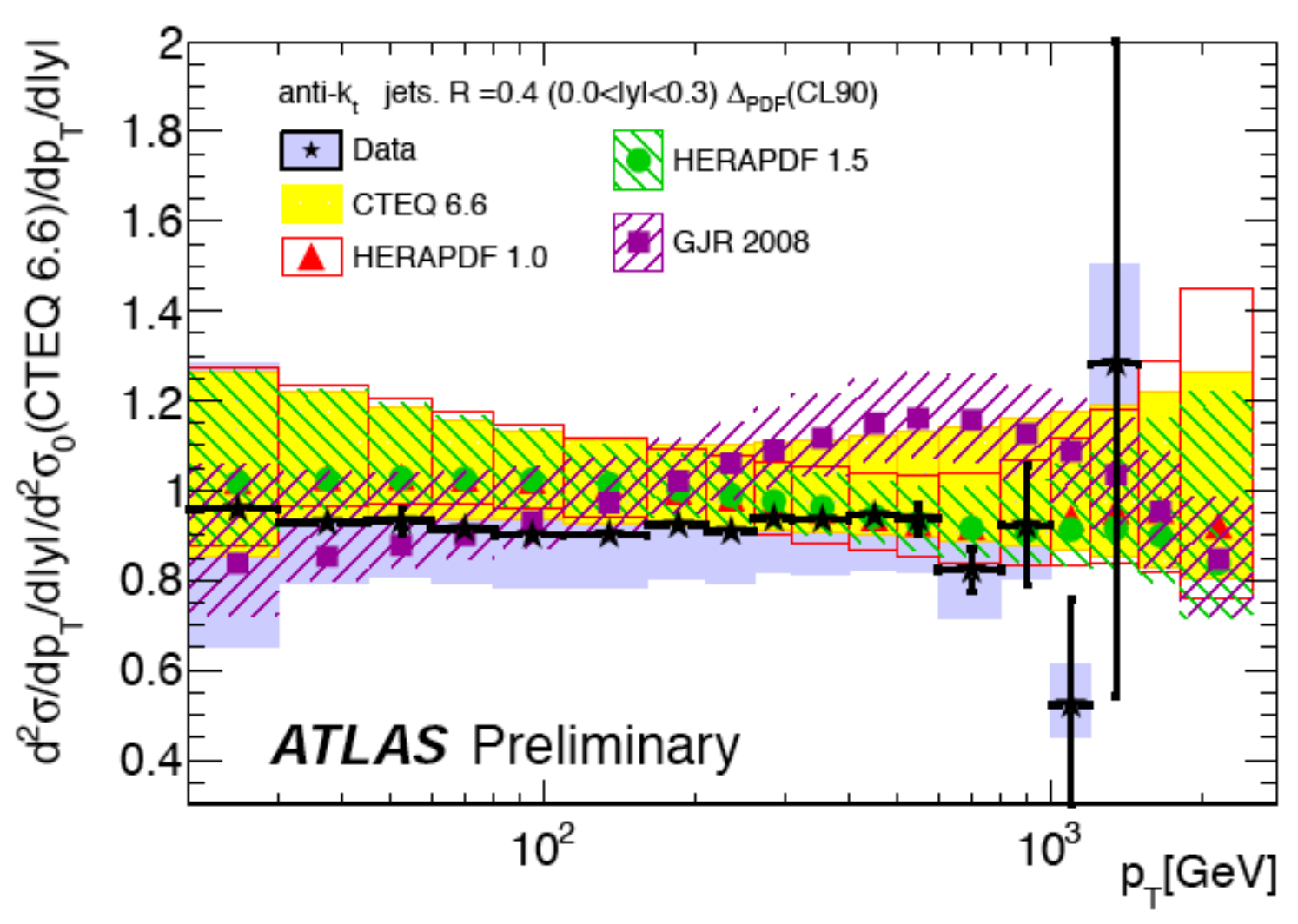}
\caption{Comparison of the inclusive jet differential 
cross section at 7 TeV for $|y| < 1.0$ with predictions from a number 
of parton distribution functions, from ~\cite{ATLASjetcomp}.  
The HERAPDF 1.5 predictions are shown as
the green circles.  }
\label{fig:HERAPDF}
\end{center}
\end{figure}
%%%%%%%%%%%%%%%%%%%%%%%%%%%%%%%%%%%%%%%%%%%%%%%%%%%%%%%%%%%%%%%%%%%%%%%%%%%
Part of the credit for the good agreement shown is an improved level of 
sophistication in the extraction of Parton Distributions Functions.  At LP11,
Katja Kr\"uger reviewed the current status of PDFs, with special 
emphasis on the new HERAPDF 1.5 set based on a combined analysis of the 
H1 and ZEUS deep inelastic scattering results~\cite{Kruger,HERAPDF}.  
The results 
from this analysis refute the expectation that direct data on gluon-gluon
scattering is needed to pin down the gluon PDF well enough to account for
the form of large $p_T$ jet cross sections.   Figure~\ref{fig:HERAPDF} shows
the HERAPDF 1.5 set in good accord with PDF sets based on global fits,
and in good accord with early ATLAS data, even though this PDF uses no 
hadron collider data at all~\cite{ATLASjetcomp}. 
 The form of the gluon distribution is 
extracted from Bjorken scaling violation at high $x$, measured using the
long lever arm provided by HERA.  

%%%%%%%%%%%%%%%%%%%%%%%%%%%%%%%%%%%%%%%%%%%%%%%%%%%%%%%%%%%%%%%%%%%%%%%%%
\begin{figure}
\begin{center}
\includegraphics[height=2.0in]{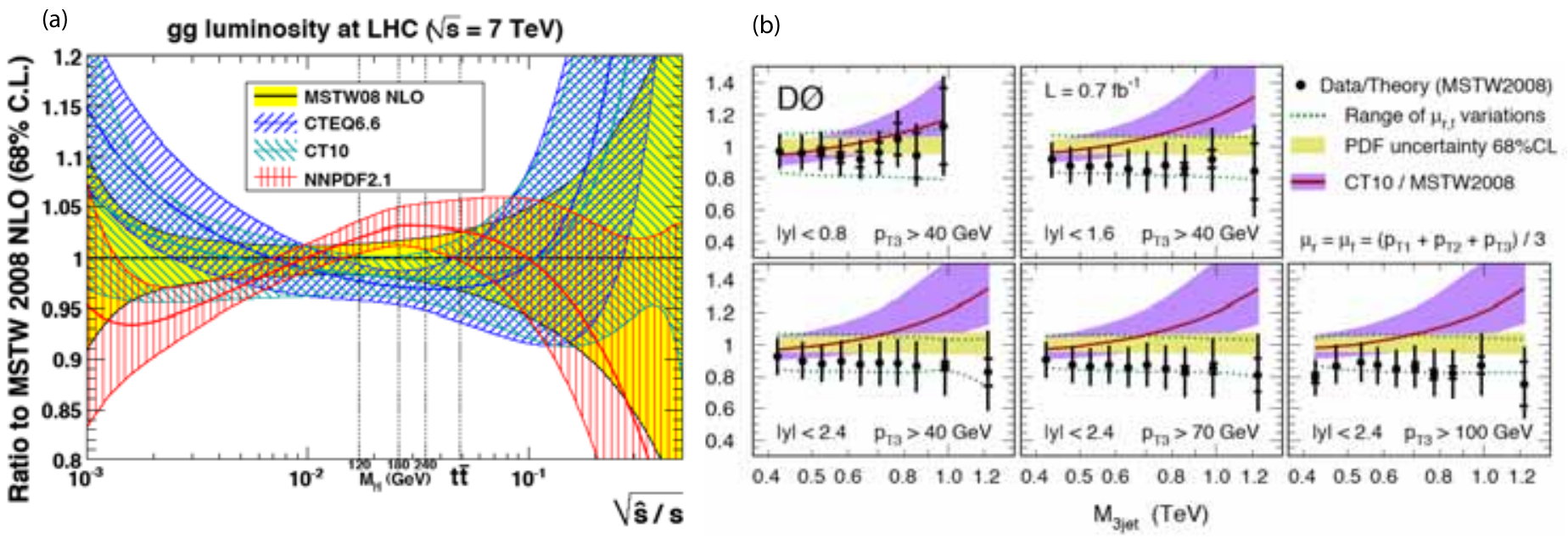}
\caption{(a) Prediction for the gluon-gluon luminosity function at the
 7 TeV LHC from various PDF sets, 
as a function of $\sqrt{\hat s/s}$.  The predictions are
normalized to the central value of the prediction from the 
 MSTW08 PDF, from \cite{Watt}.  (b) Measurements of the $p\bar p\to 3$~jet 
cross section by 
the D\O \ experiment, and predictions from the CT10 PDF set, 
normalized to the to the central value of the prediction from the 
 MSTW08 PDF, from \cite{Dzerothreejet}.} 
\label{fig:threejet}
\end{center}
\end{figure}
%%%%%%%%%%%%%%%%%%%%%%%%%%%%%%%%%%%%%%%%%%%%%%%%%%%%%%%%%%%%%%%%%%%%%%%%%%%

Pilcher showed, however, that this is not
the end of the story for the determination of the gluon PDF.  Recent 
Tevatron data from D\O\ on the 3-jet cross section makes a more sensitive
probe of the high-$x$ behavior of this distribution and highlights the 
different behavior predicted by the two leading global fits, MSTW08~\cite{MSTW}
and CTEQ 2010~\cite{CTEQ}.  Figure~\ref{fig:threejet}(a) shows the 
ratio of predicted values for the gluon-gluon luminosity at 7~TeV for these 
and other PDF sets~\cite{Watt}.  
The behavior above $\sqrt{\hat s/s} = 0.1$ is 
constrained only a little by data and depends more on the way that the
gluon PDF is parametrized as $x\to 1$. 
These differences can be compared to the PDF dependence
of the observable studied by D\O, shown in 
Fig.~\ref{fig:threejet}(b)~\cite{Dzerothreejet}.
The trend of the D\O\ data favors MSTW08.  But this behavior will need to 
be pinned down further by data before we can make the most sensitive searches
for high-mass gluon resonances and quark compositeness that will be possible
at the LHC. 

%%%%%%%%%%%%%%%%%%%%%%%%%%%%%%%%%%%%%%%%%%%%%%%%%%%%%%%%%%%%%%%%%%%%%%%%%
\begin{figure}
\begin{center}
\includegraphics[height=3.7in]{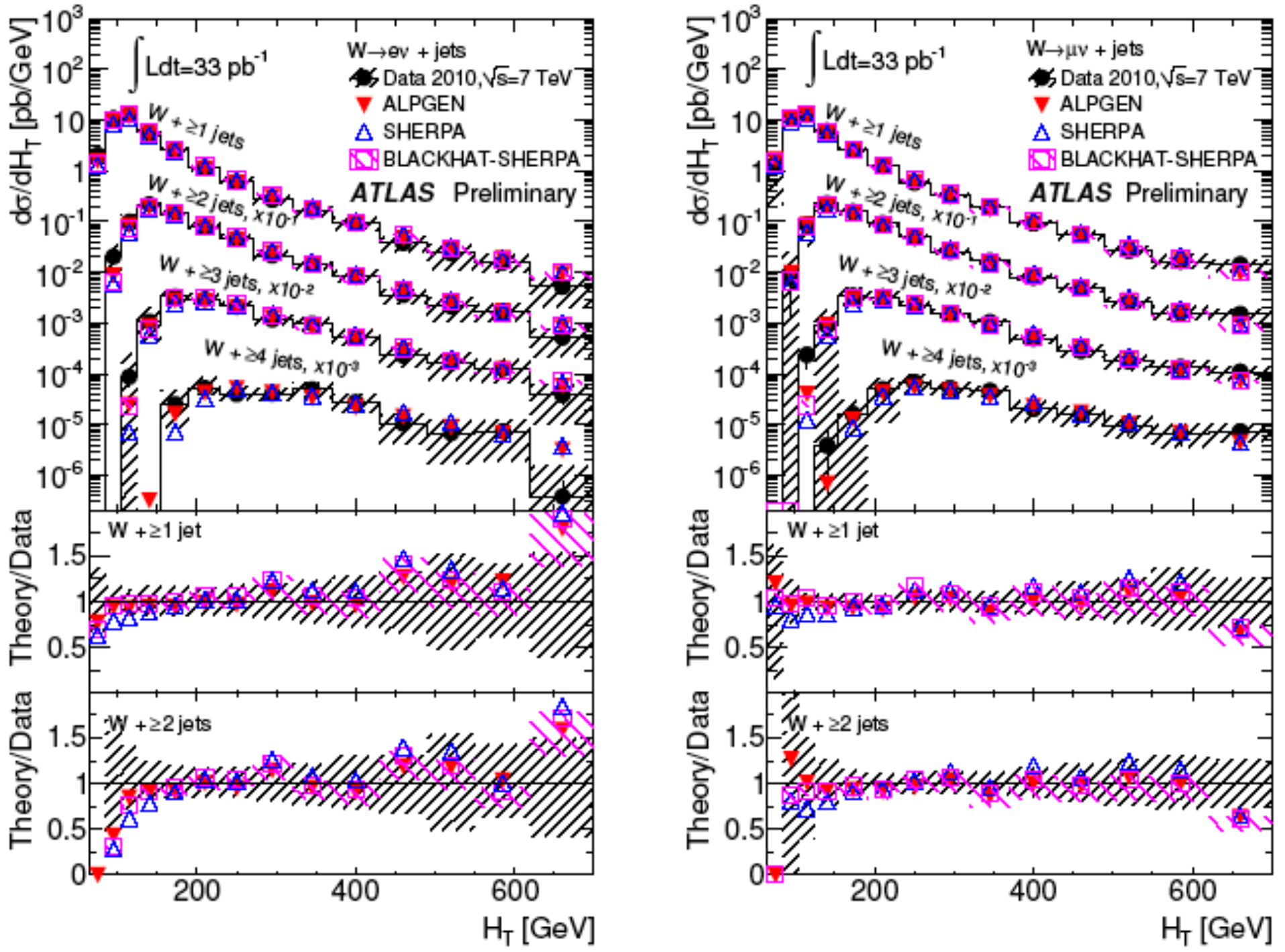}
\caption{Measurement by the ATLAS experiment of the cross section for 
production of $W + n$ jet events as a function of the total transverse
energy $H_T$, for $n = 1,2,3,4$, and comparison to the 
predictions of
QCD event generators, from \cite{ATLASWjets}.}
\label{fig:Wjets}
\end{center}
\end{figure}
%%%%%%%%%%%%%%%%%%%%%%%%%%%%%%%%%%%%%%%%%%%%%%%%%%%%%%%%%%%%%%%%%%%%%%%%%%%

Figure~\ref{fig:Wjets}, from ATLAS~\cite{ATLASWjets}
 shows the remarkable depth of the study of the 
$W +$ jets process that is already available at the LHC.  This process
is the major background to new physics searches and must be understood with
high precision.

%%%%%%%%%%%%%%%%%%%%%%%%%%%%%%%%%%%%%%%%%%%%%%%%%%%%%%%%%%%%%%%%%%%%%%%%%
\begin{figure}
\begin{center}
\includegraphics[height=2.0in]{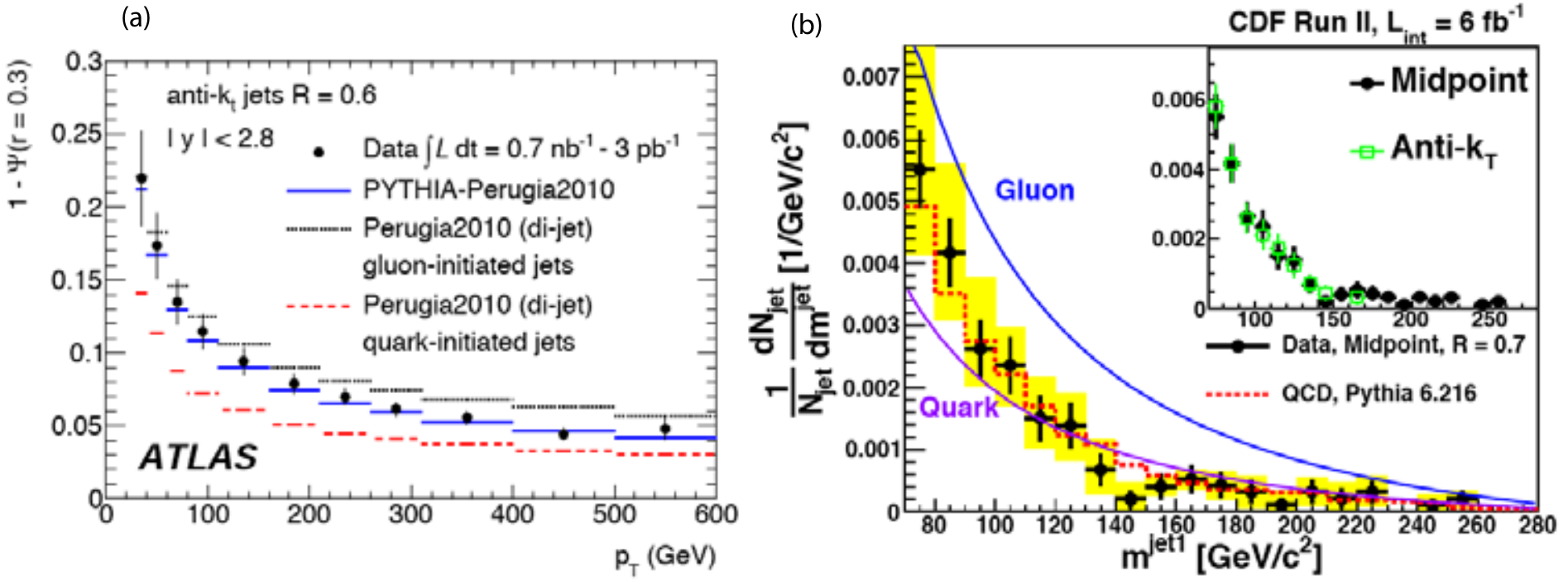}
\caption{Two 2011 measurements of jet shape in hadron-hadron collisions: 
(a) from ATLAS, the fraction of the jet energy contained within a cone 
of radius $R = 0.3$ as a function of the jet $p_T$~\cite{ATLASjetshape};
(b) from CDF, the distribution of single-jet masses~\cite{CDFjetmass}.}
\label{fig:jetshape}
\end{center}
\end{figure}
%%%%%%%%%%%%%%%%%%%%%%%%%%%%%%%%%%%%%%%%%%%%%%%%%%%%%%%%%%%%%%%%%%%%%%%%%%%

Figure~\ref{fig:jetshape}(a) shows a new result from ATLAS on the internal
structure of jets observed at the LHC~\cite{ATLASjetshape}.  
The quantity on the vertical 
axis is a measure of the width of the jet.  Jets of larger $p_T$ are 
narrower for two reasons, first, because the QCD radiation from the parent
quark or gluon is more collimated, second, because, quark jets, which are
narrower than gluon jets, become a larger fraction of the sample at large
$p_T$.  Both behaviors are modelled by PYTHIA in good agreement with the
data.  The analysis in Fig.~\ref{fig:jetshape}(a) repeats a classic
analysis from CDF~\cite{CDFjetshape}.  At LP11, we saw a more sophisticated
version of this analysis from CDF, showing the spectrum of the measured
jet mass~\cite{CDFjetmass}. This result is reproduced
 in Fig.~\ref{fig:jetshape}(b).

Given the excellent understanding of jet shapes shown in the figure, it is
realistic  to use jet shape variables to identify unusual jets that are 
unlikely to arise from QCD.  Reactions that produce heavy particles, including
the Higgs boson and the top quark, often give these particles boosts large
enough that all of the decay products of the particle end up within a single
jet cone.  Many years ago, Michael Seymour suggested that the properties of 
heavy particle decay could be used to search for these particles in 
high-$p_T$ jet samples~\cite{Seymour}.  More recently, a number of groups have
built effective algorithms to search for heavy particles against the 
forbidding background of QCD gluon 
jets~\cite{Brooijmans,Butterworth,Kaplan,Ellis}. This rapidly developing 
field is reviewed in~\cite{Boostedreview}. In the next few years, the
selection of heavy particles by these methods will be a major theme in studies
of reactions involving Higgs bosons and in searches for new physics effect
from beyond the Standard
Model.

%%%%%%%%%%%%%%%%%%%%%%%%%%%%%%%%%%%%%%%%%%%%%%%%%%%%%%%%%%%%%%%%%%%%%%%%%
\begin{figure}
\begin{center}
\includegraphics[height=2.3in]{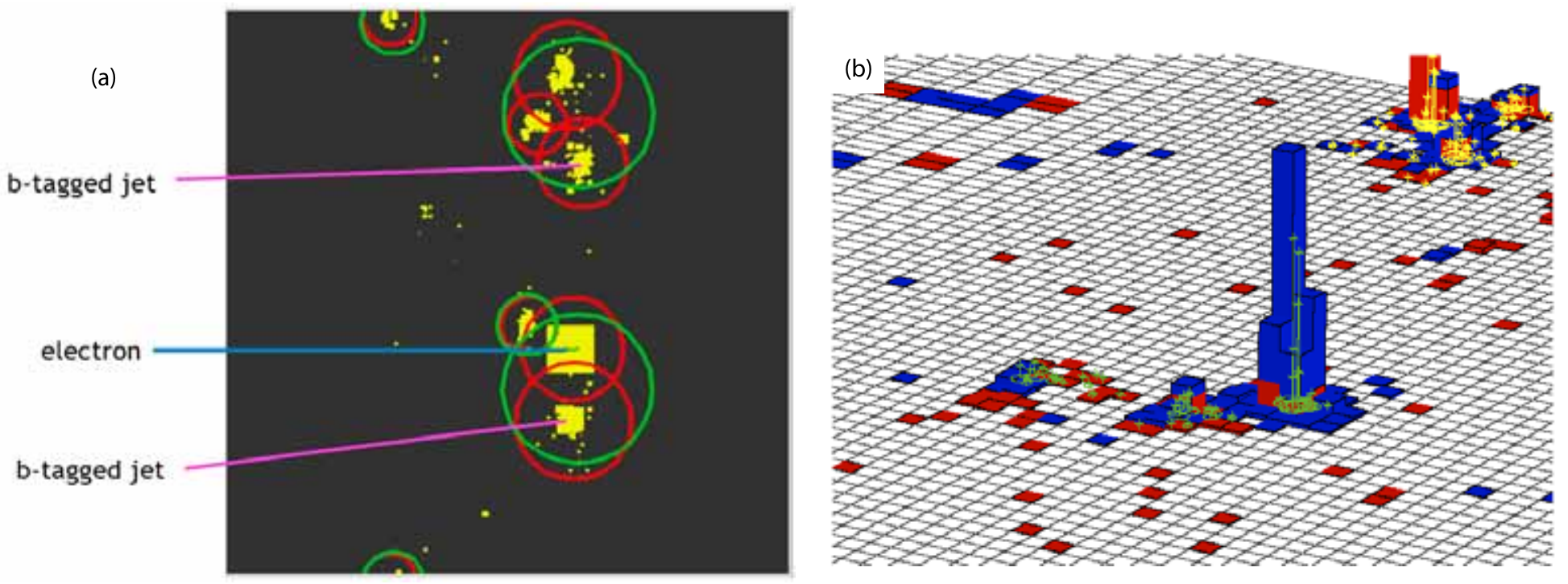}
\caption{Events from (a) ATLAS and (b) CMS showing candidate $t\bar t$
events with highly boosted top quarks~\cite{Millerthesis,CMSboosted}.  
Both events resolve the internal structure of the top quark jet in 
terms of three subjets corresponding to the products of top quark decay.}
\label{fig:boosted}
\end{center}
\end{figure}
%%%%%%%%%%%%%%%%%%%%%%%%%%%%%%%%%%%%%%%%%%%%%%%%%%%%%%%%%%%%%%%%%%%%%%%%%%%

As enticement to this study, I show in Fig.~\ref{fig:boosted} some of the 
first examples of boosted top quarks~\cite{Millerthesis,CMSboosted}, 
showing the internal structure
that distinguishes the heavy quark origin of these exotic jets.

The complexity of QCD processes at the Tevatron and the LHC that must be 
explained by QCD have stimulated new developments in the art of QCD 
computation.  Following the invention in the 
early 1990's of methods that allowed the computation of the rates of 
$2\to 3$ processes at the next-to-leading order (NLO), the art of QCD
computation progressed in deliberate stages.  The pace of the advances changed
in 2003, when a remarkable paper of Witten~\cite{Witten} introduced a
radical new perspective.  The new perspective
 quickly led to new methods for the computation
of tree amplitudes~\cite{BCFW}.  Another outcome was the transformation of the 
unitarity method 
pioneered in the 90's by Bern, Dixon, Dunbar, and Kosower~\cite{BDDK} into
a robust method applicable to reactions of practical interest in 
QCD~\cite{BCFunitarity,BDunit,EGZ}.  The final result of that development, the 
method of Ossola, Papadapoulos, and Pittau (OPP)~\cite{OPP}, has given us 
a effective method not only for NLO computations of $2\to 4$ processes but
also for processes of arbitrarily higher complexity. This set of theoretical
developments, was reviewed
 at LP11 in the lectures of Frank Petriello and Giulia
Zanderighi~\cite{Petriello,Zanderighi}.  It is, in my opinion, the most 
important development in theoretical particle physics of the past few years.

Many authors have contributed to this progress.  I hope that I may be 
excused by referring to Petriello's lecture~\cite{Petriello}
 for a more complete bibliography
and concentrating here on three projects on the leading edge: BlackHat,
HELAC, and the POWHEG Box.  

%%%%%%%%%%%%%%%%%%%%%%%%%%%%%%%%%%%%%%%%%%%%%%%%%%%%%%%%%%%%%%%%%%%%%%%%%
\begin{figure}
\begin{center}
\includegraphics[height=3.0in]{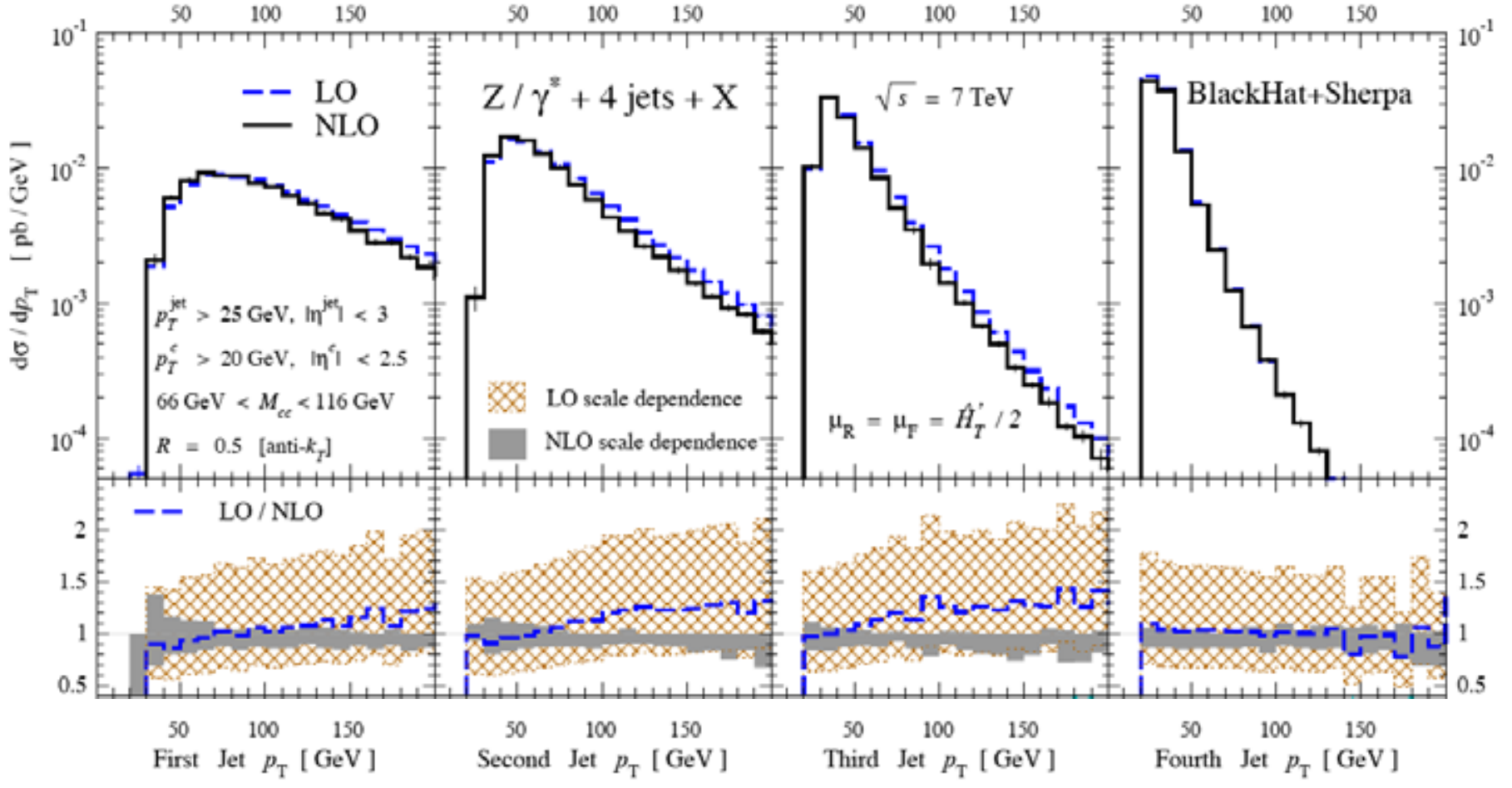}
\caption{NLO QCD calculation by the BlackHat Collaboration of the 
  $p_T$ spectrum of the leading four jets in $pp\to Z +$ jets, from 
    \cite{BHpTZ}.  In the lower boxes, the comparison of the 
 brown hatched and grey shaded regions shows the decrease in the 
    theoretical error in going from LO to NLO. }
\label{fig:pTZ}
\end{center}
\end{figure}
%%%%%%%%%%%%%%%%%%%%%%%%%%%%%%%%%%%%%%%%%%%%%%%%%%%%%%%%%%%%%%%%%%%%%%%%%%%
BlackHat, the product of a collaboration led by Zvi Bern, Lance Dixon, and
David Kosower, has been the most aggressive in incorporating the new 
theoretical methods into phenomenological computation.  As a result, this
collaboration has presented results for the most complex processes, most
recently the NLO contributions to the production of $Z^0 +$ 4 jets at the
LHC.  The computed $p_T$ spectra for the first, second, third, and fourth
jets are shown in Fig.~\ref{fig:pTZ}~\cite{BHpTZ}. The difference between 
the hatched and grey
shaded areas in the insets shows the reduction in theoretical error with the
inclusion of the NLO amplitudes.  The calculation reflects a smooth merger
of the BlackHat methods for loop amplitude computation with the sophisticated
handling of multi-leg tree amplitudes in the Sherpa Monte Carlo 
program~\cite{Sherpa}.  This combination of methods will eventually lead to 
the integration of multijet NLO computations with the complete event 
description by parton showers~\cite{Menlops}.

%%%%%%%%%%%%%%%%%%%%%%%%%%%%%%%%%%%%%%%%%%%%%%%%%%%%%%%%%%%%%%%%%%%%%%%%%
\begin{figure}
\begin{center}
\includegraphics[height=2.0in]{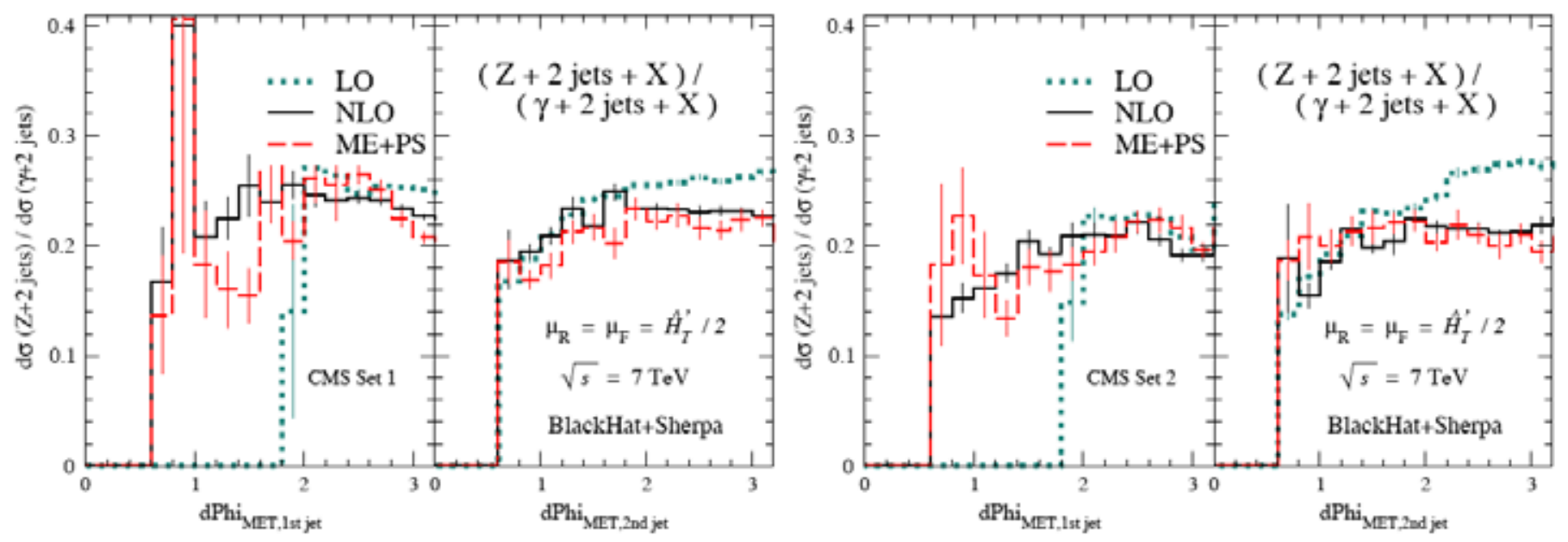}
\caption{Correction factors for the conversion of measured differential cross 
 sections for $pp\to \gamma + $ jets to cross sections for 
$pp\to Z^0 + $ jets, computed in NLO QCD by the BlackHat Collaboration, from
 \cite{BHgamma}.}
\label{fig:photonZ}
\end{center}
\end{figure}
%%%%%%%%%%%%%%%%%%%%%%%%%%%%%%%%%%%%%%%%%%%%%%%%%%%%%%%%%%%%%%%%%%%%%%%%%%%
Figure~\ref{fig:photonZ} shows a second interesting output of BlackHat.
An important background to supersymmetry searches is the process of 
$Z^0 +$ jets production with the $Z^0$ decaying invisibly to $\nu\bar \nu$.
In principle, it is possible to estimate this background from data by 
measuring the same process with $Z^0$ decay to charged leptons.  However, 
this reaction has a cross section 6 times less than the invisible decay.
An alternative is to measure the cross section for production of a 
direct photon plus jets.  This process has much higher statistics, but the
kinematics is somewhat different from that of $Z^0$ production, requiring a 
nontrivial correction factor to convert the measured cross section to a 
background estimate.  The figure shows this correction factor, computed to
NLO accuracy by BlackHat~\cite{BHgamma}.

%%%%%%%%%%%%%%%%%%%%%%%%%%%%%%%%%%%%%%%%%%%%%%%%%%%%%%%%%%%%%%%%%%%%%%%%%
\begin{figure}
\begin{center}
\includegraphics[height=2.2in]{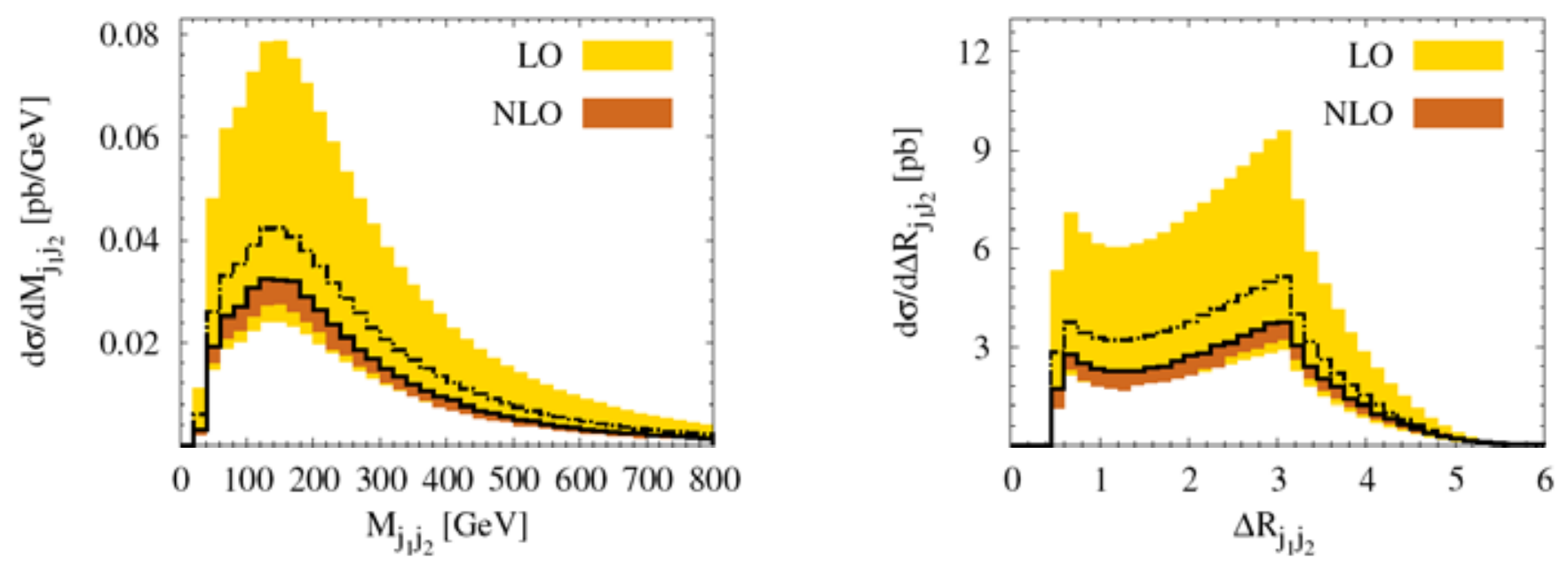}
\caption{Observables of the 2-jet mass distribution in $pp\to t\bar t + 2$
jets, computed in NLO QCD by the HELAC group~\cite{HELACttjj}.}
\label{fig:ttjj}
\end{center}
\end{figure}
%%%%%%%%%%%%%%%%%%%%%%%%%%%%%%%%%%%%%%%%%%%%%%%%%%%%%%%%%%%%%%%%%%%%%%%%%%%
HELAC is a program that combines the OPP method with a more robust 
recursive method for tree amplitude generation, developed by a collaboration
led by Papadopoulos and Pittau~\cite{HELAC}.  
The products of this method have so far 
been limited to $2\to 4$ processes, but have included processes with heavy
particles such as top quarks in intermediate states.  Figure~\ref{fig:ttjj}
shows the HELAC computation of observables for the process 
$pp\to t\bar t jj$~\cite{HELACttjj}.
This is important progress toward the complete understanding of $t\bar t +$
jets reactions that dominate LHC backgrounds to new physics searches with 
leptons and large numbers of jets.

The POWHEG Box is an innovative supplement to these computational methods.
Nason originally developed the POWHEG method for integrating NLO calculations
with parton shower Monte Carlo programs~\cite{POWHEG,POWHEGFNO}.  
With this method, one uses standard NLO QCD
to calculate the inclusive distribution of the hardest partons, then adds
further partons from an iterative parton shower, restricting the emission
phase space so that the computed inclusive distributions are not affected.
In  an 
important insight, Alioli, Nason, Oleari, and Re conceived
 the idea that the method could be formulated as an 
{\it interface} between NLO computations and existing parton shower programs
such as PYTHIA and HERWIG~\cite{POWHEGbox}.  They built a computational
environment
that is open for any NLO calculator to turn his or her calculations into a 
simulation code useful for experimenters.   In \cite{ZandWW}, for example,
the NLO 
calculation of a very complex
$2\to 4$ process is successfully integrated with parton showers using 
the POWHEG Box.

These advances in QCD calculation are of practical importance, but, as I have 
already emphasized, they are
also intertwined with developments on the formal mathematical side of 
quantum field theory.  One way to express this is to explain that computations
in QCD are simplified if QCD is replaced by a similar theory with higher
symmetry, the $N=4$ supersymmetric version of Yang-Mills theory.  
$N=4$ supersymmetric Yang-Mills is a fascinating theory in its own right
with many special properties.  Shiraz Minwalla reviewed these developments
in an energetic and exciting lecture at LP11~\cite{Minwalla}.

For all values of its  parameters, 4-dimensional 
$N=4$ supersymmetric Yang-Mills theory is a theory of 
gluons and fermions with exact scale invariance.  However, in the limit in 
which the number of QCD  colors becomes large, it has additional unexpected
properties.  The theory has a dual description as a gravity theory in 
5-dimensional anti-de Sitter space~\cite{Malda}.
  The duality maps the strong-coupling
region of the original theory into a region of the gravity theory corresponding
to small background curvature and classical dynamics.  Under duality,
scattering amplitudes map to Wilson loops with lightlike edges.   The
full symmetry of the theory includes conformal symmetry in the original
space, plus a second conformal invariance acting on Wilson loops in the 
dual description~\cite{Henn}.   The weak-coupling scattering amplitudes
seem to have another alternative, nonlocal description in terms of 
integrals over a space of Grassmannians~\cite{Arkani}. 

By making use of these features, it is possible to solve for the S-matrix
of theory both at weak and at strong coupling~\cite{Alday,Gaiotto}.  It is
very likely that the exact S-matrix of $N=4$ super-Yang-Mills theory for a 
large number of colors will be found in the next few years.  This is an 
exciting prospect that will undoubtedly bring surprising and useful insights
both for mathematical quantum field theory and for practical application
to  QCD.

\section{The Higgs Boson}

I now turn to searches for new phenomena at the LHC.  First of all, I will
discuss the search for the Higgs boson.

At the first Lepton Photon conference that I attended---1981---the final 
talk was given by Lev Okun~\cite{Okun}. Hearing Okun was a memorable
experience for me, and one
important for the development of my own ideas about particle physics.
Here are some excerpts from the talk:

\begin{quotation}
Instead of giving a general overview of the prospects, I decided to choose and
discuss in some details just one problem, which could be considered as 
problem No. 1 in particle physics.  To be No. 1 this problem has to be 
theoretically advanced and urgent.  It should also be experimentally 
accessible.

It seems to me that the problem No. 1 of high energy physics are scalar
particles.

Painstaking search for light scalars should be considered as the highest 
priority for the existing machines ... and even more so for the next 
generation of accelerators ... 
\end{quotation}
Thirty years later, this question is still unresolved, and still No. 1.

This summer, we have seen tantalyzing hints of a Higgs boson of mass
about 140~GeV.  Both ATLAS and CMS have seen excess events in the 
channel $pp\to \ell^+\nu\ell^-\bar\nu$ that might signal Higgs production
with the decay $h^0\to WW^*$.  The excesses are not yet significant, but
ATLAS and CMS expect that a Higgs boson in this  mass range could actually
be discovered (at $5 \sigma$) with about 5 fb$^{-1}$ of data.  If the LHC 
continues to perform as it has, data samples of this size should be available
before the end of this year.

It is hard not to be impatient.  However, as I discussed above in the 
section on neutrino physics, it is important not to allow our expectations 
to substitute hints for discovery.  This is especially true for the
discovery of the long-awaited Higgs boson. As a way to analyze this issues I 
would like to propose five criteria for the discovery of a Higgs boson of mass
140~GeV:
\begin{enumerate}
\item A clear excess of events in $pp\to \ell^+\nu\ell^-\bar\nu + $ (0,1) jets,
   with careful attention to the background from $t\bar t$ production.
\item Correlation of the excess with variables indicating that the parent is
      a $0^{++}$ state.
\item Corroboration by a pileup of $pp\to ZZ^*\to 4\ell$ events  in a single
     bin of width 4 GeV.
\item  Corroboration by an excess of events in $pp\to \gamma\gamma$.
\item  Corroboration by observation of a similar excess of 
          $\ell^+\nu\ell^-\bar\nu$ at the Tevatron.
\end{enumerate}
If the Higgs boson mass turns out to be lower, for example, 120~GeV, the
list of criteria for discovery will be different but, I hope, no less
rigorous.

%%%%%%%%%%%%%%%%%%%%%%%%%%%%%%%%%%%%%%%%%%%%%%%%%%%%%%%%%%%%%%%%%%%%%%%%%
\begin{figure}
\begin{center}
\includegraphics[height=3.7in]{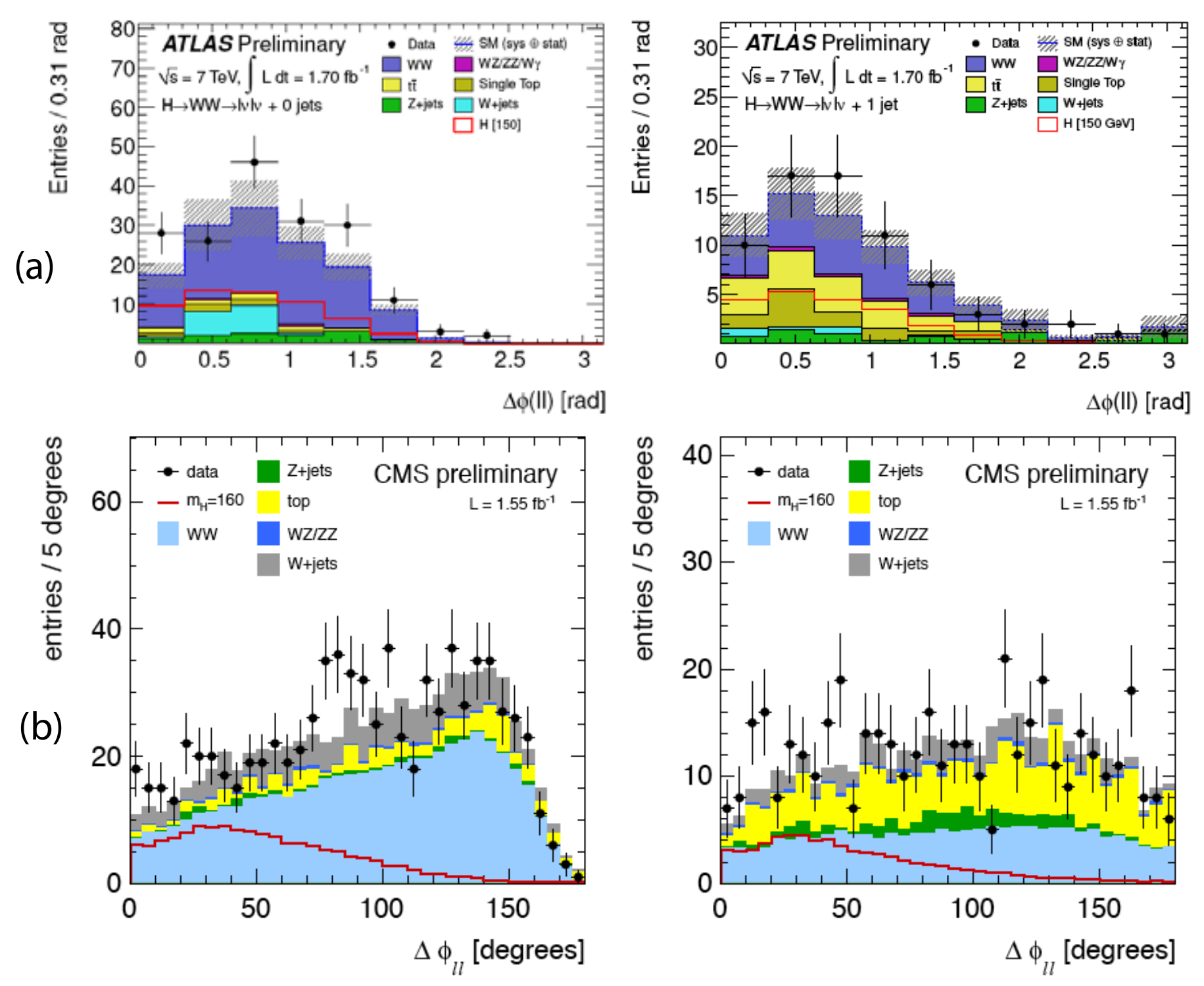}
\caption{Current status of the measurment of the cross section for
$pp\to \ell^+\ell^-\nu\bar\nu + 0,1$ jets, from (a) ATLAS and (b) CMS, from 
\cite{ATLASHiggs,CMSHiggs}.  The colored histograms show various 
backgrounds expected from QCD.  The red line histogram shows the expectation
for a Standard Model Higgs boson of mass 150~GeV.  The data is plotted as 
a function of the azimuthal angle between the two leptons.}
\label{fig:llnn}
\end{center}
\end{figure}
%%%%%%%%%%%%%%%%%%%%%%%%%%%%%%%%%%%%%%%%%%%%%%%%%%%%%%%%%%%%%%%%%%%%%%%%%%%

\begin{figure}
\begin{center}
\includegraphics[height=1.8in]{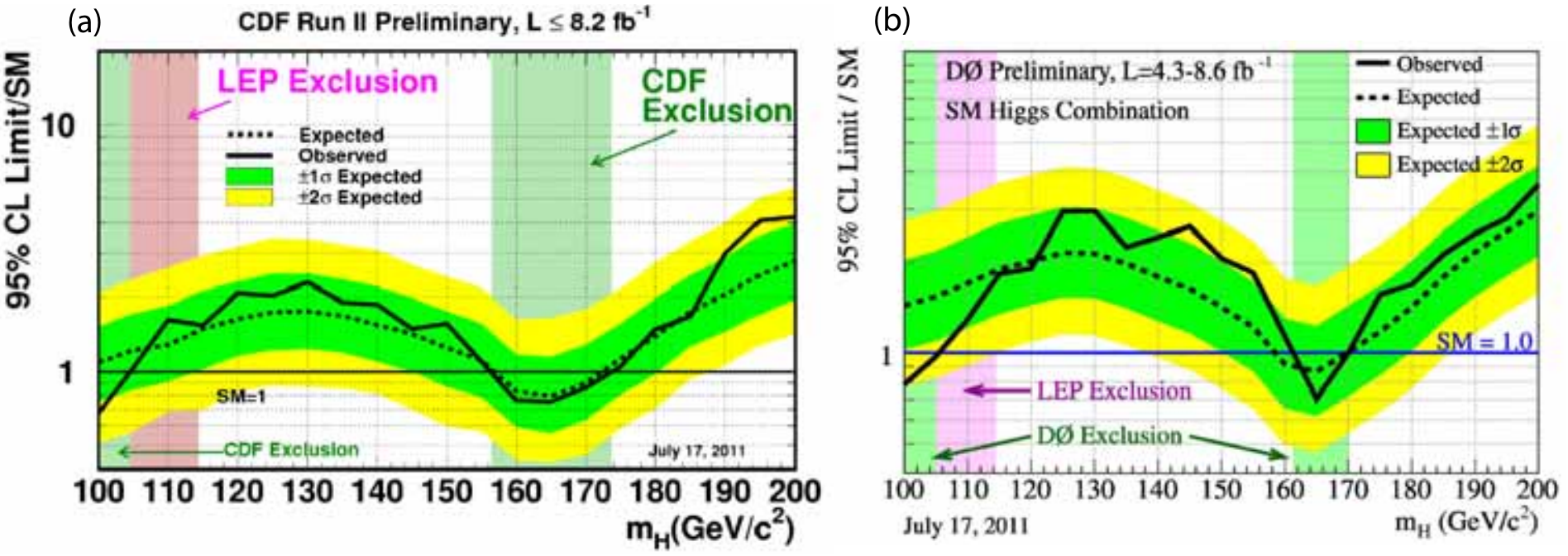}
\caption{Upper limits on the cross section for production of a Standard 
Model Higgs boson at the Tevaton, as a ratio to the Standard Model expectation,
from (a) CDF and (b) D\O, from \cite{Verzocchi}.}
\label{fig:THiggs}
\end{center}
\end{figure}

How far had we come toward the goals of discovering a 140~GeV Higgs boson
by LP11?  Figure~\ref{fig:llnn} shows the
$pp\to \ell^+\nu\ell^-\bar\nu$ events recorded at the LHC, as presented by 
Aleandro Nisati
for ATLAS~\cite{Nisati,ATLASHiggs} and Vivek Sharma for 
CMS~\cite{Sharma,CMSHiggs},plotted in the 
variable $\Delta\phi(\ell\ell)$.  The distribution in this 
variable, the polar angle between the 
two charged leptons, peaks at low values for a spin 0 Higgs boson but at 
high values for at least one of the relevant backgrounds, continuum 
$WW$ production.  The data is presented in 0-jet and 1-jet bins.  Bins with
higher numbers of jets are seriously contaminated by background from 
$t\bar t$. The data sets presented are of size 1.7 fb$^{-1}$ for ATLAS, 
1.55 fb$^{-1}$ for CMS. 
Small excesses are seen:  93 events over $76\pm 10$ expected 
background for ATLAS, 140 events over $120.3 \pm 10.8$ for CMS.  The 
correlation of these excesses with small values of $\Delta\phi(\ell\ell)$
is certainly not obvious from the figures.  Both ATLAS and CMS see
$pp\to ZZ^*$ events roughly covering the range of masses from 120~GeV to 
450~GeV.  Near 140~GeV, ATLAS has a candidate at 144., CMS has candidates
at 139.3 and 144.9.  There is no significant peak in $pp\to \gamma\gamma$.
The Tevatron limits on the Standard Model Higgs boson, presented by 
Marco Verzocchi~\cite{Verzocchi,TevatronHiggs}, are 
shown in Fig.~\ref{fig:THiggs}.
D\O\ sees a substantial excess of events in the mass region $140-150$ GeV;
CDF does not.  

%%%%%%%%%%%%%%%%%%%%%%%%%%%%%%%%%%%%%%%%%%%%%%%%%%%%%%%%%%%%%%%%%%%%%%%%%
%%%%%%%%%%%%%%%%%%%%%%%%%%%%%%%%%%%%%%%%%%%%%%%%%%%%%%%%%%%%%%%%%%%%%%%%%%%
It is not possible now to make a persuasive case for the 
appearance of the Higgs boson.  But the current situation is quite consistent
with the expected signals a Higgs boson of mass about 140 GeV.  We can hope 
that the picture will become clearer when the full 2011 data set has been 
accumulated.

ATLAS and CMS have also {\it excluded} the Standard Model Higgs boson over
a broad range of higher masses.  In most of the higher-mass range, the on-shell
decays $h^0 \to W^+W^-, Z^0Z^0$ would be obvious in several channels if they 
were in fact present. These same channels typically dominate in the 
high-mass region in models with multiple or non-standard Higgs.
 The union of the ATLAS and CMS 95\% exclusion regions
covers the entire range from 145 to 446~GeV, except for a small interval
from 288 to 296~GeV.  Thus, we already know from the LHC that {\it either}
the Higgs boson is light, with a mass within 30~GeV of the direct lower
limit from LEP, {\it or} the Higgs boson is very heavy and strongly
self-coupled.

The precision electroweak predictions strongly favor a light Higgs boson, 
with mass close to the LEP limit.  The Minimal Supersymmetric Standard 
Model predicts that the Higgs boson mass is below 130~GeV, and many other
theoretical models also favor a light Higgs boson.  So, as a theorist, I 
see no reason to be worried that the Higgs boson will be excluded.  The
situation is quite the reverse:  The Higgs boson is now being restricted 
by the LHC results to the narrow mass region in which it is most strongly
expected to appear.  If the true Higgs boson of Nature is similar to the
one in the Standard Model, and if it is in the low-mass range, this
particle should be discovered using the 15
fb$^{-1}$ LHC data set that, hopefully, will be accumulated in 2012.
If the Higgs boson is not found in this search, the reason can only be 
because of other new physics, outside of the Standard Model, that should
also be discoverable at the LHC.

There is one further point to be made about the Higgs boson.  This was 
nicely expressed by Abdelhak Djouadi in his theoretical review of 
Higgs physics~\cite{Djouadi}: ``Finding the Higgs is only the first part of
our contract.''  Much more work will be needed to establish experimentally
that the particle discovered in Higgs boson searches is indeed the Higgs 
boson and that it plays the role required in the Standard Model of 
generating the masses of quarks, leptons, and gauge bosons.

\section{The Top Quark}

It is remarkable that 
  the LHC experiments have now accumulated as many top quarks as the 
Tevatron experiments.  So far, the properties of the top quark observed 
at the LHC have been in accord with the Standard Model.  That picture
could well change, though, as we go to still higher statistics.

%%%%%%%%%%%%%%%%%%%%%%%%%%%%%%%%%%%%%%%%%%%%%%%%%%%%%%%%%%%%%%%%%%%%%%%%%
\begin{figure}
\begin{center}
\includegraphics[height=2.8in]{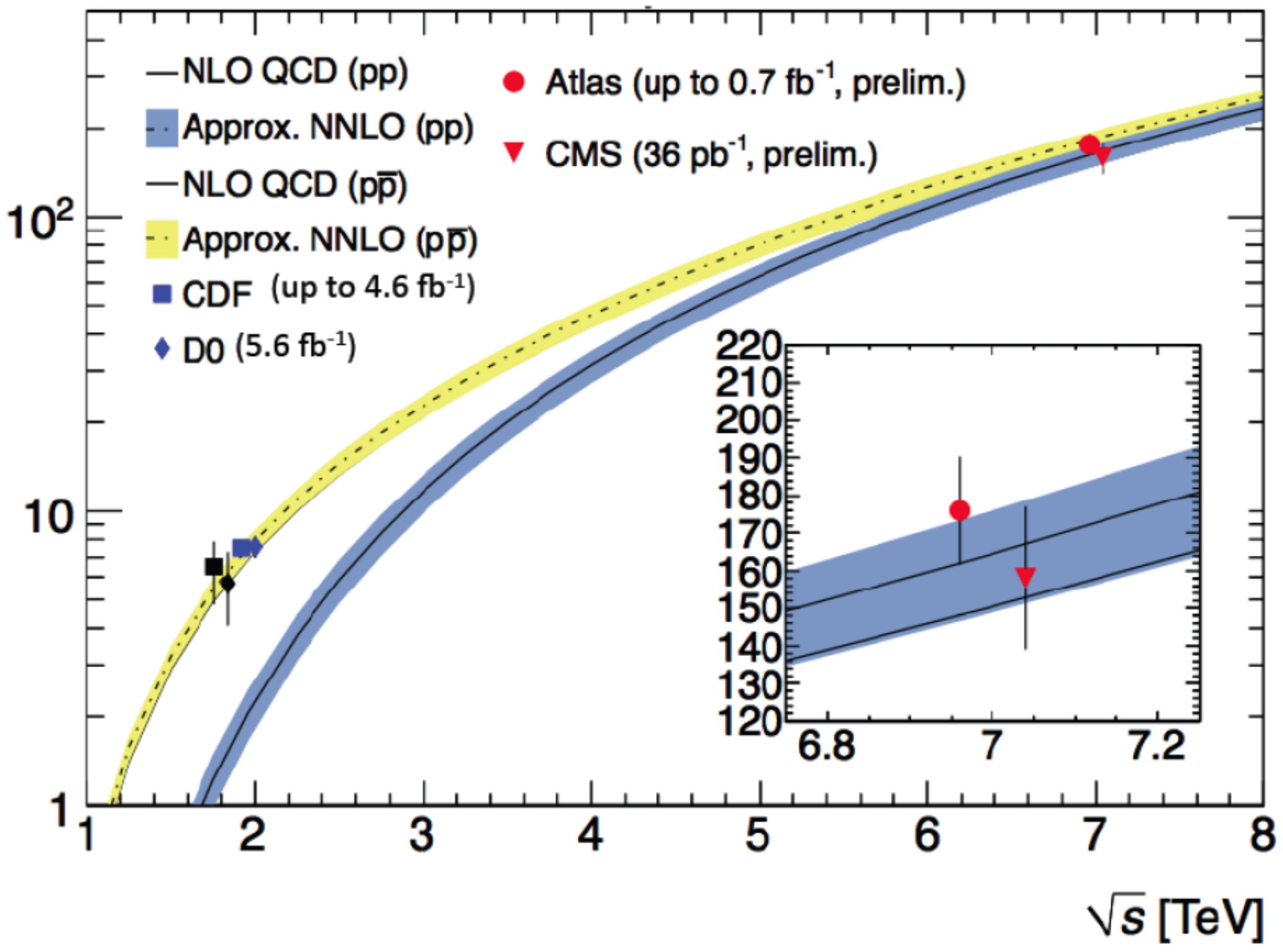}
\caption{Measured values of the cross section for $t\bar t$ production at the
Tevatron and the LHC, compared to QCD theory, from \cite{deRoeck}.}
\label{fig:ttbarcs}
\end{center}
\end{figure}
%%%%%%%%%%%%%%%%%%%%%%%%%%%%%%%%%%%%%%%%%%%%%%%%%%%%%%%%%%%%%%%%%%%%%%%%%%%
The measurement of the top quark pair production total cross section at the
LHC was presented in the talk of Albert de Roeck~\cite{deRoeck}; the
result is shown in Fig.~\ref{fig:ttbarcs}.  The result is a remarkable one.
The LHC at 7 TeV has an energy only a factor of 3.5 higher than that of the
Tevatron and has the relative disadvantage of using $pp$ rather than 
$p\bar p$ collisions, yet the cross section is higher than that at the 
Tevatron by a factor of 25.  The effect is understood within the Standard
Model as signalling the transition from $q\bar q$ to $gg$ production, and
the measured result is in good agreement with QCD 
estimates~\cite{topQCDcs,topQCDcstwo}.

The Tevatron experiments have reported an anomalously large forward-backward
asymmetry in $t\bar t$ production.  The results deserve discussion.  They 
represent a large deviation from the Standard Model expectation.  It is
difficult to imagine an experimental problem that would lead to a false 
measurement.  Among several Tevatron anomalies presented at LP11, I consider
this the most likely to survive, and the most puzzling.  

%%%%%%%%%%%%%%%%%%%%%%%%%%%%%%%%%%%%%%%%%%%%%%%%%%%%%%%%%%%%%%%%%%%%%%%%%
\begin{figure}
\begin{center}
\includegraphics[height=2.4in]{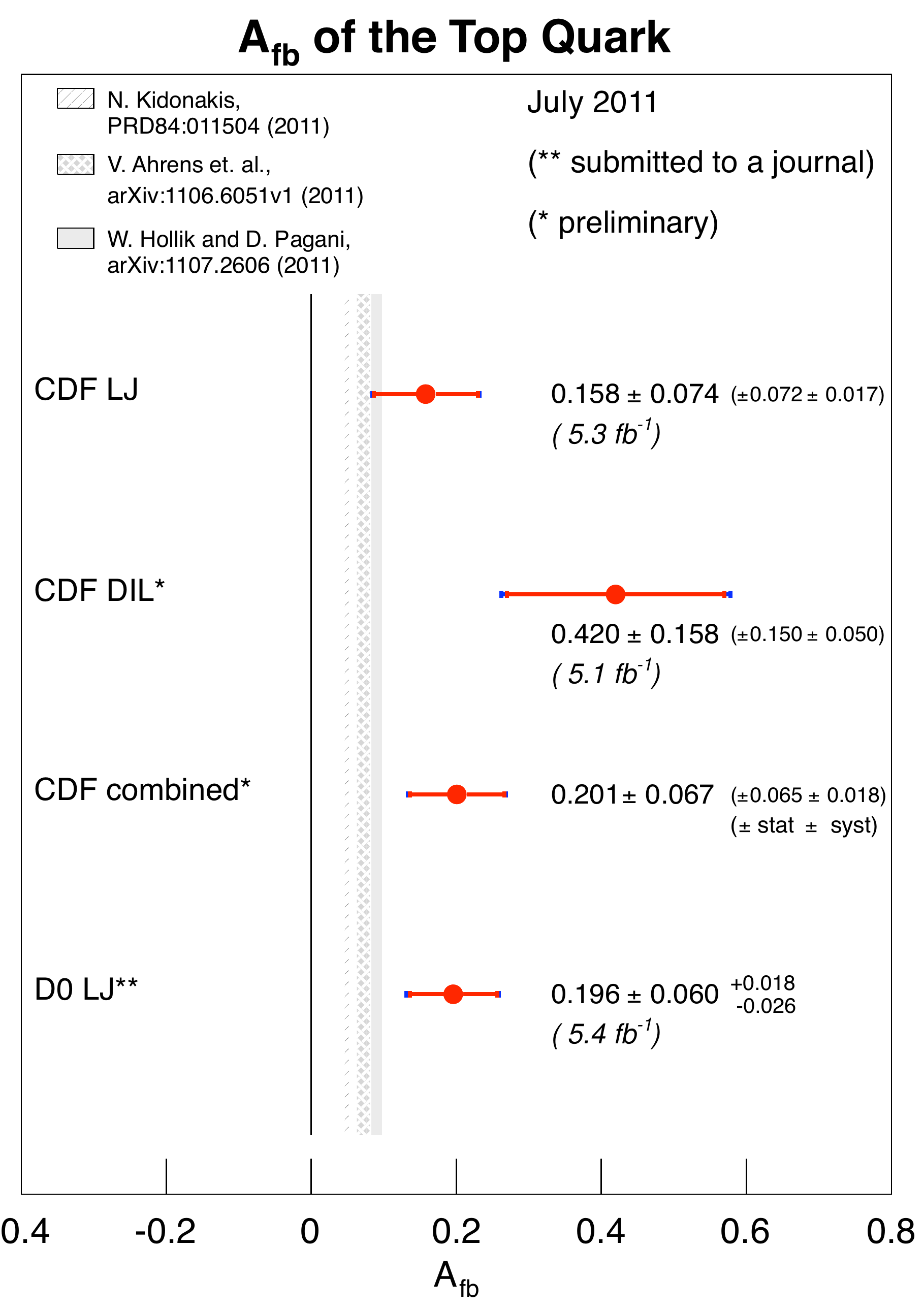}
\caption{Measurements of the forward-backward asymmetry in 
 $t\bar t$ production at the Tevatron, and comparison to theoretical
predictions, from \cite{CDFAfbcomp}.}
\label{fig:AFBtt}
\end{center}
\end{figure}
%%%%%%%%%%%%%%%%%%%%%%%%%%%%%%%%%%%%%%%%%%%%%%%%%%%%%%%%%%%%%%%%%%%%%%%%%%%
A summary of the
measurements is shown in Fig.~\ref{fig:AFBtt}~\cite{CDFAfbcomp}.
The numbers in the figure are to be compared to an expectation from QCD
of 6\%, plus an extra 1\% from electroweak effects.  The effect is strictly
zero in $gg\to t\bar t$, but it can appear in  $q\bar q\to t\bar t$
from the interference between 
1- and 2-gluon production diagrams.   The measured value of the asymmetry
is ($20\pm 6$)\% in each of the two Tevatron experiments.

It is tempting to say that the effect must be due to new physics in the 
$q\bar q t \bar t$ interaction.  This necessarily implies one or more new
particles contributing to this interaction in the $s$, $t$, or $u$ channel.
The various possibilities are reviewed comprehensively by Gresham, Kim,
and Zurek~\cite{GKZu}.  Many of these possibilities are already excluded
by the higher-energy measurements at the LHC.  For example, a new particle
in the $s$-channel would be a color octet vector particle coupling to 
$q\bar q$ and $t\bar t$.  If the coupling to $q\bar q$ is sufficiently 
large---unitarity in the $t\bar t$ system gives a lower bound---this particle
should appear as a resonance in the 2-jet mass distribution at the LHC 
similar to the signature of an axigluon~\cite{BHKR}. ATLAS and CMS exclude 
resonances in the dijet mass distribution up to 
3.3~TeV~\cite{axigluonATLAS,axigluonCMS} and thus, at least for a narrow
resonance, exclude the entire relevant parameter space.
  A new particle in the $t$ channel would be a 
$Z^\prime$ boson coupling to $q\bar t$.  If there is only one such particle,
the exchange of this boson mediates the reaction $qq\to tt$.   A CMS
search for $tt$ production at 7 TeV eliminates the entire parameter
space of this model~\cite{CMStt}.  Some possibilities do remain, including an
interesting suggestion by Tavares and Schmaltz that the anomaly is due to a 
broad resonance
 at 400~GeV that decays mainly to 3-jet final states~\cite{Schmaltz}.
These should be tested in the near future.

In his talk at LP11, Frank Petriello expressed the opinion that the 
anomaly could be explained within the Standard Model~\cite{Petriello}.
  As I have already
explained, the forward-backward
asymmetry in $q\bar q\to t\bar t$ appears for the first time at NLO, and
therefore the estimate cited above, from one-loop diagrams, is only a 
leading-order result.  It is not uncommon that QCD effects are enhanced by 
a factor of 2 due to next-to-leading corrections.  To verify this 
hypothesis, an NNLO calculation of the $t\bar t$ cross section is needed.
This  has not yet been done.  However, several groups, beginning with 
Almeida, Sterman, and Vogelsong~\cite{ASV}, have done calculations that 
resum terms with large logarithms that typically give the largest 
contributions to the next order result.  These calculations give only small
corrections to the forward-backward asymmetry.

I find, at this moment, no persuasive explanations of the $t\bar t$ 
forward-backward asymmetry either from Standard Model physics or from 
new physics.  This is a problem that needs attention and new ideas.

\section{Beyond the Standard Model}

The first 1 fb$^{-1}$ of data from the LHC has shown no evidence for 
new physics beyond the Standard Model.  Experimenters, I have noticed, are
not disturbed by this.   We are still at a very early stage in the LHC
program.  We have seen only 0.1\% of the eventual LHC data set, and only 
the first year of a program that will last 15 years or longer.

And, it should be said, many experimenters believe that the Standard Model
is literally correct.  (``If only we could find that Higgs boson ...'')  
I will comment on
this possibility in the next section.

Theorists feel differently.  This statement applies doubly to those of us
who have written for many years about the incompleteness of the Standard
Model and the necessity of its extention, and to those of us who have 
championed specific models of new physics such as supersymmetry.  As the 
LHC experiments become sensitive to hypothetical new particles with TeV
masses, we are reminded of the phrase from the Latin Requiem Mass.

\begin{quotation}
 \noindent
  Confutatis maledictis, flammis acribus addictis, voca me cum benedictus.
\end{quotation}

A loose translation is:  Thousands of theory papers are being tossed into the 
furnace.  Please, Lord, not mine!

The territory explored by ATLAS and CMS in 2011 has already excluded many 
possibilities for physics beyond the Standard Model.  I will now discuss
what it is that, in my opinion, we have learned from these exclusions.  
At LP11, Henri Bachacou
gave a beautiful summary of the new LHC limits on a variety of new physics
models~\cite{Bachacou}.  I will discuss only a few of these limits that 
follow the main lines of my argument.

First, the idea of a sequential fourth generation of quarks and leptons is
in serious trouble.  If there exist new heavy quarks $U$ and $D$ that 
couple to the Standard Model as a conventional quark doublet, the cross 
section for the production process $gg\to h^0$ is multiplied by a factor of
9.  Given the fact that Higgs limits are now within a factor
of a few of the Standard Model expectation, this excludes fourth generation
models over  the entire range of Higgs mass, excepting only high
values above 550~GeV.  It is important to note that other types of exotic
fermions are still in play and are even interesting; I will discuss one
particular example below.

\begin{figure}
\begin{center}
\includegraphics[height=2.2in]{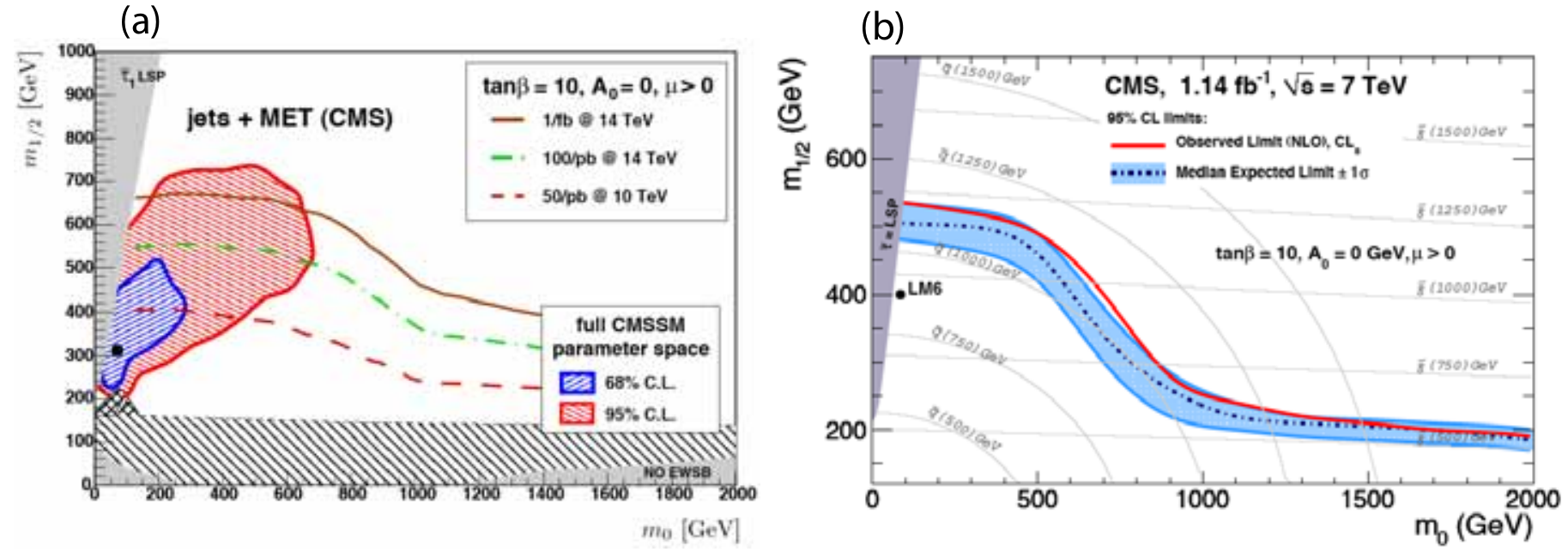}
\caption{(a) Prediction of \cite{Buch} at 68\% and 95\% confidence of 
the unified supersymmetry-breaking parameters $(m_0, m_{1/2})$ of the
constrained Minimal Supersymmetric Standard Model.  (b) 95\% confidence
exclusion region in the parameters $(m_0, m_{1/2})$ in the $\alpha_T$
search for supersymmetry presented by CMS at LP11~\cite{CMSSUSYex}.}
\label{fig:cSUSY}
\end{center}
\end{figure}

Before the startup of the LHC, I expected early discovery of events with
the jets + missing transverse energy signature of supersymmetry.  It did
not happen.   A particularly striking comparison is shown in 
Fig.~\ref{fig:cSUSY}.  On the left, I show the expectation given in 2008 by 
De Roeck, Ellis, and their collaborators for the preferred region of the 
parameter space of the constrained Minimal Supersymmetric Standard Model
(the cMSSM, also known as MSUGRA)~\cite{Buch}.  The red region is the 95\%
confidence expectation.  On the right, I show the
95\% confidence {\it excluded region}
 from one of the many supersymmetry search analyses
presented by CMS at LP11~\cite{CMSSUSYex}.
  No reasonable person could view these
figures together without concluding that we need to change our
 perspective.  But, 
what new perspective is called for?

%%%%%%%%%%%%%%%%%%%%%%%%%%%%%%%%%%%%%%%%%%%%%%%%%%%%%%%%%%%%%%%%%%%%%%%%%
%%%%%%%%%%%%%%%%%%%%%%%%%%%%%%%%%%%%%%%%%%%%%%%%%%%%%%%%%%%%%%%%%%%%%%%%%%%

There was a good reason to expect that SUSY would be found at a 
relatively low energy. We postulate supersymmetry to provide a natural 
explanation of the mass scale of electroweak symmetry breaking.  In 
most models of supersymmetry, the SUSY mass scale is very closely
tied to the mass scale of the Higgs vacuum expectation value.  
In 
SUSY models with the minimal content of Higgs fields, there is a
relation
\beq
     m_Z^2 =  2 { M^2_{H_d} - \tan^2\beta M^2_{H_u}\over \tan^2\beta - 1} - 
        2 \mu^2 \ ,
\eeq{murelation}
where $M^2_{H_d}$, $M^2_{H_u}$ are the SUSY-breaking masses of the 
two Higgs doublets and $\mu$ is the Higgsino mass term.   It is difficult
to understand how the $\mu$ parameter could be even as large as 1 TeV
if the Higgs vacuum expectation value is small enought to put the $Z$ boson
mass at 91 GeV.

The simplest possibility is that all SUSY mass terms are of the 
order of a few hundred GeV.  In some types of SUSY models, it is
possible that only a few of the SUSY mass terms are small.  But this is 
not an option over most of the parameter space of the cMSSM.  In the cMSSM, 
$\mu$ is an output parameter and typically is computed to be one of the largest
masses in the theory.  Then the squarks and gluinos have masses comparable 
to $\mu$ or smaller, leading to very large production cross sections 
at the LHC.  These predicted large cross sections are not observed.

Despite the lack of evidence for SUSY at the LHC so far, SUSY remains a 
very attractive possibility for physics at the TeV scale.  The important 
questions about the TeV scale---the mechanism of electroweak symmetry breaking
and the origin of dark matter---have not gone away.  Supersymmetry offers
solutions to these problems within a theory that includes only weak couplings,
so that all aspects of the theory can be computed in detail.  Supersymmetry
has the additional advantage of providing simple avenues for connecting to 
grand unified theories and string theory at very high mass 
scales~\cite{Martin,mySUSY}. 

However, realizations like the cMSSM or MSUGRA, or, in another class of 
models, minimal gauge mediation,  are tiny subspaces within 
the whole class of supersymmetry models.  They reflect very simple assumptions
about the parameters that break SUSY---assumptions that represent 
the first guesses people made decades ago.  Now the data tells us that these
guesses were incorrect.  We thus need to abandon them 
and acknowledge that, to test SUSY,
 we must search over the full parameter space of the model.

I find strong motivation to consider SUSY 
 parameter regions in which the natural 
expectation from 
\leqn{murelation} that the Higgs mass parameters are near the 100 GeV
scale does not force the strongly interacting superpartners to have masses
of a few hundred GeV.  Actually, SUSY models with this property have been
studied for many years.  In 1996, Cohen, Kaplan, and Nelson introduced the
`more minimal supersymmetric
 Standard Model'~\cite{CKN}, in which only the third
generation sfermions and the gauginos are light, while the first and 
second generation quark and lepton partners are very heavy.  Additional 
models that
share the property of having naturally light Higgsinos along with very 
heavy squarks have been introduced regularly ever since that 
time~\cite{FM,Martincomp,Perelstein}. 

%%%%%%%%%%%%%%%%%%%%%%%%%%%%%%%%%%%%%%%%%%%%%%%%%%%%%%%%%%%%%%%%%%%%%%%%%
\begin{figure}
\begin{center}
\includegraphics[height=2.0in]{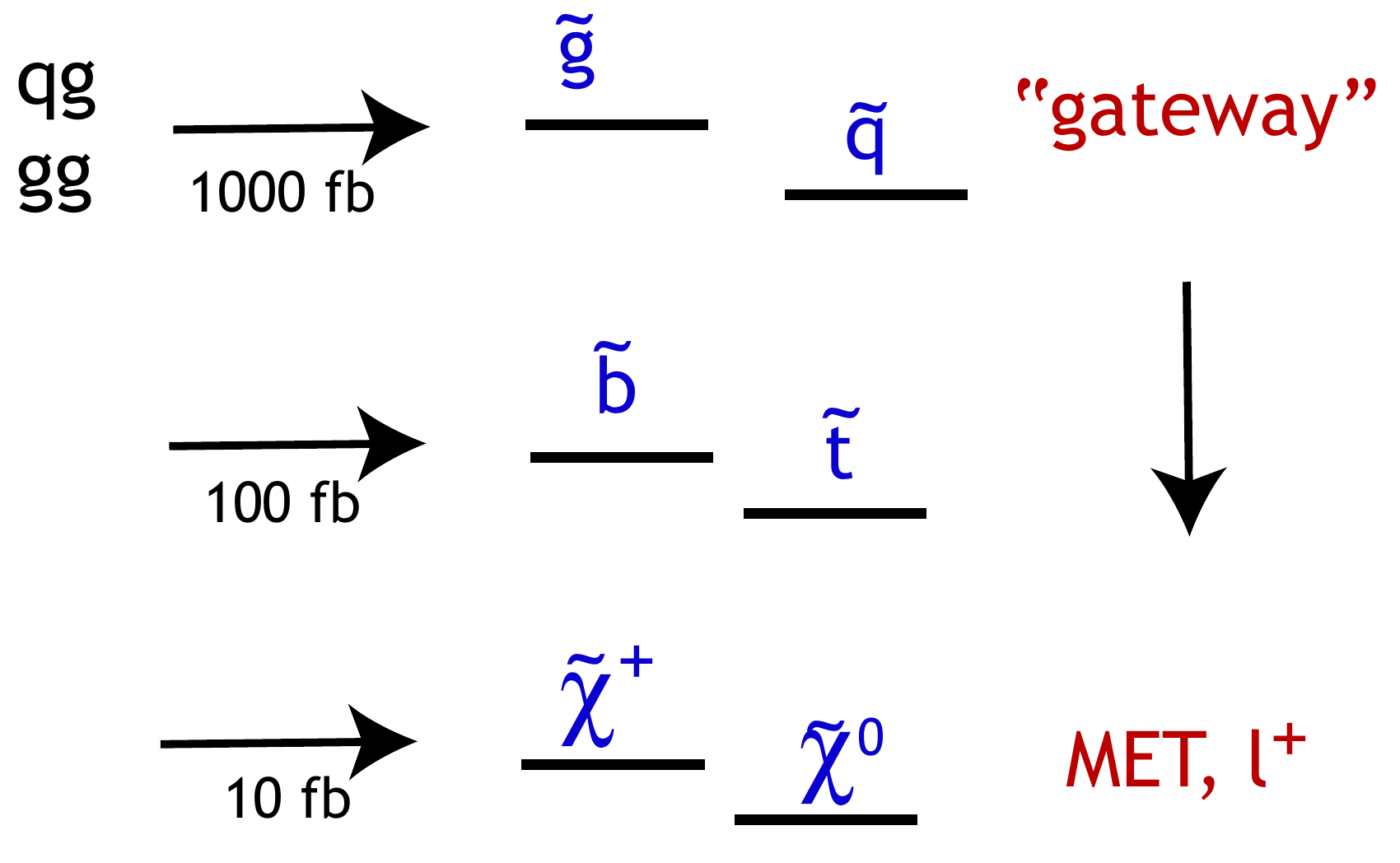}
\caption{Schematic diagram illustrating the SUSY phenomenology at the 
7 TeV LHC.  Quark-gluon reactions can create pairs of SUSY particles,
dominated by particles the highest cross sections that are kinematically
allowed. Eventually, all of these particles decay to charginos and neutralinos,
generating missing transverse energy and other characteristic signatures.}
\label{fig:SUSYmodel}
\end{center}
\end{figure}
%%%%%%%%%%%%%%%%%%%%%%%%%%%%%%%%%%%%%%%%%%%%%%%%%%%%%%%%%%%%%%%%%%%%%%%%%%%

To discuss the experimental signatures of this and related regions of 
SUSY parameter space, it is useful to view SUSY models using the general
schema shown in Fig.~\ref{fig:SUSYmodel}.  There are three ways to 
produce SUSY particles.  First, if the squarks and gluinos are sufficiently
light, these particles can be pair-produced in $gg$ and $qg$ collisions
with cross sections of the order of 1 pb.  The squarks and gluinos 
provide a {\it gateway} into the SUSY world.  Other, color-singlet SUSY
particles are found in the decays of these particles.  By this mechanism,
 the  exotic 
signatures of SUSY 
such as missing transverse energy and like-sign dileptons are
also produced with pb cross section.  This  large production is apparently 
not taking place at the 7 TeV LHC, although we could still see it at the
14 TeV LHC.

Second, the primary means of SUSY particle production could be the production
of heavy particles with suppressed cross sections.  The examples relevant
here are top or bottom scalars.  It is possible that only a single spin-0
color triplet boson (for example, the $\tilde t_L$) is light.  Then the 
cross section for SUSY particle production would be at the 100 fb level.
At luminosities sufficient to access this cross section, SUSY signatures
such as jets plus missing energy will appear.

Finally, it is possible that all of the possible gateway particles are 
too heavy to be produced at the 7 TeV LHC.  Then the primary reactions 
for production of SUSY particles will be the electroweak production of
gauginos and Higgsinos in $q\bar q$ annihilation.  The relevant cross sections
are 10-100 fb.  If the Higgsinos are light, below 200~GeV as we expect from
\leqn{murelation}, we should see pair production of charginos and neutralinos,
yielding trilepton and like-sign dilepton signals, in the 2012 LHC data 
set.

%%%%%%%%%%%%%%%%%%%%%%%%%%%%%%%%%%%%%%%%%%%%%%%%%%%%%%%%%%%%%%%%%%%%%%%%%
\begin{figure}
\begin{center}
\includegraphics[height=2.3in]{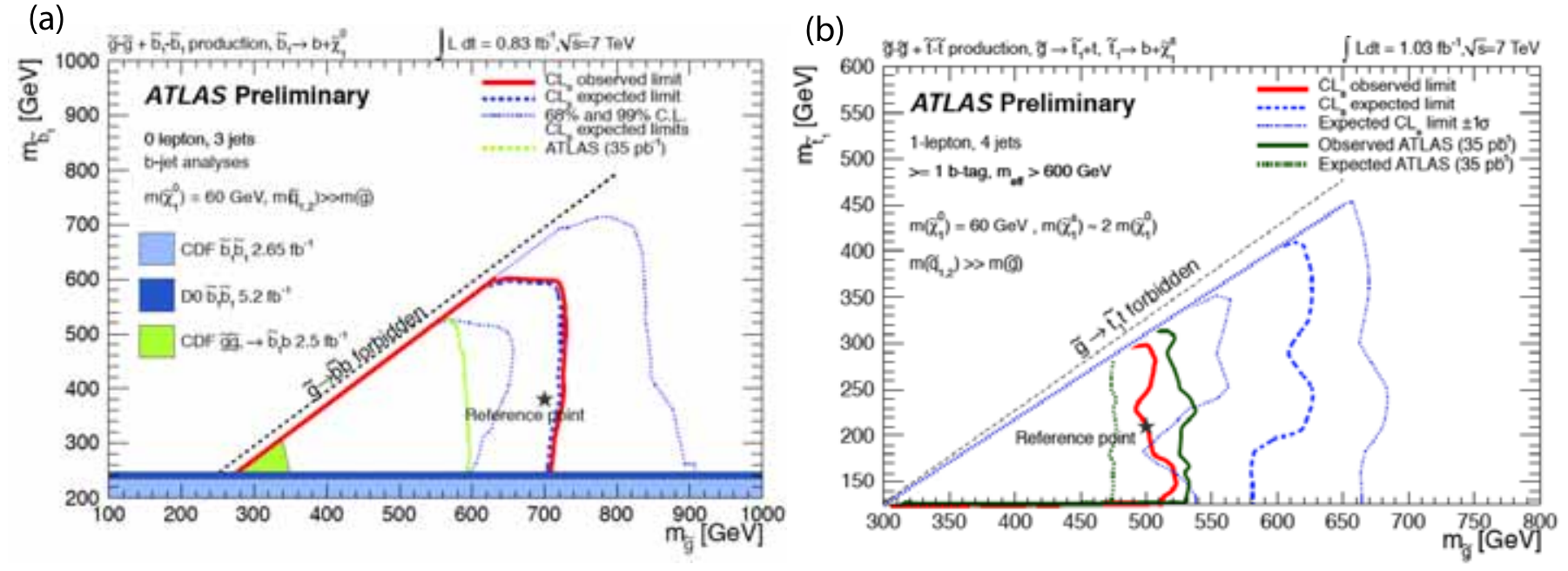}
\caption{95\% exclusion regions in the plane of the gluon 
mass $m_{\tilde g}$ and
the bottom or top quark mass presented by ATLAS at 
LP11~\cite{ATLASSUSYb,ATLASSUSYt}.}
\label{fig:SUSYb}
\end{center}
\end{figure}
%%%%%%%%%%%%%%%%%%%%%%%%%%%%%%%%%%%%%%%%%%%%%%%%%%%%%%%%%%%%%%%%%%%%%%%%%%%

The current limits on models  of the second type just described, 
with only third generation squarks light, are 
quite weak.  The limits presented by ATLAS are shown in
Fig.~\ref{fig:SUSYb}.
For models in which gluinos decay to a $b$ jet plus missing energy, 
Fig.~\ref{fig:SUSYb}(a), the
gluino mass can be anywhere above 700~GeV~\cite{ATLASSUSYb}.  
The vertical cutoff of the 
exclusion region shown in the figure indicates that we are not yet 
sensitive to direct squark pair production.  In scenarios with a lepton
in the final state (for example, from gluino decay to $\tilde t + \bar t$),
the limits on the gluino mass are even weaker, only 500 GeV, as shown in 
Fig.~\ref{fig:SUSYb}(b)~\cite{ATLASSUSYt}.  
The current LHC limits do not 
yet probe the expected regions of parameter space in these scenarios.

As an alternative to exploring new regions of SUSY parameter space, 
we can explore alternatives to SUSY.  As I have explained above, the main
advantage of SUSY is that it gives a complete description of the physics of
electroweak symmetry breaking by weak-coupling interactions that can be 
treated using perturbation theory.  If we give up this property and allow
composite or strongly interacting Higgs bosons, more possibilities open 
up.  These were described at LP11 in the lecture of Gautam 
Bhattacharyya~\cite{Bhattacharyya,Bhattreview}.

It is difficult to make a model of  strong Higgs interactions at the TeV scale
consistent with the constraints of precision electroweak measurements.  It is
easier to make a model in which Higgs bosons are composite at the 10 TeV
scale.  There are several schemes for the construction of such models. 
The most direct approach is that of Little Higgs models~\cite{LH,LHtwo}.  
Alternatively, one can construct Randall-Sundrum models~\cite{RS}, in which a
fifth dimension of space, with constant negative curvative, serves as a 
dual representation of new strong interactions above the 
TeV scale~\cite{RSdual}.   Extra-dimensional theories with
 {\it gauge-Higgs unification}, in which the Higgs bosons arise as the 
higher-dimensional components of gauge fields~\cite{GaugeHiggs}, share
many features and can be considered within the same class of models.  In
all of these models, there is a light composite Higgs boson with properties
similar to those of the Standard Model Higgs, plus new exotic particles
with masses about 1~TeV.

Bhattacharyya explained the general structure of the Higgs potential
that is  found in 
 models of these types.  At the first level of approximation, the composite
Higgs has a flat potential.  A perturbation generates a 
stabilizing $|\phi|^4$ potential for the Higgs field.  The mass term for the
Higgs is generated in the second order of perturbation theory.  There are
specific mechanisms---based on loop corrections involving the top quark
and the large value of the top quark Yukawa coupling---that produce a 
{\it negative} value for the induced mass term~\cite{LHtwo,Hosotani}, 
giving a dynamical explanation for electroweak symmetry breaking.
The final Higgs potential has the form
\beq 
  V =    - {m_t^4\over (16\pi^2)^2 F^2}
      |\phi|^2   + {\alpha_w\over 4\pi} |\phi|^4 \ ,
\eeq{Vfromcomposite}
where $F \sim 1$ TeV is a parameter similar to pion decay constant
$f_\pi$ arising from the 
strong interactions that bind the composite Higgs particles.

All of these models possess symmetries that forbid the generation of a 
Higgs boson mass in leading order.  This means that they must include
new particles that cancel the usual Standard Model quadratic divergences
in the Higgs mass due to $W$, $Z$, and top quark loops.  The models we
are discussing now have no fermion-boson symmetry, so these are massive
particles with the quantum numbers of the Standard Model states.  In
extra-dimensional models, they are the Kaluza-Klein excitations of 
the $W$, $Z$, and $t$ fields.

Among these particles, a very interesting target for searches is the 
partner $T$ of the top quark.  This particle is not a member of a sequential
fourth generation; instead, it is a vectorlike massive fermion of 
hypercharge $\tthird$, or a member of a vectorlike doublet.  Because this
particle does not obtain its mass from electroweak symmetry breaking, 
there is no 
rigorous upper bound to the mass, in contrast to the expectation for a 
fourth-generation quark.  However, 
 to cancel divergences in the Higgs boson self-energy, the mass 
of the $T$ should be 
of the order of 1 TeV.  A typical decay pattern found in models is
\beq 
    T\to b W^+, \ tZ^0 , \  t h^0 \ ,
\eeq{Tdecays}
with branching ratios close to (50\%, 25\%, 25\%).  It is possible that the
Higgs boson could be discovered in decays of the $T$ by searching for 
events with $W \to \ell\nu$ and multiplet $b$-tagged jets.  At the moment,
the LHC experiments have only weak constraints on such a particle; for 
example, CMS constrains a particle decaying by $T\to t Z^0$ only to be 
above 400~GeV~\cite{CMSTtoZ}.

  The $W$ and $Z$ partners predicted by these theories
have a suppressed coupling to light fermions.
Current search limits on heavy $W$ bosons are usually quoted for 
sequential $W$s~\cite{ATLASWsearch,CMSWsearch}.
 In the models I am discussing now, the $W$ and $Z$ partners typically
couple to light quarks with
 10\% of the $SU(2)$ coupling strength, giving single resonance 
production rates of the order of 1\% of those for full-strength 
coupling.
The ATLAS and CMS limits for a $W$ boson with a cross section at 1\% of the 
standard value are about 1.0~TeV, leaving much parameter space to explore
but  giving a good prospect 
for thorough searches for these particles.

The supersymmetric and non-supersymmetric scenarios that I have discussed
in this section are just a few examples of plausible model of electroweak
symmetry breaking that are just beginning to be constrained by the 
LHC data.  In the next year, we will explore possibilities far beyond the
most commonly discussed models of new physics.  We can only guess what is
there.

Before concluding this section, I would like to describe a useful innovation
in the presentation of limits on searches for new particles. Searches based
on specific signatures constrain or, in time, will give evidence for, a wide
variety of models.  It would be of great advantage if the limits or
nonzero signals could be presented in a manner that is as model-independent
as possible.  In a very 
interesting paper, Alwall, Schuster, and Toro have proposed creating 
{\it simplified models} that contain only the particles responsible for the 
particular signature being studied~\cite{AST}.  They showed that this 
approach can be used to analyze more complete and realistic models of 
new physics by demonstrating that the full complexity of supersymmetry
models can be built up by constructing
a sequence of simplified 
models that increase in complexity as more signatures are
included.   In the current situation, where there are no observed signals
of new physics and no preferred models, it makes sense to analyze searches
in simplified models with the very minimum number of particles and parameters.
A compilation of simplified models covering a wide variety of 
possible signatures
of new physics has been presented in \cite{Rouven}.

%%%%%%%%%%%%%%%%%%%%%%%%%%%%%%%%%%%%%%%%%%%%%%%%%%%%%%%%%%%%%%%%%%%%%%%%%
\begin{figure}
\begin{center}
\includegraphics[height=2.2in]{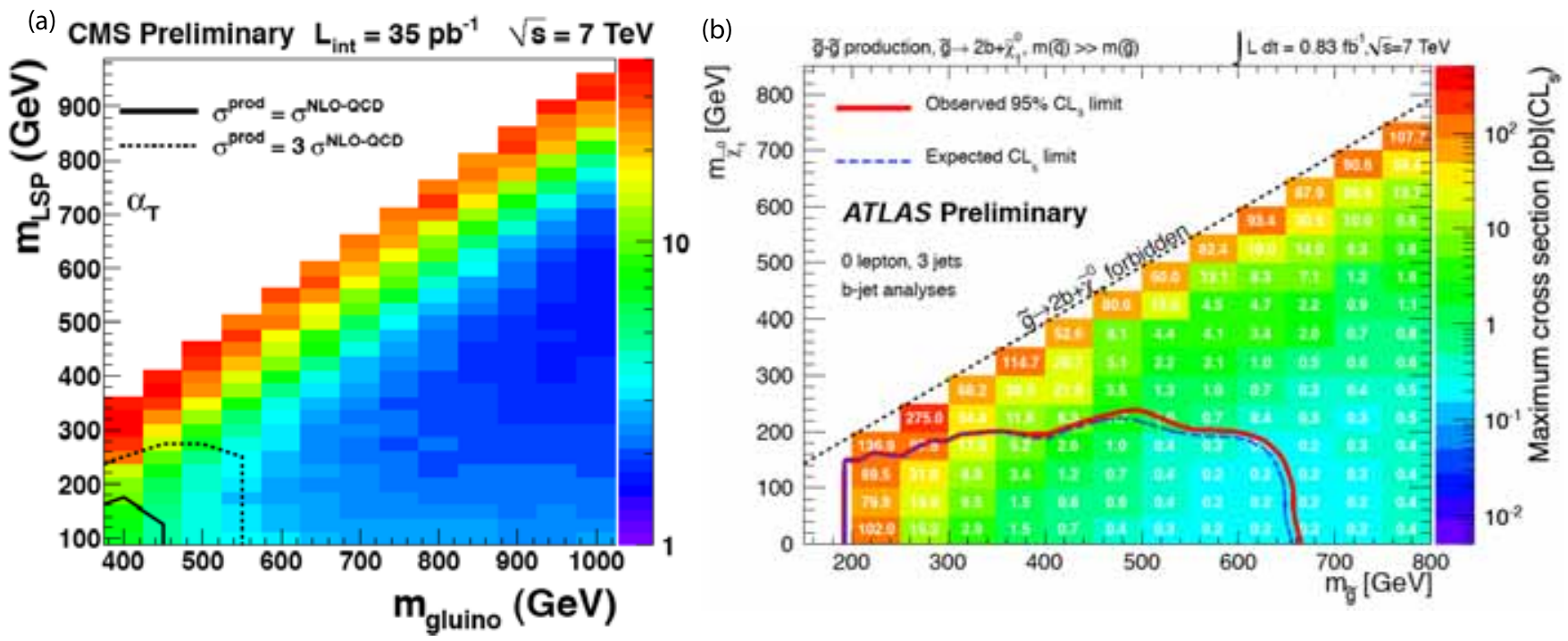}
\caption{Two examples of LHC search analyses giving cross section limits
using the parameters of simplified models: (a) CMS missing energy plus
jets search~\cite{CMSsimple}, (b) ATLAS missing energy plus 
$b$ jet search~\cite{ATLASSUSYb}.}
\label{fig:simple}
\end{center}
\end{figure}
%%%%%%%%%%%%%%%%%%%%%%%%%%%%%%%%%%%%%%%%%%%%%%%%%%%%%%%%%%%%%%%%%%%%%%%%%%%

Figure~\ref{fig:simple} shows the presentation of limits on supersymmetry
cross sections given in the context of simplified models, from 
CMS~\cite{CMSsimple} and ATLAS~\cite{ATLASSUSYb} analyses.  The very 
simple models considered have two relevant masses. The figures show
limits on the observed cross section in the various mass intervals.  The 
CMS result is reported in detail as a set of Root files
 posted at \cite{CMSroot}.
Results presented in this way can be translated into limits on any other
model, supersymmetric or not, that leads to final states of the particular
topology under study.

\section{The Path Forward}

There is one more issue to discuss:  {\it Couldn't it just be the 
Standard Model?}

The answer to this question is, unfortunately, yes.  We know that there 
are phenomena in Nature that the Standard Model cannot explain, most 
prominently, the dark matter of the universe and the nonzero values of 
neutrino masses.  However, there exist very plausible models in which 
the explanations for those phenomena lie ten or more orders of magnitude above
the reach of current accelerators.  Within the realm that we can 
explore directly with particle physics experiments, there is no hint
or anomaly we see today that could not simply disappear with improved
measurements, leaving the 
Standard Model unchallenged.  If the Higgs boson mass is above the 
LEP lower bound of 114~GeV
and below the upper limit from the LHC quoted in Section~9, the 
Standard Model is self-consistent up to very high energies, all the way to 
the Planck scale~\cite{unstable}.
 Thus, a possible outcome of the LHC experiments could 
be the end of experimental particle physics. The LHC experiments would
confirm the 
predictions of the Standard Model at accessible energies, and would 
give no guidance 
as to the location of a higher energy scale at which the Standard Model
might break down.

This would leave us in a terrible situation. All of the questions that we have
today about the properties of particles within the Standard Model would not 
only be left unanswered but would be unanswerable.  In the Standard Model,
the 
 parameters of the theory are renormalizable couplings.  This means that, as a 
matter of principle, these parameters are not computable within the model.
The most obvious difficulty of the model is our inability to understand
the value of the Higgs field vacuum expectation value, or even, as the 
famous naturalness problem is stated~\cite{Susskind}, to understand why 
this value is not 16 orders of magnitude larger. However, within the 
Standard Model, we also cannot understand the values of the gauge boson
couplings, or the values of any of the quark and lepton masses and mixing
angles.

Those who choose to believe
 that the Standard Model is literally true should understand
that this is what they are buying.  There is no fundamental objection to 
this point of view.  As I have explained in Section 2, this
viewpoint is made 
intellectually respectable by theory of inflation, in particular, 
the idea that the equations of motion that we observe
are 
true only locally in our small patch of the universe.  The path that led to
the Standard Model could have been determined in the very early history of the 
universe through mechanisms that are invisible to us today.

This is a thoroughly deplorable prospect. But every particle 
physicist needs to confront these ideas  and ask: Do I think that this 
is how Nature works?

There is an alternative point of view.  We do not know whether it correct.
Nature will choose.

That point of view is the optimism that the physics of the Higgs field
and electroweak symmetry breaking has a mechanism, and that that mechanism
will be visible to our experiments at the TeV scale.  There is a 
compelling justification 
for accepting 
this idea:  Only people who believe in it can make the
discovery that it is true.

It is a good time for optimism.  With the LHC, we have a new engine that
will produce the data that we need to find the explanations.  The LHC
gives us the power, over the course of its long program to 1000 fb$^{-1}$
luminosity samples and beyond, to uncover evidence of any variant of 
physics explanations for the TeV scale.  We can harness this power
to look in every corner, under every stone, to find the clue that will 
unlock the secrets of the Higgs boson and its partners.

The measurements that will 
turn out to be the most important might not be the 
ones that we expect.  We will need persistence and patience.  But, if we
are right, we have the chance before us to discover an entirely new level
in the fundamental laws of physics.

\begin{figure}
\begin{center}
\includegraphics[height=3.0in]{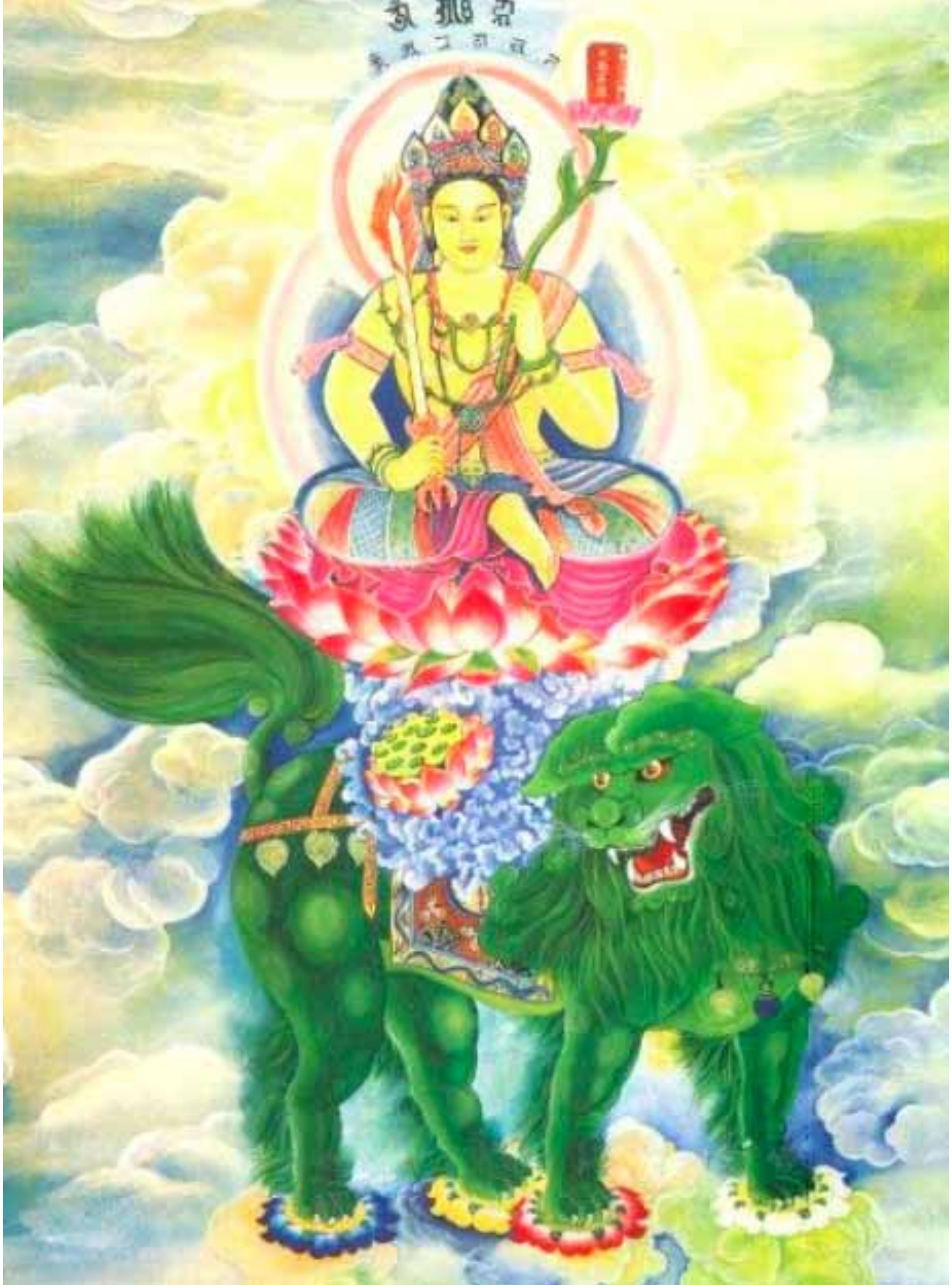}
\caption{Manjusri, the bodhisattva of indomitable wisdom,
protrayed riding on a lion~\cite{Manjusricredit}.}
\label{fig:Manjusri}
\end{center}
\end{figure}

I leave you with the image in Fig.~\ref{fig:Manjusri}.  This is Manjusri,
in Buddhist teaching,
the bodhisattva associated with wisdom triumphant over all obstacles.
In his traditional portrait, he has an impassive expression, above all
trivial concerns, and he
rides on a lion.  We also have a lion, the LHC, to carry us forward.

%%%%%%%%%%%%%%%%%%%%%%%%%%%%%%%%%%%%%%%%%%%%%%%%%%%%%%%%%%%%%%%%%%%%%%%%%
%%%%%%%%%%%%%%%%%%%%%%%%%%%%%%%%%%%%%%%%%%%%%%%%%%%%%%%%%%%%%%%%%%%%%%%%%%%

We will meet again at the Lepton Photon conference of 2013.  We look 
forward to the prospect that the many issues left unsettled in this 
year's report will be informed by the new data that is about to appear.
We wait expectantly to see what next year's experiments will say.

\Acknowledgements

I am grateful to Naba Mondal, Rohini Godbole, and the other organizers
of LP11 for inviting me to this very stimulating conference and for 
providing very pleasant arrangements in Mumbai for all of the participants.
I thank the speakers at LP11 and others whose work is reviewed here for 
patiently answering my questions, however persistent.  I am grateful to 
my colleagues at SLAC for discussions of many of the points raised in this
lecture.  These people tried 
as hard as they could to give me the best perspective, so whatever 
errors remain are my responsibility.  Finally, I thank the
US Department of Energy for supporting me and the SLAC HEP Theory Group,
under  contract DE--AC02--76SF00515.

\end{document}

%% file: mymacros.tex
\def\beq{\begin{equation}}
\def\eeq#1{\label{#1}\end{equation}}
\def\eeqn{\end{equation}}

\newenvironment{Eqnarray}%
   {\arraycolsep 0.14em\begin{eqnarray}}{\end{eqnarray}}
\def\beqa{\begin{Eqnarray}}
\def\eeqa#1{\label{#1}\end{Eqnarray}}
\def\eeqan{\end{Eqnarray}}

\def\leqn#1{(\ref{#1})}

\let\bar=\overbar

\def\lsim{\mathrel{\raise.3ex\hbox{$<$\kern-.75em\lower1ex\hbox{$\sim$}}}}
\def\gsim{\mathrel{\raise.3ex\hbox{$>$\kern-.75em\lower1ex\hbox{$\sim$}}}}

\def\tthird{\frac{2}{3}}

\def\del{\partial}
\def\Dslash{\not{\hbox{\kern-4pt $D$}}}
\def\dslash{\not{\hbox{\kern-2pt $\del$}}}
\def\pslash{\not{\hbox{\kern-2pt $p$}}}
\def\qslash{\not{\hbox{\kern-2pt $q$}}}
\def\kslash{\not{\hbox{\kern-2pt $k$}}}
\def\oneslash{\not{\hbox{\kern-2pt $1$}}}
\def\twoslash{\not{\hbox{\kern-2pt $2$}}}
\def\threeslash{\not{\hbox{\kern-2pt $3$}}}
\def\fourslash{\not{\hbox{\kern-2pt $4$}}}

\def\Pl{{\mbox{\scriptsize Pl}}}

\def\ee{e^+e^-}

\def\msb{{\bar{\scriptsize M \kern -1pt S}}}

\def\drb{{\bar{\scriptsize D \kern -1pt R}}}

\def\apb#1 {  \langle #1 ] }

\makeatletter
\def\section{\@startsection{section}{0}{\z@}{5.5ex plus .5ex minus
 1.5ex}{2.3ex plus .2ex}{\large\bf}}
\def\subsection{\@startsection{subsection}{1}{\z@}{3.5ex plus .5ex minus
 1.5ex}{1.3ex plus .2ex}{\normalsize\bf}}
\def\subsubsection{\@startsection{subsubsection}{2}{\z@}{-3.5ex plus
-1ex minus  -.2ex}{2.3ex plus .2ex}{\normalsize\sl}}

\renewcommand{\@makecaption}[2]{%
   \vskip 10pt
   \setbox\@tempboxa\hbox{\small #1: #2}
   \ifdim \wd\@tempboxa >\hsize     % IF longer than one line:
       \small #1: #2\par          %   THEN set as ordinary paragraph.
     \else                        %   ELSE  center.
       \hbox to\hsize{\hfil\box\@tempboxa\hfil}
   \fi}

 \def\citenum#1{{\def\@cite##1##2{##1}\cite{#1}}}
\newcount\@tempcntc
\def\@citex[#1]#2{\if@filesw\immediate\write\@auxout{\string\citation{#2}}\fi
  \@tempcnta\z@\@tempcntb\m@ne\def\@citea{}\@cite{\@for\@citeb:=#2\do
    {\@ifundefined
       {b@\@citeb}{\@citeo\@tempcntb\m@ne\@citea\def\@citea{,}{\bf ?}\@warning
       {Citation `\@citeb' on page \thepage \space undefined}}%
    {\setbox\z@\hbox{\global\@tempcntc0\csname b@\@citeb\endcsname\relax}%
     \ifnum\@tempcntc=\z@ \@citeo\@tempcntb\m@ne
       \@citea\def\@citea{,}\hbox{\csname b@\@citeb\endcsname}%
     \else
      \advance\@tempcntb\@ne
      \ifnum\@tempcntb=\@tempcntc
      \else\advance\@tempcntb\m@ne\@citeo
      \@tempcnta\@tempcntc\@tempcntb\@tempcntc\fi\fi}}\@citeo}{#1}}
\def\@citeo{\ifnum\@tempcnta>\@tempcntb\else\@citea\def\@citea{,}%
  \ifnum\@tempcnta=\@tempcntb\the\@tempcnta\else
  {\advance\@tempcnta\@ne\ifnum\@tempcnta=\@tempcntb \else\def\@citea{--}\fi
    \advance\@tempcnta\m@ne\the\@tempcnta\@citea\the\@tempcntb}\fi\fi}
\makeatother